\documentclass[11pt]{article}

\usepackage{calc}
\usepackage{xspace}
\usepackage{subfig}
\usepackage{amsmath}
\usepackage{amssymb}
\usepackage{amsfonts}
\usepackage{mathrsfs}
\usepackage{enumerate}
\makeatletter
\@addtoreset{equation}{section}

\makeatother
\usepackage{feynmf}
\usepackage{graphicx}
\usepackage{amssymb,amsfonts,amsmath}
\usepackage{accents}

\def\beq{\begin{equation}}
\def\eeq{\end{equation}}
\def\bea{\begin{eqnarray}}
\def\eea{\end{eqnarray}}

\newcommand{\gd}{\mathrm{gd}^{-1}}

\newcommand{\ai}{\mathrm{Ai}}

\begin{document}

\renewcommand{\thefootnote}{\fnsymbol{footnote}}
\newcommand{\inst}[1]{\mbox{$^{\text{\textnormal{#1}}}$}}
\begin{flushright}
EPHOU 17-009\\
June, 2017
\end{flushright}
\begin{center}
{\LARGE  An Alternative Lattice Field Theory Formulation Inspired by Lattice Supersymmetry }\\[4ex]
%
{\large
Alessandro D'Adda\inst{a}\footnote{\texttt{dadda@to.infn.it}},
Noboru Kawamoto\inst{b}\footnote{\texttt{kawamoto@particle.sci.hokudai.ac.jp}}\\[1ex]
{\normalsize and}
Jun Saito\inst{c}\footnote{ \texttt{jsaito@obihiro.ac.jp}}. }\\[3ex]
%
{\large\itshape
\inst{a} INFN Sezione di Torino, and\\
Dipartimento di Fisica Teorica,
Universita di Torino\\
I-10125 Torino, Italy\\[3ex]
\inst{b} Department of Physics, Hokkaido University\\
Sapporo, 060-0810 Japan\\[3ex]
\inst{c} Department of Human Science,\\
Obihiro University of Agriculture and Veterinary Medicine\\
Obihiro, 080-8555 Japan}
\end{center}
\bigskip
\setcounter{footnote}{0}
\renewcommand{\thefootnote}{\arabic{footnote}}

\begin{abstract}
We propose an unconventional formulation of lattice field theories which
is quite general, although originally motivated by the quest of exact
lattice supersymmetry. Two long standing problems have a solution in this
context: 1) Each degree of freedom on the lattice corresponds to $2^d$
degrees of freedom in the continuum, but all these doublers have (in the
case of fermions) the same chirality and can be either identified, thus
removing the degeneracy, or, in some theories with extended supersymmetry,
identified with different members of the same supermultiplet. 2) The
derivative operator, defined on the lattice as a suitable periodic
function of the lattice momentum, is an addittive and conserved quantity,
thus assuring that the Leibniz rule is satisfied. This implies that the
product of two fields on the lattice is replaced by a non-local ``star
product'' which is however in general non-associative. Associativity of the
``star product'' poses strong restrictions on the form of the lattice
derivative operator (which becomes the inverse Gudermannian function of
the lattice momentum) and has the consequence that the degrees of freedom
of the lattice theory and of the continuum theory are in one-to-one
correspondence, so that the two theories are eventually equivalent. 
We can show that the non-local star product of the fields effectively 
turns into a local one in the continuum limit. 
Regularization of the ultraviolet divergences on the lattice is not
associated to the lattice spacing, which does not act as a regulator, but
may be obtained by a one parameter deformation of the lattice derivative,
thus preserving the lattice structure even in the limit of infinite
momentum cutoff. However this regularization breaks gauge invariance and a
gauge invariant regularization within the lattice formulation is still
lacking.
\end{abstract}




\section{Introduction}
\label{0.1}
We have been asking ourselves the question: ``If we stick to keeping  exact supersymmetry (SUSY) on the lattice, what kind of lattice formulation is required ?" 
We have reached the following conclusion: ``We need a non-local lattice field theory formulation which does not have the lattice chiral fermion problem." 
This formulation must be very general  in character and  it must be applicable to non-SUSY lattice theories, so we end up with eventually  proposing an alternative lattice field theory formulation which does not have the chiral fermion problem. There is a general belief that a non-local field theory is meaningless and thus this possibility is never considered seriously. However here we propose a lattice field theory which is non local on the lattice but is well defined, and it maintains exactly the symmetries of the corresponding continuum theory.  

Since the fundamental lattice chiral fermion problem \cite{NiNi, NiNi2, KarSm} was 
posed it took a many years struggle to find the complete solution for 
lattice QCD \cite{chiral-ferm, chiral-ferm2, chiral-ferm3, chiral-ferm4}. Avoiding the difficulty of the No-Go theorem 
of lattice chiral fermion, a modified lattice version of chiral transformation 
which is compatible with Ginsparg-Wilson relation \cite{GW-relation} was proposed 
as a solution. The overlap fermion operator satisfying Ginsparg-Wilson 
relation was found and shown to be local \cite{Neuberger,HJL}. 

In the lattice chiral fermion formulation the Ginsparg-Wilson relation played a crucial 
role in providing a criteria of how much  breaking of the lattice chiral transformation is 
allowed for the chiral symmetry breaking formulation in the renormalization group 
flow of ultraviolet regime. It turned out that the breaking effects from the continuum 
chiral symmetry can be confined in local irrelevant terms. 

A long time has gone past since the problem of lattice supersymmetry(SUSY) were posed    
\cite{Dondi-Nicolai} for the first time and we still have not  reached  a complete solution. 
We may wonder: ``Why is the solution for lattice 
SUSY so difficult ?" We consider that there are several reasons for this 
difference.  

There are two fundamental difficulties for a realization of exact SUSY on the 
lattice: 
\begin{itemize} 
\renewcommand{\labelitemi}{$A):$}
\item Breakdown of Leibniz rule for the difference operator. 
\renewcommand{\labelitemi}{$B):$}
\item Chiral fermion species doubler problem. 
\end{itemize}
$A):$ In the SUSY algebra a bilinear product of supercharges is equal to 
a differential operator which should be replaced by a local difference operator 
on the lattice. The difference operator breaks the distributive law with respect to the product, namely the Leibniz rule,  while supercharges satisfy the same rule, thus leading 
to a breakdown of the SUSY algebra \cite{Dondi-Nicolai, Nojiri, Nojiri2, Fujikawa1}. 

It was shown that there is no lattice derivative operator which is locally defined and 
satisfies the Leibniz rule exactly \cite{ Kato-Sakamoto-So1}. 
A Ginsparg-Wilson type analyses of blocking transformation for lattice SUSY gave 
a similar result \cite{Bergner-Bruckmann}: the only solution which is consistent 
with lattice SUSY version of Ginsparg-Wilson relation is the SLAC 
derivative \cite{SLAC} which is non-local. 
These results suggest that the breaking effects of SUSY algebra with the local 
difference operator are considered to be non-local in nature. These breaking 
effects, however, may appear only in the part of SUSY algebra that include 
the derivative operator, not in the nilpotent part of an extended SUSY algebra. 
This consideration suggests that we need to accept non-local formulation of 
lattice SUSY if we stick to find a exact lattice SUSY formulation for all super 
charges of extended SUSY. We think that this would be a most prohibited barrier 
which one may not dare to go over.  
In this paper we first establish to formulate a non-local lattice field theory 
which is equivalent to a corresponding local continuum theory. The formulation 
was inspired by the lattice SUSY formulation \cite{DFKKS,DKKS}.  

As far as the nilpotent part of extended SUSY algebra is concerned 
exact lattice SUSY formulations have been successfully constructed by 
various methods:  1) Nicolai mapping \cite{Nicolai-map, Nicolai-map2}, 
2) Orbifold construction \cite{Kaplan1, Kaplan1-2, Kaplan1-3, phyrep}, 3) Q-exact topological field 
theory \cite{Catterall1, Sugino, Sugino2, Sugino3}.   It was shown that Nicolai mapping is closely 
related to the Q-exact formulation of topological field theory \cite{phyrep}. 
The nilpotent part of super algebra can thus be realized exactly on the lattice 
within a local lattice field theory.   

There was a challenge to realize exact lattice SUSY for all supercharges 
by modifying the Leibniz rule of super charges in such a way  to be compatible with the 
breaking terms of the lattice difference operator \cite{DKKN, DKKN2, DKKN3}. 
Although an ordering ambiguity of this formulation (link approach) was pointed out 
in \cite{Bruckmann-Kok, Bruckmann-Kok2}, it was recognized later that the introduction 
of noncommutativity solves this problem \cite{DFKKS}. 
 Algebraic consistency of this formulation with noncommutativity was confirmed 
 in the framework of Hopf Algebra \cite{Dadda-Kawamoto-Saito1}. 
This link approach with a particular choice of parameter coincides with the 
orbifold construction of lattice super Yang-Mills. 
The relation between these two formulations was made clear in 
\cite{Damgaard-Matsuura, Damgaard-Matsuura2, Damgaard-Matsuura3, Damgaard-Matsuura4}. 
It has been explicitly shown in the link approach that Q-exact lattice SUSY 
formulation is essentially the lattice version of continuum twisted super 
Yang-Mills formulation via Dirac-K\"{a}hler twisting procedure \cite{twist-susy, twist-susy2, twist-susy3}. 
Although the lattice SUSY formulation of the link approach was based on the 
local formulation, noncommutativity is needed for the Hopf algebraic consistency.

There may still be a room to modify the local difference operator in such a way 
that surface terms can be canceled out for lattice total 
derivatives to preserve an action symmetry \cite{Kato-Sakamoto-So2}.  
In this sense modified SUSY can be realized in some model for the nilpotent 
part of extended SUSY algebra. 

In any of these local lattice field theory formulations only the nilpotent part of 
extended SUSY algebra is exactly kept. We claim that a non-local lattice field theory 
formulation is unavoidable to formulate an exact lattice SUSY for all supercharges. 

$B):$ In general the lattice chiral fermion problem is considered to be a different issue 
from lattice SUSY problem. One may naively expect that the lattice chiral fermion solution 
of lattice QCD can be used for lattice SUSY formulation of chiral 
fermions \cite{So-Ukita, So-Ukita2, Kikukawa-Nakayama}. 
From exact lattice SUSY point of view we consider that it is not so simple and 
the exact lattice SUSY may not be 
realized since the fermion propagator does not have simple relation with the boson propagator 
in contrast with the continuum case. In fact bosonic Wilson terms were needed to match 
the Wilson term of the fermionic sector and get a correct quantum level Ward-Takahashi identity for 
lattice SUSY Wess-Zumino model \cite{Bartels-Kramer,GKPY}. These examples show that 
the modification of the fermion propagator requires the modification of the corresponding boson 
propagator to fulfill a quantum level consistency of lattice SUSY. 

In this paper we take a totally different point of view from the common approach on the lattice 
chiral fermion problem.     
The lattice regularization of chiral fermions unavoidably generates species doublers, but in our non local lattice formulation, that follows from the request of keeping the Leibniz rule, they have the same chirality and are matched by a similar doubling phenomenon in the bosonic sector. 
So the doublers can be either identified, thus removing the degeneracy and providing a consistent truncation procedure for doublers for non-SUSY lattice formulation,  or, in the case of some extended supersymmetric theories, they can be interpreted 
as different members of a supermultiplet.
Closely related to the existence of the species doublers is the half lattice structure introduced in this paper. This has a 
geometrical correspondence 
with lattice SUSY algebra where  a half lattice translation generates  a SUSY transformation \cite{DFKKS,DKKS}. 
In fact also the link approach of the lattice super Yang-Mills was based on this 
geometric and algebraic correspondence \cite{DKKN, DKKN2, DKKN3}.   

In this paper we formulate lattice field theory in the momentum representation since the 
lattice Leibniz rule and the species doubler d.o.f. can be more easily described in  
momentum space. This is not new as
there have been already some examples of formulation of lattice theories in momentum representation \cite{Kadoh-Suzuki}.
Within our formulation we find a particular choice of blocking transformation of the Ginsparg-Wilson type which surprisingly corresponds to a blocking transformation from continuum to lattice. 
In this way we obtain a lattice formulation which is equivalent to the corresponding continuum 
theory and thus all the symmetries are kept on the lattice including lattice SUSY.
However the formulation is non-local in the coordinate space and not yet regularized 
even though it is a lattice formulation, in fact the lattice spacing does not act here as a regulator.
 Regularization can however be obtained  by modifying  the lattice derivative,
and the lattice structure of the theory can be preserved even in the limit of infinite
momentum cutoff. However in gauge theories  this regularization breaks gauge invariance and a gauge invariant regularization within the lattice formulation is still
lacking.

This paper is organized as follows: In section \ref{1.1} we review, from a slightly unconventional point of view, how the chiral fermion problem and the violation of the Leibniz rule arises in the standard approach.   In section \ref{1.2}  we explain the basic ideas of this paper, namely how the Leibniz rule may be restored and the doubling problem avoided by replacing the usual local product on the lattice with a non local product, the star product, that originates from requiring the conservation of the derivative operator on the lattice. In section \ref{1.3}  we study the properties of the star product in particular with respect to the issues of associativity and locality. We prove that an associative star product can be defined with a suitable choice of the derivative operator on the lattice, and that in that case an invertible map exists between the degrees of freedom in the continuum and the ones on the lattice.
In section \ref{1.4} we show how a lattice action can be obtained from the one of the continuum theory by a blocking transformation induced by the aforementioned map. In this action, which is classically equivalent to the continuum action, the lattice spacing does not act as a regulator and a renormalization procedure for the ultraviolet divergences is needed as in the continuum theory. This is discussed in section \ref{section6} for two simple examples: the non interacting supersymmetric Wess-Zumino model in four dimensions and the $\Phi^4$ theory in four dimensions. It is shown that in the latter case a renormalization scheme can be defined that preserves the lattice structure. Some conclusions and discussions are given in section \ref{1.7}.

\section{Conventional lattice: violation of the Leibniz rule and the doublers problem}
\label{1.1}
The approach to lattice theories that we develop in the present paper was motivated by the attempt to construct a lattice theory in which supersymmetry is exactly realized. In ordinary lattice theories there are two major obstacles to exact supersymmetry. The first is that on a lattice the derivative operator is replaced by a finite difference operator ( or some other ultra-local operator ) which does not satisfy the Leibniz rule. Since supersymmetry transformations contain derivatives, the violation of the Leibniz rule poses a serious problem to an exact formulation of supersymmetry on the lattice. The second obstacle is the so called doubling of fermions on the lattice. This is essentially the chiral fermion problem: a chiral fermion cannot be put on a $d$ dimensional cubic lattice without introducing  $2^d - 1$  copies  of it  ( the doublers). Of the resulting $2^d$ states, half have the same chirality of the original fermion, and half the opposite one.  This proliferation of fermions in a supersymmetric theory would upset the balance between bosons and fermions making exact supersymmetry on the lattice impossible. 

In this section we shall review how these problems arise in the conventional lattice formulation, and look at them from a slightly different point of view with the aim of understanding how and under what conditions they could be overcome.
We shall mostly make use of the momentum representation where the root of the above problems can be better understood.

Consider a set of fields $\varphi_A(x)$ defined on a regular lattice\footnote{Here and in the following we shall use small greek letters like $\varphi$ to denote fields on a lattice and capital letters like $\Phi$ to denote fields in the continuum} with lattice spacing $l$, namely\footnote{The use of the letter $l$ to denote the lattice spacing in place of the standard notation $a$ is not accidental and will be explained shortly}:
\beq
x:~~~~~~~x^\mu  = n^\mu l,       \label{regularlattice} \eeq
with $n^\mu$  integer numbers.
The discrete  Fourier transform of $\varphi_A(x)$ produces the momentum representation $\tilde{\varphi}_A(p)$ of the fields. In the momentum representation the lattice structure appears as a periodicity in momentum space, namely all lattice fields are invariant under
\beq
p_\mu \rightarrow p_\mu + \frac{2\pi}{l} k_\mu~~~~~~~~\mu=(1,2,\cdots,d), \label{momentumperiodicity} \eeq
with $k_\mu$ arbitrary integer.
Similarly all physical operators, such as for instance the derivative operator, must be described by functions with the same periodicity (\ref{momentumperiodicity}).
This means that  a d-dimensional regular lattice with lattice spacing $l$ is described by a momentum space which is a d-dimensional torus with period $\frac{2 \pi}{l}$ in each dimension.
Each momentum component on the lattice is then an angular variable, and momenta that differ by multiples of $\frac{2 \pi}{l}$ are indistiguishable.
Instead in continuum theories each component of the momentum is an arbitrary real number ranging from $-\infty$ to $+\infty$  and momentum space is the non-compact $\mathbb{R}^d$ variety. 

The momentum space  corresponding to a regular lattice and the one corresponding to a continuum space-time are  then topologically different varieties, which means that there is no smooth one-to-one  map between the two. 
A map however, albeit not a one-to-one smooth correspondence,  should be established as the continuum theory should   be recovered from the lattice theory as the lattice spacing goes to zero. 
We discuss  below how the topological obstruction to a one-to-one smooth correspondence between lattice and continuum momentum space is  at the root of both of the chiral fermion problem and of the impossibility of finding on the lattice a derivative operator that satisfies the Leibniz rule. 

In the continuum theory momentum conservation follows from the invariance of the theory under translations.  The natural lattice counterpart of  translational invariance is the invariance under the descrete group of displacements that map the lattice into itself, namely the displacements which are integer multiples of $l$ in each direction.  It is natural then to assume that the invariance of the lattice theory under such displacements reproduces the ordinary translational invariance in the limit $l \rightarrow 0$. There is however an obstruction to such naive correspondence: the derivative operator is replaced on the lattice by a finite difference operator, and if we require the latter to be hermitian it has to be left-right symmetric and necessarily involves a finite difference over two lattice spacings, namely in a one dimensional example: 
\beq
 \Delta_s\varphi(x) = \frac{1}{2l} \left( \varphi(x+l) - \varphi(x-l) \right). \label{symmderiv} \eeq
If we take $\Delta_s$ as the lattice correspondent of the derivative operator, namely of the generator of infinitesimal translations, then the smallest displacement that corresponds to a translation in the continuum limit has not spacing $l$ but $2l$.

We are led then to introduce two distinct concepts: the lattice spacing $l$, that denotes the spacing between two neighboring sites of the lattice, and the ``effective lattice spacing", for which we shall use the standard notation $a$, that denotes the smallest displacement on the lattice that corresponds to an infinitesimal translation in the continuum limit. 
In general we have:
\beq
a = n l, \label{latticespacing} \eeq
with $n$ integer. With the symmetric choice (\ref{symmderiv}) of the lattice derivative operator we have $n=2$, while the value $n=1$  occurs if the derivative operator on the lattice is defined as a finite difference over one lattice spacing.  This however leads to an ambiguity, since with a finite difference over one lattice spacing it is possible to define two  hermitian conjugate operators, the right and left difference operators, which we shall denote by $\Delta_{\pm}$ and are given by:
\beq
\frac{\partial \Phi(x)}{\partial x} \rightarrow \Delta_{\pm}\varphi(x) =     
  \pm \frac{1}{l} \left( \varphi(x\pm l) - \varphi(x) \right). \label{rightleftderiv} \eeq

 Although $\Delta_{\pm}$ are not hermitian ( unlike their correspondent operator $i\partial$ in the continuum)  we can construct a hermitian quadratic operator $\Delta_{+\mu}\Delta_{-\mu}$ which becomes $\partial_\mu\partial^\mu$ in the continuum limit and can be used to construct a lattice  lagrangian of a free boson.
So, as far as free bosons are concerned,  $\Delta_{\pm}$ (with $n=1$ ) is a possible choice for the derivative operator on the lattice.

Instead, the fermionic inverse propagator  is linear in the derivatives, and the only linear hermitian combination of $\Delta_{\pm}$ is $\Delta_s = \frac{1}{2} (\Delta_+ + \Delta_- )$. So  for any theory containing fermions  the symmetric difference operator  $\Delta_s$ has to be used as derivative operator on the lattice\footnote{More general choices for the derivative operator on the lattice will be introduced further in the paper as an essential ingredient of the present formulation, but in all of them hermiticity will enforce the condition $n=2$.}. Hence  $n=2$ is required  in (\ref{latticespacing}),  and this leads  to the so called fermion doubling phenomenon as it  will be discussed shortly.

The correspondence between translations in the continuum and displacements of multiples of $a=nl$ on the lattice determines the map between the momentum $p_\mu$ on the lattice and the momentum $\hat{p}_\mu$ in the continuum.
In fact while translational invariance implies momentum conservation in the continuum, on the lattice the invariance under displacements of $a$ in each direction also implies the conservation of the momentum on the lattice but only modulo $\frac{2 \pi}{a}$,  because of the discrete nature of the translational symmetry. 
On the other hand the momentum $p_\mu$ on the lattice and the momentum $\hat{p}_\mu$ in the continuum are both conserved quantities associated to the invariance respectively under descrete and continuum translations and should then be identified modulo  $\frac{2 \pi}{a}$. This provides the following relation between $p_\mu$ and $\hat{p}_\mu$:
\beq
\hat{p}_\mu - p_\mu  = \frac{2\pi}{a} k_\mu~~~~~~~~~~-\frac{ \pi}{l} < p_\mu < \frac{ \pi}{l},  \label{momentumrel} \eeq
with $k_\mu$ arbitrary integers.
In eq. (\ref{momentumrel}) the lattice momentum $p_\mu$, being an angular variable according to (\ref{momentumperiodicity}), is restricted to take values in the fundamental region (the Brillouin zone) of size $\frac{2\pi}{l}$ .  Eq. (\ref{momentumrel}) defines a map between the momentum space $\hat{\mathcal{P}}$ of the continuum theory and the momentum space $\mathcal{P}$ on the lattice defined by (\ref{momentumperiodicity}).
In $d$ dimensions  $\mathcal{P}$ is a $d$-dimensional torus, whereas $\hat{\mathcal{P}}$ is a non-compact $\mathbb{R}^d$ variety, so the map defined by  (\ref{momentumrel}) is not a one-to-one correspondence. It is clear in fact from  (\ref{momentumrel}) that  a point of $\mathcal{P}$, which is defined by the set of coordinates $p_\mu$ with  $-\frac{ \pi}{l} < p_\mu < \frac{ \pi}{l}$, has an infinite number of images  in   $\hat{\mathcal{P}}$ which are labeled by the integers $k_\mu$.
This means that, given a configuration on the lattice with momentum $p_\mu$ within the Brillouin zone, the corresponding configuration in the continuum is in general the superposition of configurations with arbitrarily high momenta corresponding to the possible choices of $k_\mu$ in (\ref{momentumrel}).

On the other hand, if we consider a point of $\hat{\mathcal{P}}$ with coordinates $\hat{p}_\mu$ the number of its images in  $\mathcal{P}$ depends on the size of the effective lattice spacing $a$ and is in fact equal to the integer $n$ in eq. (\ref{latticespacing}).
This is because the manifold $\mathcal{P}$ is defined by the periodicity condition (\ref{momentumperiodicity}) with period $\frac{2\pi}{l}$, whereas  $\frac{2\pi}{a}$ is involved in (\ref{momentumrel}). 

In fact,  given a configuration in the continuum with momentum  $\hat{p}_\mu$, if $a=l$, namely if $n=1$, there is only one value of the lattice momentum $p_\mu$ within the Brillouin zone for which  (\ref{momentumrel}) is satisfied with a suitable choice of $k_\mu$.
On the other hand if $a=2l$, namely if $n=2$, given an arbitrary momentum   $\hat{p}_\mu$  in the continuum, for each value of $\mu$ there are  two different values of $p_\mu$, separated  by $\frac{\pi}{l}$ and both in the interval $-\frac{ \pi}{l} < p_\mu < \frac{ \pi}{l} $  for which (\ref{momentumrel}) is satisfied. 

Since this is true independently for all values of $\mu$, in $d$ dimensions a point in $\hat{\mathcal{P}}$ has in this case $2^d$ distinct images in $\mathcal{P}$.
As an example let us consider the case  $\hat{p}_\mu=0$, which corresponds to a  translational invariant configuration, namely to a constant field in coordinate space. 
 For $a=l$  $\hat{p}_\mu=0$ is mapped according to  (\ref{momentumrel}) onto the lattice configuration $p_\mu=0$, which corresponds in coordinate space to a constant field on the lattice.

According to the previous discussion, for $a=2l$, namely for $n=2$, the vanishing momentum configuration in the continuum  $\hat{p}_\mu=0$ is mapped through  (\ref{momentumrel}) onto $2^d$ distinct momentum configurations on the lattice which we shall denote as $p_\mu^{(\mathcal{A})}$ where the labels  $\mathcal{A}$ run over the $2^d$ subsets of the possible values of the space-time index $\mu$:
\beq
\mathcal{A} \subseteq  \{ 1,2, \dots,d \}. \label{subset} \eeq
From (\ref{momentumrel}) we find:
\beq
\hat{p}_\mu=0  ~~~~~~\Longrightarrow ~~~~~ p_\mu^{(\mathcal{A})}= \left\{\begin{array}{ll}  0 & \mbox{if} ~~~\mu \not\in \mathcal{A} \\  \pm \frac{\pi}{l} & \mbox{if}~~~ \mu \in \mathcal{A} \end{array} \right.,  \label{pa} \eeq
where of course the sign in $ \pm \frac{\pi}{l}$ is irrelevant due to the $\frac{2\pi}{l}$ periodicity.
In coordinate representation zero momentum corresponds to a translationally invariant constant field configuration. The field configurations in coordinate space that correspond to a state of momentum $ p_\mu^{(\mathcal{A})}$ can be easily obtained from (\ref{momentumrel}) by taking the Fourier transform of a field $\tilde{\varphi}^{(\mathcal{A})}(p_\mu)$  given in momentum space by:
\beq  \tilde{\varphi}^{(\mathcal{A})}(p_\mu)= c^{(\mathcal{A})} \prod_\mu \sum_{k_\mu} \delta\left( p_\mu - p_\mu^{(\mathcal{A})} + \frac{2\pi}{l} k_\mu \right), \label{moma} \eeq
where $c^{(\mathcal{A})}$ are arbitrary constants.
The Fourier transform of (\ref{moma}) gives:
\beq
\varphi^{(\mathcal{A})}(x_\mu) = c^{(\mathcal{A})} \left( -1 \right)^{\sum_{\mu \in \mathcal{A}} n_\mu}
.\label{constconf} \eeq
 Here the integers $n_\mu$ are labeling the lattice sites according to (\ref{regularlattice}).
All the field configurations of eq. (\ref{constconf}) are invariant under $ x_\mu \rightarrow x_\mu + m_\mu a $ with $m_\mu$ arbitrary integers, namely:
\beq
\varphi^{(\mathcal{A})}(x_\mu+ m_\mu a ) = \varphi^{(\mathcal{A})}(x_\mu)
.\label{ivn} \eeq
 This stems from the fact that a shift on the lattice of an integer multiple of $a$ corresponds, for $a=2l$ to a shift of an even number of lattice spacing that leaves the signs at the r.h.s. of (\ref{constconf}) invariant. 

Since $a$ is the smallest shift on the lattice that corresponds to a translation in the continuum, all the  $2^d$ field configurations correspond to a translationally invariant (constant) field configurations in the continuum. This is obviously in agreement with (\ref{pa}) and implies that in $d$ dimensions there are $2^d$ distinct configurations on the lattice that correspond to the constant field configuration of the continuum.

Fluctuations around a translational invariant configuration correspond to a degree of freedom, so the existence of $2^d$ distinct translationally invariant configurations on the lattice also implies that a single field on the lattice describes  $2^d$ distinct degrees of freedom in the continuum in the case $n=2$.
This is the origin of the doubling of fermions on the lattice, since in the case of fermions the $n=2$ choice is unavoidable. 
Bosons on the lattice on the other hand can be consistently described by choosing $n=1$. However a different choice of $n$ for boson and fermions would inevitably break supersymmetry and the choice $n=2$ for bosons as well as for fermions seems unavoidable in supersymmetric theories. This is a crucial point in our approach, and it will be discussed in the following sections.

Before further discussing the doubling of fermions on the lattice, we need to introduce another key ingredient in defining a theory on the lattice: namely the  derivative (or finite difference) operator.

Let $\Phi(x)$  be a field in coordinate representation of a $d$-dimensional continuum space, and $\tilde{\Phi}(\hat{p})$ its Fourier transformed representation in momentum space. Acting on $\Phi(x)$ with the derivative operator $\partial_\mu$ amounts in momentum space to multiplying the field by the momentum itself $\hat{p}_\mu$:
\beq
i \partial_\mu \Phi(x)~~~~   \Longrightarrow~~~~ \hat{p}_\mu   \tilde{\Phi}(\hat{p}).  \label{momderi} \eeq 
Notice that  the derivative operator is local in momentum representation, namely it is a multiplicative function of  $\hat{p}_\mu$, and we shall work under the assumption that the same property is valid also on the lattice. So if we denote by $\Delta_\mu$ the derivative operator on the lattice, $\varphi(x)$ and $\tilde{\varphi}(p)$ a field on the lattice respectively in coordinate and momentum representation, then eq. (\ref{momderi}) is replaced on the lattice by:
\beq
\Delta_\mu \varphi(x) ~~~~\Longrightarrow~~~~\Delta(p_\mu) \tilde{\varphi}(p),  \label{momderilatt} \eeq
where $x=n l$ and $\tilde{\varphi}(p) =\tilde{\varphi}(p+\frac{2\pi}{l})$. 
We want the derivative of a lattice field to be still a lattice field, so the quantity at the r.h.s. of (\ref{momderilatt}) must still be periodic in all the $p_\mu$ variables with period $\frac{2\pi}{l}$.
The derivative operator must then be periodic itself, and $\Delta(p_\mu)$ must satisfy the condition:

\beq
 \Delta(p_\mu) = \Delta(p_\mu + \frac{2\pi}{l}) .    \label{der} \eeq 
As a consequence of (\ref{der}) the quantity $\Delta(p_\mu)$, which represents in momentum space the lattice derivative,  cannot coincide with the momentum $p_\mu$, unlike the continuum case, because the choice $\Delta(p_\mu) = p_\mu$ would be in contradiction with (\ref{der}).

The derivative operator in the continuum satisfies the Leibniz rule. 
This is a consequence of the fact that in momentum representation the derivative is the momentum itself (\ref{momderi}) and that the  momentum is a conserved and additive quantity.
 Additivity of momentum is on the other hand  related to   locality. In fact  in local field theories the product of two fields is defined by the standard local product of two functions
\beq
\Phi_{12}(x) \equiv \Phi_1\cdot\Phi_2(x) = \Phi_1(x) \Phi_2(x) \label{localproduct}
, \eeq
which becomes in momentum representation a convolution stating that the momentum of the composite field is the sum of the momenta of the component fields\footnote{In the present discussion we restrict for notational simplicity to a one dimensional case, extension to higher dimensions is trivial.}:
\beq
\tilde{\Phi}_{12}(\hat{p}) = \int_{-\infty}^{+\infty}~d\hat{p}_1 ~d\hat{p}_2  ~\tilde{\Phi}_1(\hat{p}_1)~\tilde{\Phi}_2(\hat{p}_2) \delta\left( \hat{p} - \hat{p}_1 -\hat{p}_2 \right)
\label{localproductp}
. \eeq
The Leibniz rule 
\beq
\partial_x \Phi_{12}(x) = \left( \partial_x \Phi_1(x) \right) \Phi_2(x) + \Phi_1(x)   \left( \partial_x \Phi_2(x) \right), \label{Lrule} \eeq
becomes in momentum representation
\beq
\hat{p} \tilde{\Phi}_{12}(\hat{p}) = \int_{-\infty}^{+\infty}~d\hat{p}_1 ~d\hat{p}_2  ~\tilde{\Phi}_1(\hat{p}_1)~\tilde{\Phi}_2(\hat{p}_2) \left( \hat{p}_1 + \hat{p}_2 \right) \delta\left( \hat{p} - \hat{p}_1 -\hat{p}_2 \right), \label{Lrulep} \eeq
which is automatically fulfilled by the delta function of momentum conservation.

If strict locality is assumed also on the lattice, namely if we assume that the product of two fields is a local product
\beq
\varphi_{12}(x) = \varphi_1\cdot\varphi_2(x) = \varphi_1(x)\varphi_2(x)~~~~~~~~~~~~x=n l,  \label{locprlatt} \eeq
we also find, as in the continuum, that the momentum is additive, but only modulo $\frac{2\pi}{l}$:
\beq
\tilde{\varphi}_{12}(p) = \int_{-\frac{\pi}{l}}^{+\frac{\pi}{l}}~dp_1 ~dp_2  ~\tilde{\varphi}_1(p_1)~\tilde{\varphi}_2(p_2) \sum_{k=-\infty}^{+\infty}\delta\left( p - p_1 -p_2 +k\frac{2\pi}{l}\right)
\label{localproductlatt}
, \eeq
with $k$ integer and all fields periodic with period $\frac{2\pi}{l}$.
We can now prove the following statement: \textit{If the product on the lattice is defined as the local product of eq. (\ref{locprlatt}) it is impossible to find a derivative operator (\ref{momderilatt}), satisfying the periodicity conditions (\ref{der}), that obeys the Leibniz rule with respect to the given product.}
This result is not new (see \cite{Bergner-Bruckmann, Bergner}), but we discuss it again in detail here, as it is the starting point of our approach.
Let us assume that a derivative operator $\Delta_\mu$ ( $\Delta$ in one dimension) exists that satisfies the Leibniz rule. Then the Leibniz rule would read:
\beq \Delta \varphi_{12}(x) = \Delta \varphi_1(x)~ \varphi_2(x) + \varphi_1(x)~ \Delta \varphi_2(x). \label{Lbl} \eeq
In momentum representation, using (\ref{momderilatt}), the Leibniz rule (\ref{Lbl}) becomes:
\beq
\Delta(p)~\tilde{\varphi}_{12}(p) = \int_{-\frac{\pi}{l}}^{+\frac{\pi}{l}}~dp_1 ~dp_2  ~\tilde{\varphi}_1(p_1)~\tilde{\varphi}_2(p_2) \left( \Delta(p_1) +\Delta(p_2) \right)  \sum_{k=-\infty}^{+\infty}\delta\left( p - p_1 -p_2 +k\frac{2\pi}{l}\right) \label{LBRcondition}
. \eeq
Equation (\ref{LBRcondition}) should be satisfied for arbitrary $\tilde{\varphi}_i(p_i)$. So by replacing $\tilde{\varphi}_{12}(p) $ in (\ref{LBRcondition}) with the r.h.s. of (\ref{localproductlatt}) and taking into account the periodicity of $\Delta(p)$ one finds that (\ref{LBRcondition}) is satisfied iff:
\beq
\Delta(p_1 +p_2 ) - \Delta(p_1) - \Delta(p_2) = 0,   \label{LBRcondition2} \eeq
which implies that the derivative $\Delta'(p)$ is a constant. So the only solution of (\ref{LBRcondition2}) is $\Delta(p) = p$ which however is not periodic, contrary to the original assumption.
In conclusion, it is impossible to define a derivative operator on the lattice that satisfies the Leibniz rule if the product of fields is the local product defined in (\ref{locprlatt}), that is if the momentum on the lattice is additive and conserved  modulo $\frac{2\pi}{l}$ . 
As we shall see in the following sections the Leibniz rule can be recovered only if the locality of the product and the translational invariance on the lattice are abandoned, at least at the lattice scale.
Notice however that with $p_1$ and $p_2$ in the fundamental interval $-\frac{\pi}{l} < p < \frac{\pi}{l}$ (\ref{LBRcondition2}) is satisfied for $|p_1+p_2| < \frac{\pi}{l}$ if $\Delta(p)$ is the ``saw tooth" function defined in the fundamental interval  by:
\beq
\Delta(p) = p~~~~~~~~~~~~-\frac{\pi}{l} < p < \frac{\pi}{l}, \label{sawtooth} \eeq
and extended by periodicity outside it.
Eq. (\ref{sawtooth}) defines the SLAC derivative\cite{SLAC}. 
Although the SLAC derivative does not satisfy the Leibniz rule, it is the best possible solution in the sense that it fulfills eq. (\ref{LBRcondition2}) for the largest possible interval in momentum space, an interval whose extension  goes to infinity as the lattice spacing goes to zero.

We shall now discuss in some more detail the origin of the fermion doubling problem on the lattice.
We already mentioned earlier in this section that the natural choice for the derivative operator $\Delta$ on the lattice, namely the finite difference over one lattice spacing, leads to an ambiguity, since it is possible to define a right or a left difference operator  $\Delta_{\pm}$   given in eq. (\ref{rightleftderiv}).

 A  symmetric finite difference on the other hand  can be defined as  $\Delta_s = \frac{1}{2} \left( \Delta_+ + \Delta_- \right)$,   but  involves a difference over two lattice spacings (see eq. (\ref{symmderiv})).

In momentum space $\Delta_+$ and $\Delta_-$ are multiplicative operators represented by complex conjugate functions of the momentum:
\beq
\Delta_\pm \varphi(x) \rightarrow \Delta_\pm(p) \tilde{\varphi}(p), \label{lrderiv} \eeq
where
\beq
\Delta_\pm(p) = \mp i \left( e^{\pm ilp} -1 \right),  \label{lrderiv2} \eeq
whereas $\Delta_s$ is just the real part of $\Delta_\pm$:
\beq  \Delta_s(p_\mu) = \frac{1}{l} \sin lp_\mu. \label{finitediff} \eeq 
In order to preserve the hermiticity of the action the inverse propagator of a free boson and of a free fermion should be  real functions of the momenta in momentum space. In the bosonic case the inverse propagator of the continuum theory is a quadratic form in the momentum, and can be written on the lattice as a real function by a combined use of $\Delta_+$ and $\Delta_-$:
\beq
\hat{p}_\mu\hat{p}^\mu \rightarrow \sum_\mu\Delta_+(p_\mu)\Delta_-(p_\mu) = \frac{2}{l^2} \sum_\mu \left( 1 - \cos(l p_\mu) \right). \label{bosonprop} \eeq
However a different form of the bosonic inverse propagator  is also possible that only involves $\Delta_s(p)$ and coincides with (\ref{bosonprop}) in the limit of small $lp_\mu$.  This can be obtained by simply replacing $\hat{p}_\mu$ with the symmetric finite difference operator $\Delta_s(p)$:
\beq
\hat{p}_\mu\hat{p}^\mu \rightarrow \sum_\mu\Delta_s(p_\mu)\Delta_s(p_\mu)  = \frac{1}{l^2} \sum_\mu \sin^2 lp_\mu. \label{bosonprops} \eeq
Instead, in the case of the fermion propagator, which is linear in the momentum, hermiticity on the lattice requires that the inverse propagator is written in terms of the symmetric difference operator, namely:
\beq
\gamma^\mu \hat{p}_\mu  \rightarrow  \gamma^\mu \Delta_s(p_\mu) = \frac{1}{l} \gamma^\mu \sin(l p_\mu). \label{fermionprop} \eeq

In standard lattice theory the form (\ref{bosonprop}) has been used for bosons and, unavoidably, the form (\ref{fermionprop}) for fermions. This avoids the appearing of extra states in the boson sector since the inverse propagator in (\ref{bosonprop}) vanishes only for $p_\mu = 0$ in the Brillouin zone. The fermion inverse propagator (\ref{fermionprop}) on the contrary vanishes for any set of $p_\mu$ that satisfies the conditions:
\beq
\Delta_s(p_\mu) = \frac{1}{l} \sin(l p_\mu) = 0~~~~~~~~~~~~\mu=1,2,\cdots,d.       \label{trconf} \eeq
The solutions of (\ref{trconf}) are the $2^d$ points in momentum space labeled by the index $\mathcal{A}$,  and whose coordinates in momentum space $p_\mu^{(\mathcal{A})}$ are given in (\ref{pa}).
All these $2^d$ momentum configurations correspond to a zero momentum configuration in the continuum, and small fluctuations around them are then interpreted as distinct degrees of freedom in the continuum.

This is the essence of the fermion doubling phenomenon.

The boson inverse propagator (\ref{bosonprops}) is obtained from the continuum case by applying the same prescription used for the fermion one, namely by replacing $\hat{p}_\mu$ with $\Delta_s(p_\mu)$. 
As a result it vanishes not just at $p_\mu=0$ but at each of $2^d$ field configurations $p_\mu = p_\mu^{(\mathcal{A})}$, leading to a doublers phenomenon also for the boson. This may be regarded as a disadvantage, but it is indeed necessary in supersymmetric theories if supersymmetry has to be kept exactly on the lattice\cite{Bartels-Kramer, GKPY}.

This argument does not depend on the particular form chosen for the derivative operator $\Delta(p)$ on the lattice as long as $\Delta(p)$ is a smooth real function of $p$ satisfying the periodicity condition (\ref{der}). In fact if $\Delta(p)$ has a simple zero at $p_\mu=0$ ( we assume that  $\Delta(p_\mu) \simeq p_\mu$ for small momenta, namely $ lp_\mu \ll 1$ ) and it is smooth and periodic it has necessarily another zero in the Brillouin zone.  This additional zero is always located at $p_\mu = \pm \frac{\pi}{l}$ if besides being periodic  $\Delta(p)$ is an odd function of $p$:
\beq
\Delta(-p)= -\Delta(p). \label{oddDelta} \eeq
The  condition (\ref{oddDelta}) is a reality condition, in the sense that it comes from the requirement that the derivative of a real field in the coordinate representation is still real. 
The double zero of $\Delta(p)$ at $p=0$ and $p=\pm \frac{\pi}{l}$  implies that the correspondence between the momentum on the lattice and the momentum in the continuum theory is given by eq. (\ref{momentumrel}) with $l=\frac{a}{2}$, where  $a$ is  the smallest translation of the lattice that corresponds to a translation in the continuum. 

It is well known that out of  the $2^d$ states arising from a lattice fermion half have positive and half negative chirality. This is discussed in all texbooks, and we review it here for comparison with the new approach introduced in the next section. 
Consider the Dirac operator in the continuum:
\beq
D(\hat{p}) = \gamma^\mu \hat{p}_\mu. \label{diracc} \eeq
On the lattice, choosing for simplicity the symmetric finite difference operator as derivative operator, the Dirac operator becomes, according to (\ref{fermionprop})
\beq
D_{l}(p) = \gamma^\mu \Delta_s(p_\mu) = \gamma^\mu \frac{1}{l} \sin(lp_\mu).  \label{diracl} \eeq
By using now the relation (\ref{momentumrel}) with $a=2l$ we replace $lp_\mu$ in (\ref{diracl}) with $l\hat{p}_\mu - \pi k_\mu$ and consider the Dirac operator on the lattice in the continuum limit $l \rightarrow 0 $ by expanding in powers of $l \hat{p}_\mu$ and keeping only the first term in the expansion:
\beq
D_{l}(p) = \gamma^\mu (-1)^{k_\mu} \hat{p}_\mu + O(l^2 \hat{p}^3_\mu).    \label{diraclc} \eeq
In the Brillouin zone the integers $k_\mu$ can only take the values $0$ and $+1$ corresponding respectively to the expansions around $p_\mu=0$ and $p_\mu = \frac {\pi}{l}$, the signature $(-1)^{k_\mu}$ arising from the fact that the slope of $\Delta_s(p)$ at $p=\frac{\pi}{l}$ has opposite sign of the one at $p=0$. The $2^d$ possible choices of the $d$ integers $k_\mu$ correspond then to the different copies of the fermion. The chirality of each copy can be derived by observing that with a redefinition of  the gamma matrices by means of a unitary transformation eq. (\ref{diraclc}) can be written  as:
\beq D_{l}(p) = \gamma^{'\mu}  \hat{p}_\mu + O(l^2 \hat{p}^3_\mu),    \label{diraclcprime} \eeq
with
\beq
\gamma^{'\mu} = (-1)^{k_\mu} \gamma^\mu. \label{gammaredef} \eeq
This also implies
\beq
\gamma^{'5} = (-1)^{\sum_\mu k_\mu} \gamma^5, \label{gamma5redef} \eeq
namely a positive or negative chirality according to the sign of  $(-1)^{\sum_\mu k_\mu}$.

The map between the  non-compact momentum space of the continuum and the compact momentum space of the lattice given in 
(\ref{momentumrel}) plays a fundamental role in defining the lattice theory. As we have seen the fermion doubling problem and the violation of the Leibniz rule are intimately connected to this map, and its modification  is at the root of the different approach that we shall develop in the following sections.

We close this section with the proof that  the correspondence (\ref{momentumrel}) is obtained if  the lattice fields are constructed starting from the fields of the continuum theory by means of a blocking transformation\footnote{We use here this term in a more general sense than usual, namely also for transformations from continuum to lattice} that  preserves the invariance of the lattice under discrete translations of  a lattice spacing $l$\footnote{In this example we restrict ourselves  to the case $a=l$ in eq. (\ref{momentumrel})}.
We consider for simplicity a one dimensional example and define the blocking transformation as:
\beq \varphi_A(n) = \int dx f(nl-x) \Phi_A(x), \label{bltr} \eeq
where the fields $\varphi_A$ and $\Phi_A$ denote respectively the lattice and the continuum fields and the function $f(y)$ is arbitrary but  in general peaked at $y=0$ so that the lattice fields $\varphi_A(n)$ are determined by the continuum fields $\Phi_A(x)$ with $x$ close to $nl$. Translational invariance under discrete displacements by $l$ of the lattice is ensured by the translational invariance of the continuum theory and by the $x-ln$ dependence in the function $f$. 
The blocking transformation (\ref{bltr}) corresponds to our intuitive notion of what a lattice theory should be: each point of the lattice is representative of the surrounding area of the continuum theory and the blocking procedure does not depend on the lattice point ( translational invariance).
From the perspective of the momentum space however the blocking transformation (\ref{bltr}) is much less intuitive. In fact, by taking the Fourier transform of both sides and denoting the transformed fields with an upper tilde we obtain:
  \beq \tilde{\varphi}_A(p) = \sum_k \tilde{f}(p + \frac{2\pi}{l}k) \tilde{\Phi}_A(p + \frac{2\pi}{l}k). \label{bltrp} \eeq
The correspondence (\ref{momentumrel}) acquires now a more precise meaning: the lattice field $\tilde{\varphi}_A(p)$ is the sum, weighed with the function $\tilde{f}$, of all the continuum fields $\Phi_A(\hat{p})$ with $\hat{p} = p + \frac{2\pi}{l} k $.  So for instance, even for very small values of $p$,  $\tilde{\varphi}_A(p)$ receives contributions from continuum fields with arbitrarily large values of the momentum $\hat{p}$.
This seems rather unnatural, and it can be avoided by restricting $\tilde{f}(\hat{q})$ to be significantly different from zero only within a region of $\hat{q}$ with size of order $\frac{1}{l}$. In coordinate representation this corresponds to a function $f(y)$ significantly different from zero in a region of order $l$ around $y=0$ in agreement with the original intuitive notion of the blocking transformation.
A well known example is obtained by choosing in eq. (\ref{bltrp}) :

\beq \tilde{f}(p) = 0 ~~~~~~~~~~~~~\text{for}~~~~~~~~~~~~~|p| \geq \frac{\pi}{l},  \label{slac} \eeq
where the equality is introduced to insure periodicity in $p$ of the lattice field: $\tilde{\varphi}_A(-\frac{\pi}{l})=\tilde{\varphi}_A(\frac{\pi}{l})$. 
With this choice the blocking transformation (\ref{bltrp}) becomes:
\beq
\tilde{\varphi}_A(p+ \frac{2\pi}{l} k) =  \tilde{f}(p ) \tilde{\Phi}_A(p )~~~~~~~~~~-\frac{\pi}{l} \leq p \leq \frac{\pi}{l}.  \label{slacp} \eeq
In (\ref{slacp}) the r.h.s. does not depend on the integer $k$, so the periodicity of the lattice field  with period $\frac{2\pi}{l}$ is guaranteed. The value of $\tilde{\varphi}_A(p) $ does not depend on the values taken by the continuum fields $\hat{\Phi}_A(\hat{p})$   for $ |\hat{p}| \geq \frac{\pi}{l}$, so that the lattice theory is effectively equivalent to introducing a momentum cutoff.  
Eq. (\ref{slacp}) is a particular form of the blocking transformation in momentum space, with the property that the lattice fields $\tilde{\varphi}_A(p)$ with $p$ in the fundamental zone $|p| \leq \frac{\pi}{l}$ are proportional through the weight function $\tilde{f}$ to the continuum fields of corresponding momentum in the continuum. A generalization of this transformation is at the root of our approach and will be discussed in the following sections.

If we replace in (\ref{slacp}) $\Phi_A$ with $\partial\Phi_A$, namely $\tilde{\Phi}_A(\hat{p})$ with $\hat{p}\tilde{\Phi}_A(\hat{p})$, then the lattice fields $\tilde{\varphi}_A(p)$ should be multiplied by the lattice derivative operator $\Delta(p)$, which can in this way be determined. In fact we have:
\beq
\Delta(p+ \frac{2\pi}{l} k) ~\tilde{\varphi}_A(p+ \frac{2\pi}{l} k) =  \tilde{f}(p )~ p~\tilde{\Phi}_A(p )~~~~~~~~~~-\frac{\pi}{l} \leq p \leq \frac{\pi}{l},  \label{slac2} \eeq
leading to the well known SLAC derivative:
\beq \Delta(p+ \frac{2\pi}{l} k) = p ~~~~~~~~~~~~~~~~~-\frac{\pi}{l} \leq p \leq \frac{\pi}{l}  \label{slac3}, \eeq
where the same result was given in \cite{Bergner-Bruckmann, Bergner}.

As discussed earlier in this section the problem of the fermion doubling is also related to the periodicity in $p_\mu$ of the derivative operator on the lattice and to the fact that any continuous and periodic function with a simple zero at $p_\mu=0$ has to vanish in some other point in the fundamental interval (Brillouin zone) giving rise to another state of opposite chirality. The only way to avoid the doubling in this context is to give up the continuity of the derivative operator as a function of the momentum.
An example is the SLAC derivative given above which only vanishes at $p=0$ in the Brillouin zone, so that the doublers problem does not arise. However, as shown in eq. (\ref{slac3}), the SLAC derivative has a discontinuity   at $p= \pm \frac{\pi}{l}$ and as a consequence of that  it is long range in coordinate space, leading to the well known problems in getting the correct continuum limit when used in gauge theories \cite{KarSm}.

\section{A new lattice. Restoring the Leibniz rule and avoiding the doubling problem.}
\label{1.2}

Locality and translational invariance are standard assumptions in conventional lattice regularization. However, as seen in the previous section, they lead to the impossibility of defining a derivative operator on the lattice that satisfies the Leibniz rule. This originates from the fact that in order to be well defined on the lattice the derivative operator should be a periodic function of the momentum (\ref{der}), whereas locality and translational invariance imply that  the additive and conserved quantity on the lattice is the momentum itself which  however is defined only modulo $\frac{2\pi}{a}$ and so  is not  suitable as a finite difference operator. 

In the present paper we shall take an entirely different  point of view which was first taken by Dondi and Nicolai in their poineering paper on lattice supersymmetry \cite{Dondi-Nicolai}.  We shall assume that  the additive and conserved quantity on the lattice is not the momentum $p_\mu$ but the operator  $\Delta(p_\mu)$,  periodic  with period $\frac{2\pi}{l}$ in $p_\mu$, which plays on the lattice  the role of the derivative operator. 
This means replacing the local product on the lattice, given in momentum representation by the convolution (\ref{localproductlatt}), with a new product, which we shall denote as star product, given by a convolution where $\Delta(p)$ is conserved, namely in one dimension:
\beq \widetilde{\varphi_1 \star \varphi_2 }(p_{12}) =\frac{1}{2\pi}  \int_{-\frac{\pi}{l}}^{\frac{\pi}{l}} dp_1 dp_2 V(p_{12};p_1,p_2) \tilde{\varphi}_1(p_1) \tilde{\varphi}_2(p_2) \delta\left( \Delta(p_{12}) -\Delta(p_1) - \Delta(p_2) \right),  \label{starprodp} \eeq
where $V(p_{12};p_1,p_2)$ defines the measure of integration, which will be assumed to be symmetric in the last two arguments
\beq
V(p_{12};p_1,p_2) = V(p_{12};p_2,p_1),  \label{simmV} \eeq
thus defining a commutative product. Further properties of the measure will be discussed later.
The derivative operator $\Delta(p)$ satisfies now the Leibniz rule by construction. In fact it is immediate to check that thanks to the delta function in (\ref{starprodp}) the relation
\beq
\Delta(p_{12}) \widetilde{\varphi_1 \star \varphi_2 }(p_{12})  =  \Delta(p_1) \tilde{\varphi}_1(p_1)  \star \tilde{\varphi}_2(p_2) + 
\tilde{\varphi}_1(p_1) \star  \Delta(p_2) \tilde{\varphi}_2(p_2),  \label{Lrulestar} \eeq
is identically satisfied.

In coordinate representation the star product (\ref{starprodp}) is non-local. This will be examined more in detail further in the paper. Moreover, since the momentum $p_\mu$  on the lattice is not anymore  additive nor conserved ( except  approximately  for momenta much smaller that $ \frac{1}{l} $),  the lattice itself is not translationally invariant. 
Translational invariance however is not lost but it is represented by infinitesimal transformations of the fields generated by $\Delta(p_\mu)$, namely:
\beq
\delta_{\epsilon} \tilde{\varphi}_A(p) = \epsilon^{\mu} \Delta(p_\mu)\tilde{\varphi}_A(p), \label{latticetranslation} \eeq
or by the corresponding finite transformations 
\beq
\tilde{\varphi}_A(p)  \rightarrow e^{ \epsilon^{\mu} \Delta(p_\mu)} \tilde{\varphi}_A(p),  \label{fintr} \eeq
where in (\ref{fintr}) $\epsilon_\mu$ are arbitrary finite parameters.

In fact, due to the conservation of $\Delta(p_\mu)$ and the validity of the Leibniz rule an action entirely constructed with the star product is automatically invariant under (\ref{latticetranslation}) and (\ref{fintr}). Notice that the symmetry (\ref{fintr}) is a continuous symmetry and not a discrete one, as it would be if it were associated to lattice displacements.

The correspondence between $p_\mu$ and $\hat{p}_\mu$, given in the conventional lattice by eq. (\ref{momentumrel}), must be modified in the present approach.  In fact, since $\hat{p}_\mu$ is an additive and conserved quantity in the continuum theory it must correspond to the additive and conserved quantity on the lattice, namely it must correspond to the derivative operator  $\Delta(p_\mu)$.  Eq. (\ref{momentumrel}) is then replaced now by
\beq \hat{p}_\mu \equiv  \Delta(p_\mu).   \label{momentumdelta} \eeq

The derivative operator $\Delta(p_\mu)$ must satisfy a number of conditions. As already discussed in the previous section it must  be a periodic function with period $\frac{2\pi}{l}$ (eq. (\ref{der}) ) and it must be an odd function of $p$ (\ref{oddDelta}). Moreover, since $\hat{p}_\mu$ is real, it has to be a real function\footnote{This fact rules out, as possible choices for $\Delta(p_\mu)$,  the difference operators over one lattice spacing $\Delta_{\pm}(p)$ given in (\ref{lrderiv2})}.   
We shall also assume that, for $lp \ll 1$, $\Delta(p)$ reduces to $p$, more precisely we shall assume that $ l \Delta(p)$ is a function of $l p$ with a simple zero at $l p = 0$:
\beq l \Delta(p) = l p + O\left( (l p)^3 \right). \label{dd} \eeq
Besides vanishing at $p=0$ the function $\Delta(p)$, being both odd and periodic,  has another zero in the Brillouin zone at $p= \pm \frac{\pi}{l}$. We shall assume in the following that these are the only two zeros of $\Delta(p)$.

The simplest function that satisfies all these constraints is the symmetric finite difference operator $\Delta_s(p)$ defined in momentum space by (\ref{finitediff}). 
With that choice the derivative of a field $\varphi(x)$ on the lattice is given in coordinate  representation by:
\beq
\Delta_{s,\mu}\varphi(x) = \frac{1}{2l} \left( \varphi(x+l \vec{n}_\mu) - \varphi(x-l \vec{n}_\mu) \right ), \label{symmfd2} \eeq
where $\vec{n}_\mu$ is the unit vector in the $\mu$ direction and $x$ is a point on the lattice whose coordinates are integer multiples of $l$. 
In momentum space $\Delta_s$ acts as a multiplicative operator:
\beq \Delta_s(p_\mu)\tilde{\varphi}(p) = \frac{1}{l} \sin(lp_\mu)\tilde{\varphi}(p).  \label{symmfdp2} \eeq

With $\Delta=\Delta_s$ the infinitesimal translations (\ref{latticetranslation}) become:
\beq
\delta_{\epsilon} \tilde{\varphi}_A(p) = \epsilon^{\mu} \frac{1}{l} \sin(lp_\mu)\tilde{\varphi}_A(p), \label{latticetranslation2} \eeq
namely in coordinate representation:
\beq
\delta_{\epsilon} \varphi_A(x) = \epsilon^{\mu} \frac{1}{2l} \left( \varphi_A(x+l \vec{n}_\mu) - \varphi_A(x-l \vec{n}_\mu) \right ), \label{latticetranscc} \eeq
which in the continuum limit $l \rightarrow 0$ reproduces the ordinaty infinitesimal translation in the continuum. It is apparent from (\ref{latticetranscc}) that an infinitesimal translation is represented on the lattice by a difference over two lattice spacings. 
If we denote by $a$, as in the previous section, the ``effective" lattice spacing, namely the smallest lattice movement that corresponds to a translation in the continuum, then we have from the previous equations that
\beq
l= \frac{a}{2}. \label{leqa2} \eeq

If in the correspondence (\ref{momentumdelta}) we replace $\Delta(p_\mu)$ with $\Delta_s(p_\mu)$ we obtain the following map:
\beq \hat{p}_\mu = \frac{1}{l} \sin(lp_\mu)= \frac{2}{a} \sin(\frac{ap_\mu}{2}). \label{sinecorr} \eeq
The correspondence (\ref{sinecorr}) is not one-to-one: for $|\hat{p}_\mu| > \frac{1}{l} $ there is no value of $p_\mu$ satisfying  (\ref{sinecorr}) whereas for $|\hat{p}_\mu| < \frac{1}{l} $ there are within the Brillouin zone $-\frac{\pi}{2l} \leq p_\mu \leq \frac{3\pi}{2l}$  two distinct solutions. In fact since $\Delta_s(p_\mu)$ is invariant under the transformation:
\beq p_\mu \rightarrow \frac{\pi}{l} - p_\mu,   \label{doubsymm} \eeq
if $p_\mu$ is a solution of (\ref{momentumdelta}) for a given $\hat{p}_\mu$, then  $ \frac{\pi}{l} - p_\mu $ is also a solution.

This implies  that  in $d$ dimensions a  constant field configuration in the continuum, namely in momentum space a configuration with $\hat{p}_\mu = 0$, has $2^d$ images on the lattice. They are labeled by the index $\mathcal{A}$ introduced in (\ref{subset}) and are given by  $p_\mu = p_\mu^{(\mathcal{A})}$ where $p_\mu^{(\mathcal{A})}$ are defined in (\ref{pa}).
The corresponding field configurations $\varphi^{(\mathcal{A})}(x_\mu)$ are the same ones introduced in the last section and given by (\ref{constconf}).
So the correspondence  between $\hat{p}_\mu$ and $p_\mu$ given in (\ref{sinecorr}) produces  the same translationally invariant configurations labeled by $\mathcal{A}$ as the correspondence (\ref{momentumrel}) introduced in the last section, provided we set $a=2l$  in the latter.
There are however two profound differences between the two cases.  In (\ref{sinecorr}) the relation $a=2l$ is fixed and the case $a=l$ is ruled out from the start as $\Delta(p)$ has to vanish an even number of times due to its perodicity and smoothness  in the fundamental region. So the doublers phenomenon is completely general and it applies  to bosons as well as to fermions \footnote{In some cases however, as already seen in ref.  \cite{DFKKS} and \cite{DKKS}, doublers of a propagating boson do not propagate and play the role of auxiliary fields in extended supersymmetric theories.} making the balance between bosonic and fermionic degrees of freedom possible in supersymmetric theories. 
Moreover, as we shall discuss shortly, with the correspondence (\ref{sinecorr}) the physical states associated to the doublers of a fermion have all the same chirality, and not opposite chirality as in the traditional lattice formulation.

Before discussing this point let us observe that for $|\hat{p}_\mu| < \frac{1}{l} $ there are  $2^d$  momentum configurations on the lattice that correspond to a given momentum $\hat{p}_\mu$ in the continuum.  These configurations are related to each other by the symmetry transformations (\ref{doubsymm}) as a result of the invariance of $\sin(l p_\mu)$ under such transformations. 
In the following we shall assume that the lattice derivative $\Delta(p)$ satisfies the same symmetries as $\sin(l p_\mu)$, namely we shall add to the list of required properties of $\Delta(p)$ the symmetry
\beq
\Delta(p) = \Delta(\frac{\pi}{l} - p).   \label{deltasymm} \eeq
This symmetry does not follow from a fundamental principle, but  is required if one wants the $2^d$ doublers to be related by the symmetry transformation (\ref{doubsymm}), as it is needed for instance if one wants to identify them with different superfield components in an extended supersymmetric theory \cite{DFKKS, DKKS}.

From (\ref{deltasymm}) it also follows that $\Delta^\prime(p)=\frac{d\Delta(p)}{dp}$ is antisymmetric with respect to (\ref{doubsymm}), namely: 
\beq
\Delta^\prime(p) = -\Delta^\prime(\frac{\pi}{l} - p),   \label{deltader} \eeq
and this implies that the points $\pm\frac{\pi}{2l}$ are extremes of $\Delta(p)$:
\beq
\Delta^\prime(\pm \frac{\pi}{2l}) = 0.   \label{extremes} \eeq
We shall assume in the following that these are the only extremes of $\Delta(p)$, then we have:
\beq
-\left|\Delta(\frac{\pi}{2l})\right| \leq \Delta(p) \leq \left|\Delta(\frac{\pi}{2l})\right|. \label{deltainterval} \eeq

In order to see the second fundamental difference with respect to the conventional approach, namely that all fermion doublers here have the same chirality, let us consider again the Dirac operator on the lattice. This is given in eq. (\ref{diracl}) for the special choice $\Delta(p) = \Delta_s(p)$, but in general it reads:
\beq 
D_l(p) = \gamma^\mu \Delta(p_\mu)
.\label{diracll} \eeq
In the case of the conventional lattice examined in the previous section,  we considered  eq. (\ref{diracl}) in the limit $l\hat{p}_\mu \ll 1$ and using the relation (\ref{momentumrel}) between $p_\mu$ and $\hat{p}_\mu$ we found eq. (\ref{diraclc}) where the sign $(-1)^{k_\mu}$ is the chirality changing signature. 
In the present approach in order to express $p_\mu$ in terms of the physical momentum $\hat{p}_\mu$ we have to use eq. (\ref{momentumdelta}) which replaced into (\ref{diracll}) gives:
\beq
D_l(p) = \gamma^\mu \hat{p}_\mu. \label{diraclll} \eeq
This means that the Dirac operator $D_l(p)$, when expressed in terms of the continuum momentum $\hat{p}_\mu$, is identical to the Dirac operator in the continuum $\gamma^\mu \hat{p}_\mu$ irrespective of the value of $p_\mu$. So all the $2^d$ copies of the fermion produce the same Dirac equation in terms of the physical momentum and have trivially the same chirality.

The main consequence of this fact is that the problem of the doublers can be avoided.  There are two options for that, which we shall denote as \textbf{A} and \textbf{B} and are explained below:
\begin{itemize} 
\item \textbf{A). Doublers identified.} Since all the $2^d$ copies of the fermionic fields on the  lattice field have now the same chirality they can be identified by requiring that the transformation (\ref{doubsymm}) is a symmetry of both bosonic and fermionic  fields.  Let $\tilde{\varphi}_A(p)$ be   the fields in the momentum representation of a one dimensional lattice theory  with lattice spacing $\frac{a}{2}$\footnote{Here and in the following we use the effective lattice spacing $a$ in preference of the lattice spacing $l$, keeping in mind that the relation $l=\frac{a}{2}$ is valid all through} . We can identify the states at $p=0$ and $p=\frac{2\pi}{a}$ by imposing the condition:
\beq
\tilde{\varphi}_A(\frac{2\pi}{a}-p) =  \tilde{\varphi}_A(p),   \label{chiralcond} \eeq
which in coordinate representation reads:
\beq
\varphi(-\frac{na}{2}) =  (-1)^n ~\varphi(\frac{na}{2}), \label{chiralcondx} \eeq 
where $\frac{na}{2}=nl=x$ is the lattice coordinate.
After this identification the number of degrees of freedom of the theory on the lattice coincides with the one of the continuum theory and the doublers problem is avoided.

It is also interesting to note that in the supersymmetric Wess-Zumino in $1$ and $2$ dimensions with $N=2$  the truncation condition (\ref{chiralcond}) follows from imposing the chiral condition on the superfields \cite{DKKS}.

A more general identification can be used instead of (\ref{chiralcond}) by imposing the condition:
\beq
\tilde{\varphi}_A(\frac{2\pi}{a}-p) =   h(p) \tilde{\varphi}_A(p),   \label{genchiralcond} \eeq
where $h(p)$ is  any periodic function satisfying for consistency the condition
 \beq
 h(p)h(\frac{2\pi}{a}-p) = 1
\label{hcond}
. \eeq
Eq. (\ref{genchiralcond}) can then be rewritten as:
\beq
\sqrt{h(\frac{2\pi}{a}-p)} \tilde{\varphi}_A(\frac{2\pi}{a}-p) = \sqrt{h(p)}\tilde{\varphi}_A(p)  
\label{hcond2}
. \eeq
According to (\ref{hcond2}) the more general condition (\ref{genchiralcond}) can always be reduced to the simple form (\ref{chiralcond}) by a rescaling of the fields in momentum space:
\beq
\tilde{\varphi}_A(p) \rightarrow \sqrt{h(p)}\tilde{\varphi}_A(p) 
\label{risch}
. \eeq
For this reason in the following we shall always refer to the symmetry condition (\ref{chiralcond}) without any substantial loss of generality.

In the case of higher dimensions the contraints (\ref{chiralcond}) and (\ref{chiralcondx}) are applied separately for each dimensions. So, for instance in $d$ dimensions:
\beq
\tilde{\varphi}(p_1,\ldots,\frac{2\pi}{a}-p_j,\ldots,p_d) =   \tilde{\varphi}(p_1,\ldots,p_j,\ldots,p_d), \label{chiralcondpd} \eeq
or, in coordinate representation:
\beq
\varphi(\frac{n_1 a}{2},\ldots,-\frac{n_j a}{2},\ldots,\frac{n_d a}{2}) =  (-1)^{n_j} \varphi(\frac{n_1 a}{2},\ldots,\frac{n_j a}{2},\ldots,\frac{n_d a}{2}). \label{chiralcondd} \eeq

Notice that since the derivative operator $\Delta(p)$ is chosen to be invariant under (\ref{doubsymm}) the field symmetries (\ref{chiralcondpd}) are not affected by derivation.
Notice also that as a result of (\ref{chiralcondd}) the value of the fields in the quadrant with $n_j\geq0$ is sufficient to determine the fields everywhere.
To summarize: one degree of freedom on the $d$ dimensional cubic  lattice with spacing $\frac{a}{2}$  corresponds to $2^d$ degrees of freedom in the  continuum theory; however, since all these degrees of freedom have, in the case of fermions, the same chirality they can be identified using eq.s (\ref{chiralcondpd}) or (\ref{chiralcondd}) and the correspondence between the lattice fields and the continuum fields can thus  be made to be a one-to-one correspondence. 

\item 
\textbf{B). Doublers as distinct degrees of freedom}. The identification (\ref{genchiralcond}) discussed above as option \textbf{A}   is not always necessary.  In some situations the $2^d$ copies of the lattice fields may be regarded as distinct degrees of freedom in the continuum, and that may even be necessary to implement some continuum symmetries on the lattice. 

An example was given  in ref.\cite{DFKKS} and  \cite{DKKS}, where it was shown that in theories with extended supersymmetries the $2^d$  doublers can be interpreted as different members of the same supermultiplet.  For instance in the $D=1$, $N=2$ supersymmetric quantum mechanics the two bosonic and the two fermionic fields of the continuum theory are represented on the lattice by a single bosonic and a single fermionic field with the exact supersymmetry being represented on the lattice in a very economic way \cite{DFKKS}.
 
The situation of  the $D=2$, $N=2$ superalgebra is apparently similar. In fact it admits a representation in terms of  a superfield with $8$ bosonic and $8$ fermionic components which can be realized on the lattice in terms of just two bosonic and two fermionic fields. However in order to write the lagrangian of the supersymmetric Wess-Zumino model chiral conditions have to be applied to the original sixteen component superfield to reduce it to a four component chiral superfield. On the lattice this corresponds exactly to identifying the  doublers as in (\ref{chiralcondpd}) , so that in the end the component fields of a chiral superfield of the Wess-Zumino model are represented on the lattice by a single lattice field satisfying the ``chiral conditions" (\ref{chiralcondpd}) and  (\ref{chiralcondd}) \cite{DKKS}.

\end{itemize}

\section{A non local product on the lattice: the star product.}
\label{1.3}
The main new feature of our approach is that the additive and  conserved quantity on the lattice is not the momentum itself, but some periodic function   $\Delta(p)$ of the momentum that plays the role of the derivative operator in the lattice  theory and that corresponds to the momentum of the continuum theory. 
 An obvious consequence of this approach is that the lattice formulation is not translational invariant and that the local product of two fields is replaced by a non local product: the star product.  This  has been defined in eq. (\ref{starprodp}), which we reproduce here for convenience, with the effective lattice spacing $a$   at the place of $l$ in the integration limits:
\beq \widetilde{\varphi_1 \star \varphi_2 }(p_{12}) =\frac{1}{2\pi}  \int_{-\frac{\pi}{a}}^{\frac{3\pi}{a}} dp_1 dp_2 V(p_{12};p_1,p_2) \tilde{\varphi}_1(p_1) \tilde{\varphi}_2(p_2) \delta\left( \Delta(p_{12}) -\Delta(p_1) - \Delta(p_2) \right).  \label{starprodp2} \eeq
 In this section we shall discuss in detail the properties of the star product defined in (\ref{starprodp2}).
\subsection{Conditions for the associativity of the star product and consequences of its violation}
\label{3.1}

Let us begin with discussing  the formal properties of the star product defined in (\ref{starprodp2}). 

Commutativity will be assumed by requiring that the integration volume is symmetric, namely that it satisfies eq.(\ref{simmV}).

Associativity of the star product requires
\beq  
\left((\tilde{\varphi_3} \star\tilde{\varphi_2})\star\tilde{\varphi_1} \right)(p) = \left(\tilde{\varphi_3} \star(\tilde{\varphi_2}\star\tilde{\varphi_1}) \right)(p)
,\label{asso} \eeq
and poses severe restrictions on both $V(p_{12};p_1,p_2)$ and $\Delta(p)$.

In fact, let us consider for instance the r.h.s. of (\ref{asso}).
With the definition (\ref{starprodp2}) it can be written explicitely as:
\bea
\left(\tilde{\varphi_3} \star(\tilde{\varphi_2}\star\tilde{\varphi_1}) \right)(p) &=&\left(\frac{1}{2\pi}\right)^2\int_{-\frac{\pi}{a}}^{\frac{3\pi}{a}} dp_1dp_2dp_3K\left(p;p_3;p_1,p_2\right)\cdot\nonumber\\  &&\cdot\delta\left(\Delta(p)-\Delta(p_1)-
\Delta(p_2)-\Delta(p_3)\right) \tilde{\varphi_1}(p_1)\tilde{\varphi_2}(p_2)\tilde{\varphi_3}(p_3),\label{compositestar} \eea
where $K\left(p;p_3;p_1,p_2\right)$ is given by:
\beq
K\left(p;p_3;p_1,p_2\right) = \int_{-\frac{\pi}{a}}^{\frac{3\pi}{a}} dp_{12} V(p_{12};p_1,p_2)~V(p;p_{12},p_3)~\delta\left(\Delta(p_{12})-\Delta(p_1)-\Delta(p_2) 
\right). \label{kernelass} \eeq
For the star product to be associative it is necessary that the kernel $K\left(p;p_3;p_1,p_2\right)$ is symmetric in the three momenta $p_1$, $p_2$ and $p_3$. 
As discussed already in \cite{WTid}, this requires in the first place that the function $\Delta(p)$   takes any value from $-\infty$ to $+\infty$ when $p$ varies in the Brillouin zone. In fact if $\Delta(p)$ is limited, namely, taking into account  eq. (\ref{deltainterval}), if
\beq
|\Delta(p)|\leq|\Delta(\frac{\pi}{a})|<\infty ~~~~~~~~~~~~~~~~~-\frac{2\pi}{a}<p<\frac{2\pi}{a}, \label{deltalimited} \eeq
 then due to the delta function in (\ref{kernelass}) we have:
\beq
K\left(p;p_3;p_1,p_2\right)=0~~~~~~~~~~~~~\text{for}~~~~~~~~~~~~\left|\Delta(p_1)+\Delta(p_2) \right| > |\Delta(\frac{\pi}{a})|.  \label{asymmetriccond} \eeq
Eq. (\ref{asymmetriccond}) is not symmetric under exchanges of $p_1$, $p_2$ and $p_3$ and that implies a violation of associativity: $ \tilde{\varphi_1} \star(\tilde{\varphi_2}\star\tilde{\varphi_3}) \neq (\tilde{\varphi_1} \star\tilde{\varphi_2})\star\tilde{\varphi_3}$. 
The only way to recover the symmetry is to make sure that (\ref{asymmetriccond})  is empty because the inequality on the right is never satisfied, which implies $|\Delta(\frac{\pi}{a})| = \infty$.
So if we assume that the points $p=\pm \frac{\pi}{a}$ are the only values where the derivative of $\Delta(p)$ vanishes ( see eq. (\ref{extremes})), associativity of the star product requires:
\beq
\Delta(-\frac{\pi}{a}) = -\infty,~~~~~~~~~~~~~\Delta(+\frac{\pi}{a}) = +\infty.               \label{extrinf} \eeq

Eq. (\ref{extrinf}) is a necessary condition for the associativity of the star product, but it is not sufficient.
In fact associativity requires that the result of the integration at the r.h.s. of (\ref{kernelass}) is symmetric in the three momenta of the component fields, and this poses severe restrictions, difficult to be met,  on the integration volume $V(p;p_1,p_2)$. 
In order to discuss this point let us first assume that (\ref{extrinf}) is satisfied and that given an arbitrary $\hat{p}$ the equation $\Delta(p)=\hat{p}$ has one and only one solution with $p$ in the interval $(-\frac{\pi}{a}, \frac{\pi}{a} )$ and, due to the symmetry $p \rightarrow \frac{2\pi}{a} - p$, one and only one solution in the interval  $(\frac{\pi}{a}, \frac{3\pi}{a} )$.  Let now $q_{12}$ be the solution of 
\beq
\Delta(q_{12}) = \Delta(p_1)+\Delta(p_2)~~~~~~~~~~-\frac{\pi}{a}\leq q_{12} \leq \frac{\pi}{a},  \label{qq} \eeq
then the integration in  (\ref{kernelass}) can be performed and gives:
\beq
K(p;p_3;p_1,p_2) = \frac{1}{\left|\frac{d\Delta(q_{12})}{dq_{12}}\right| } \left( V(q_{12};p_1,p_2)V(p;q_{12},p_3)+V(\frac{2\pi}{a}-q_{12};p_1,p_2)V(p;\frac{2\pi}{a}-q_{12},p_3)\right). \label{kernelintegr} \eeq 
The associativity condition is then given by 
\beq
K(p;p_3;p_1,p_2) = K(p;p_2;p_1,p_3),  \label{asscond} \eeq
or equivalently the same with different permutations of the momenta, with the kernel $K$ given by (\ref{kernelintegr}) and (\ref{qq}).
This is a non trivial functional equation, whose most general solution is unknown to us as yet. 
The most general case is the one labelled as \textbf{B} in the previous section, in which the fields $\tilde{\varphi}_A(p)$ have no symmetry with respect to the transformation $p \rightarrow \frac{2\pi}{a} - p$, and hence all the $2^d$ doublers on the lattice correspond to distinct degrees of freedom in the continuum.
In this case also the integration volume $V(p;p_1,p_2)$ does not have any symmetry, and we do not know if any solution of (\ref{asscond}) exists at all.
Indeed there is an almost trivial solution to (\ref{asscond}), discussed below, that corresponds however to the case \textbf{A} of the previous section, namely to the case where all doublers are identified.
This solution is given by:
\beq
V(p;p_1,p_2) = \left|\frac{d\Delta(p)}{dp}\right| \frac{f(p_1)f(p_2)}{f(p)}
,\label{symmetricvolume} \eeq
where $f(p)$ has to be periodic but is otherwise arbitrary. The function $f(p)$ amounts to a momentum dependent rescaling of all fields and can always be absorbed by a field redefinition. 
Let us relax for a moment the first associativity condition (\ref{extrinf}), then with the volume element  (\ref{symmetricvolume})  the kernel $K\left(p;p_3;p_1,p_2\right)$ is given by (\ref{asymmetriccond}) and by:
\beq
K\left(p;p_3;p_1,p_2\right)=2\left|\frac{d\Delta(p)}{dp}\right| \frac{f(p_1)f(p_2)f(p_3)}{f(p)}~~~~~~~~\text{for}~~~~~~~~\left|\Delta(p_1)+\Delta(p_2) \right| < |\Delta(\frac{\pi}{a})|.  \label{kernel2} \eeq
The first equation in (\ref{kernel2}) is symmetric in the momenta $p_i$ of the constituent fields; moreover, if we  reinstate eq. (\ref{extrinf}), namely $|\Delta(\frac{\pi}{a})| = \infty$, the inequality in (\ref{kernel2}) is valid for all values of $p_i$ and so the corresponding star product is associative. 
The final form of the associative star product corresponding to (\ref{symmetricvolume}) is then given by:
\beq
f(p) ~\widetilde{\varphi_1 \star \varphi_2 }(p) =\frac{1}{2\pi}  \int_{-\frac{\pi}{a}}^{\frac{3\pi}{a}} dp_1 dp_2  \left|\frac{d\Delta(p)}{dp}\right| f(p_1) \tilde{\varphi}_1(p_1)~ f(p_2) \tilde{\varphi}_2(p_2) \delta\left( \Delta(p) -\Delta(p_1) - \Delta(p_2) \right) 
, \label{assstarprodp} \eeq
with
\beq
|\Delta(\pm\frac{\pi}{a})| = \infty
.\label{deltainfinite} \eeq

The products $f(p_i) \tilde{\varphi}_i(p_i)$ in (\ref{assstarprodp}) should be chosen symmetric under $p_i \rightarrow \frac{2\pi}{a} - p_i$:
\beq
f(\frac{2\pi}{a} - p_i) \tilde{\varphi}_i(\frac{2\pi}{a} - p_i)  = f(p_i) \tilde{\varphi}_i(p_i), \label{doubid} \eeq
because the antisymmetric part of $f(p_i) \tilde{\varphi}_i(p_i)$  gives a vanishing contribution to the star product. 
This follows from the symmetry of $\Delta(p_i)$ under such transform, and can be checked by splitting the integration interval $(-\frac{\pi}{a},\frac{3\pi}{a})$ in half, and performing the  change of variable $ p'_i = \frac{2\pi}{a} - p_i$ on the integrals between $\frac{\pi}{a}$ and $\frac{3\pi}{a}$ .
On the other hand, due to the symmetry of $\Delta(p)$ also $f(p) \widetilde{\varphi_1\star \varphi_2 }(p)$ satisfies the same  condition, so that eq.(\ref{doubid})  is valid for all fields\footnote{The same is true, with obvious generalization, in higher dimensions.  }.
Eq. (\ref{doubid}) coincides with eq. (\ref{hcond2}) and can be reduced to (\ref{chiralcond}) by fields redefinition. So the associative integration volume (\ref{symmetricvolume}) implies the identification of the doublers, as discussed in the case \textbf{A} of the previous section.
 The symmetry (\ref{doubid}) can be used to reduce the integrals in  (\ref{assstarprodp}) to integrals over the $(-\frac{\pi}{a}, \frac{\pi}{a})$ interval: 
\beq
f(p) ~\widetilde{\varphi_1 \star \varphi_2 }(p) =\frac{2}{\pi}  \int_{-\frac{\pi}{a}}^{\frac{\pi}{a}} dp_1 dp_2  \frac{d\Delta(p)}{dp} f(p_1) \tilde{\varphi}_1(p_1) ~f(p_2) \tilde{\varphi}_2(p_2) \delta\left( \Delta(p) -\Delta(p_1) - \Delta(p_2) \right) 
, \label{assstarprodpr} \eeq
where we have assumed, as discussed earlier,  that $p=\pm \frac{\pi}{a}$ are the only extremes of $\Delta(p)$ and hence dropped the absolute value within the integral.
The arbitrary function $f(p)$  defines the  integration volume over the momentum $p$ and  it amounts to a rescaling of the fields in momentum space. So different choices of $f(p)$ define, modulo such rescaling, the same star product.

The associative product defined in (\ref{assstarprodpr}) is equivalent to the standard local product of the continuum theory, provided  the lattice fields $\tilde{\varphi}(p)$ and the continuum fields $\tilde{\Phi}(\hat{p})$ are identified in a suitable way. 
In fact if we define 
\beq
  \tilde{\varphi}(p)= \frac{1}{f(p)}\frac{d\Delta(p)}{dp}  \tilde{\Phi}(\Delta(p))~~~~~~~~~~~~~-\frac{\pi}{a}\leq p \leq \frac{\pi}{a},   \label{Phivarphi} \eeq
the star product (\ref{assstarprodpr}) becomes:
\beq
\widetilde{\Phi_1 \star \Phi_2 }(\hat{p}) = \frac{2}{\pi}  \int_{-\Delta(\frac{\pi}{a})}^{\Delta(\frac{\pi}{a})} d\hat{p}_1 d\hat{p}_2  \tilde{\Phi}_1(\hat{p}_1)\tilde{\Phi}_2(\hat{p}_2) \delta \left( \hat{p} - \hat{p}_1 - \hat{p}_2 \right), \label{contpr} \eeq
where, as in (\ref{momentumdelta}), we have put $\hat{p} = \Delta(p)$. 
If $\Delta(\frac{\pi}{a})$ in (\ref{contpr}) is finite, the star product written in terms of the $\tilde{\Phi}$ fields looks just like the ordinary product  of the continuum theory, written in momentum representation, but where a cutoff on the momenta has been introduced.
On the other hand if  $\Delta(p)$ satisfies the conditions (\ref{extrinf}), namely it  becomes $\pm \infty$  at the extremes $\pm \frac{\pi}{a}$ ( an explicit example of $\Delta(p)$ satisfying this condition will be considered and studied in the next section) the star product written in terms of  $\tilde{\Phi}$ is associative and  coincides with the standard local product written  in momentum representation.

The fields $\Phi_i(\hat{p})$ in eq. (\ref{contpr}), with $-\infty \leq  \hat{p} \leq \infty$, can be interpreted as fields of a continuum theory in the momentum representation.     Eq. (\ref{Phivarphi}) is then a kind of ``blocking transformation" from the continuum to the lattice\footnote{A similar transformation within the context of conventional lattice theory is given in (\ref{bltrp}).}, in the sense that given a field configuration in the continuum it produces  a corresponding lattice field configuration. If $\Delta(p)$ is not limited, namely $|\Delta(\frac{\pi}{a})|= \infty$, this correspondence is one-to-one and the map between lattice and continuum field configuration  is invertible. In fact in that case $\Delta^{-1}(\hat{p})$ exists and is uniquely determined for any real $\hat{p}$ with values  in the $(-\frac{\pi}{a}, \frac{\pi}{a})$ interval. Hence we can write:
\beq
  \tilde{\Phi}(\hat{p})=f\left(\Delta^{-1}(\hat{p})\right) \frac{d\Delta^{-1}(\hat{p})}{d\hat{p}} \tilde{\varphi}\left(\Delta^{-1}(\hat{p})\right)~~~~~~~~~~~~~-\frac{\pi}{a}\leq \Delta^{-1}(\hat{p}) \leq \frac{\pi}{a}.   \label{varphiPhi} \eeq
 In this case the lattice fields  and the continuum fields describe the same degrees of freedom, and the lattice field amounts to a discrete relabeling of the continuum degrees of freedom. 
On the other hand if $\Delta(p)$ is limited to a finite range of values, namely if  $|\Delta(\frac{\pi}{a})|< \infty$ such relation is incomplete and the blocking transformation is not invertible because $\Delta^{-1}(\hat{p})$ does not exists for $\hat{p}> |\Delta(\frac{\pi}{a})|$ and the corresponding degrees of freedom have no  correspondence on the lattice. 

It would be important at this stage to establish if (\ref{symmetricvolume}) is the only solution of (\ref{asscond}), namely the only integration volume compatible with the associativity of the star product. We already mentioned that for case \textbf{B}, namely the case with independent doublers, the question is still open as whether a solution exists at all. In the case \textbf{A}, namely when all doublers are identified and the fields satisfy eq. (\ref{chiralcond}) modulo a rescaling of the fields, we can produce, if not a rigorous proof, a solid argument based on an algebraic investigation performed with the aid of Mathematica, that  (\ref{symmetricvolume}) is the only solution. 

The argument goes as follows: suppose the fields satisfy, after a suitable rescaling, the symmetry condition given in eq.(\ref{chiralcond}), then also the integration volume $V(q;p_1,p_2)$ must satisfy the same symmetry in all its variables and eq. (\ref{kernelintegr}) simplifies to:
\beq
K(p;p_3;p_1,p_2) =2 \frac{1}{\left|\frac{d\Delta(q_{12})}{dq_{12}}\right| }  V(q_{12};p_1,p_2)V(p;q_{12},p_3), \label{kernelintegrsimpl} \eeq
where all variables may now be restricted to the interval $(-\frac{\pi}{a}, \frac{\pi}{a})$.
Given the one-to-one correspondence between $p$ in the above interval and $\hat{p} = \Delta(p)$ ranging from $-\infty$ to $+\infty$, we can use $\hat{p}$ as independent variable and define:
\beq
\mathcal{V}(\hat{p}_1+\hat{p}_2,\hat{p}_1,\hat{p}_2) = \frac{1}{\left|\frac{d\Delta(p)}{dp}\right| } V(p;p_1,p_2),  \label{calV} \eeq
where $p$ on the r.h.s. is given by $\Delta(p)=\Delta(p_1)+\Delta(p_2)$. The associativity condition (\ref{asscond}) in terms of $\mathcal{V}(\hat{p},\hat{p}_1,\hat{p}_2)$ reads:
\beq
\mathcal{V}(\hat{p}_1+\hat{p}_2,\hat{p}_1,\hat{p}_2) \mathcal{V}(\hat{p},\hat{p}_1+\hat{p}_2,\hat{p}_3) = \mathcal{V}(\hat{p}_1+\hat{p}_3,\hat{p}_1,\hat{p}_3) \mathcal{V}(\hat{p},\hat{p}_1+\hat{p}_3,\hat{p}_2) \label{ass3}
, \eeq
with $\hat{p}=\hat{p}_1+\hat{p}_2+\hat{p}_3$.
The solution of (\ref{ass3}) that corresponds to (\ref{symmetricvolume}) is given by:
\beq
\mathcal{V}(\hat{p}_1+\hat{p}_2,\hat{p}_1,\hat{p}_2) = \frac{\hat{f}(\hat{p}_1) \hat{f}(\hat{p}_2)}{\hat{f}(\hat{p}_1+\hat{p}_2)}, \label{solg} \eeq
where $\hat{f}(\hat{p})=f(p)$.
The ansatz is that (\ref{solg}) is the most general solution of the associativity condition (\ref{ass3}). We do not have a rigorous proof, but the ansatz can be checked order by order by expanding $\mathcal{V}(\hat{p}_1+\hat{p}_2,\hat{p}_1,\hat{p}_2) $ in power series of $\hat{p}_1$ and $\hat{p}_2$ and determining the restrictions on the coefficients of the expansion imposed by (\ref{ass3}). The resulting expansion matches exactly the one obtained by expanding (\ref{solg}) in powers series, the residual arbitrariness of the coefficients corresponding exactly to the arbitrariness of the function $\hat{f}(\hat{p})$ in (\ref{solg}). This has been checked with Mathematica up to the sixth order.
In conclusion, if the doublers are identified according to the case \textbf{A} of the last section, the only associative star product coincides with the ordinary local product of the continuum theory provided the lattice and the continuum degrees of freedom are related by the identity (\ref{Phivarphi}). If the doublers on the other hand are kept as independent degrees of freedom (case \textbf{B}) no associative star product has been found, but its existence has not been ruled out so far. 
\subsection{Some provisional conclusions on star product associativity and the building of lattice theories}
\label{3.1.1}
The star product was introduced to replace  the local product on the lattice in a way that preserves the Leibniz rule for the derivative operator. If the star product enjoyed the same formal properties as the standard local product it would be possible to formulate a lattice theory starting  from one in the continuum by simply replacing the product with the star product and the derivative with the lattice derivative operator $\Delta(p)$.  

We have seen that the only known case in which this is possible is the one discussed in the last  subsection, namely the star product with the integration volume given by  (\ref{symmetricvolume}) and an unbound derivative operator $\Delta(p)$ satisfying (\ref{extrinf}). This however leads to a lattice theory where the degrees of freedom are in one-to-one correspondence with the ones of the continuum theory (see eq.s (\ref{Phivarphi}) and (\ref{varphiPhi})), hence to a reformulation of the continuum theory in lattice language, where the lattice spacing doesn't act as a regulator but  merely as an arbitrary unit used to make momenta dimensionless.   

The fact that a continuum theory (in fact any continuum theory) may be reformulated on a lattice by replacing the product with a non local star product and the derivative by a non local operator $\Delta(p)$ is in itself interesting, but it does not correspond to the original purpose of the lattice regularization scheme.

On the other hand if we give up associativity we are faced with two different types of problems. The first occurs in the interaction terms. If we insist in using the star product as the building block, an ambiguity arises with terms  in the Lagrangian higher than quadratic in the fields. For instance in a $\varphi^4$ interaction with identical fields, interaction terms of the form $\left(\varphi\star\varphi\right)\star\left(\varphi\star\varphi\right)$ and $\left(\left(\varphi\star\varphi\right)\star\varphi\right)\star\varphi$ are essentially different if associativity is violated. This problem can be overcome by defining the interaction terms in a symmetric way without making use of the star product, for instance in the $\varphi^4$ interaction by writing:
\beq
I_4 = g \int dp_1dp_2dp_3dp_4 V(p_1,p_2,p_3,p_4) \tilde{\varphi}(p_1)\tilde{\varphi}(p_2)\tilde{\varphi}(p_3)\tilde{\varphi}(p_4) \delta\left( \Delta(p_1)+\Delta(p_2)+\Delta(p_3)+\Delta(p_4) \right), \label{symmphi4} \eeq
where $V(p_1,p_2,p_3,p_4)$ is a suitable integration volume. 

The second problem is more serious. The original idea of this approach is to obtain a lattice theory directly from the continuum theory by simply replacing the derivative with the lattice derivative operator $\Delta(p)$ and the product with the star product. If the lattice derivative and the star product enjoy the same formal properties as the corresponding objects in the continuum then all continuum symmetries would automatically be preserved on the lattice.
 
Lack of associativity in the star product would then break all those symmetries that rely upon it in the continuum. Supersymmetry is not among those. In fact supersymmetry is a global symmetry: the parameters of the transformations are constants and do not carry momentum hence supersymmetry transformations are  local in momentum representation, namely they do not involve any convolution in momentum space.
As a consequence associativity of the product does not enter in the proof of invariance under supersymmetry and so exact supersymmetry is not affected by the lack of associativity of the star product. On the other hand supersymmetry  crucially depends on the validity of the Leibniz rule for the derivative, which is why keeping the Leibniz rule is a key ingredient of our approach.
This is consistent with the results of ref.s  
 \cite{DFKKS} and \cite{DKKS} where lattice actions with exact $N=2$ supersymmetry were constructed for supersymmetric Wess Zumino models in $D=1$ and $D=2$.  The star product used in these papers was based on a lattice derivative operator of the form $\Delta(p)= \frac{2}{a} \sin \frac{ap}{2} $ , and hence, according to the previous discussion, certainly non associative. Nevertheless the resulting lattice theories had exact supersymmetry, even at the quantum level \cite{WTid} .

The violation of associativity is instead fatal for gauge invariance. In fact gauge transformations  are represented in coordinate space by local products of the fields with the $x$ dependent paramenters of the gauge transformations. These products would consistently become star products on the lattice, and, for instance, an infinitesimal gauge transformation of a charged $U(1)$ field would take the form:
\beq
\delta_{\alpha} \varphi(x) = i (\alpha \star \varphi )(x). \label{gaugestar} \eeq
 For the would be gauge invariant quantity $(\varphi^{\dagger}\star\varphi)(x)$ this implies:
 \beq \delta_{\alpha}(\varphi^{\dagger} \star\varphi)(x) = -i \left( (  \varphi^{\dagger} \star  \alpha) \star  \varphi\right)(x) + i \left( \varphi^{\dagger} \star ( \alpha\star\varphi )\right)(x)
,   \label{gauge} \eeq
and the r.h.s. does not vanish unless the star product is associative.
As a consequence, unless some new associative star product is found in the \textbf{B} case, the only star product which can preserve gauge invariance on the lattice is the one given in (\ref{symmetricvolume}) which coincides with ordinary local product in the continuum modulo the field identification (\ref{Phivarphi}).  This merely corresponds to a relabeling of the degrees of freedom, although a non trivial one, and does not involve  any regularization procedure. Within this scheme the lattice constant $a$ is simply a dimensional quantity used to turn the compact momenta on the lattice into dimensionless angular variables. If such dimensionless variables are used, the lattice constant never appears and hence it does not play any role as regulator. 
The last result can then be presented as a kind of ``no go theorem": within the scheme \textbf{A} of the previous section (doublers identified) there is no lattice regularization (in the sense described above) that preserves gauge invariance.
 
\subsection{An associative star product}
\label{3.2}
In the lattice  formulation  of the $N=2$ Wess Zumino model in $D=1$ and $D=2$ with exact supersymmetry given in ref.s \cite{DFKKS} and \cite{DKKS} we used the symmetric lattice  difference operator $\Delta_s(p_\mu)$  given by (see eq. (\ref{finitediff}) with $l=\frac{a}{2}$):
\beq
\Delta_s(p_\mu) = \frac{2}{a} \sin\frac{ap_\mu}{2}. \label{finitediff2} \eeq 

This choice satisfies all the conditions discussed in sec. \ref{1.2} and has the advantage of being local in coordinate representation. However $\Delta_s(p_\mu)$ does not satisfy the necessary condition (\ref{extrinf}) for the associativity of the star product, so that the corresponding star product is not associative irrespective of the choice of the integration volume. 
In order to construct an associative star product we have to choose a lattice derivative operator $\Delta(p_\mu)$ that satisfies eq. (\ref{extrinf}). 
Moreover we shall also maintain the requirements for $\Delta(p_\mu)$ already stated in the previous section, namely that it should be an odd function of $p_\mu$, symmetric under the transformation $p_\mu \rightarrow \frac{2\pi}{a} -p_\mu$ and of course periodic with period $\frac{4\pi}{a}$.  We shall also assume that its normalization is such that it satisfies eq. (\ref{dd}) for small values of $p_\mu$.

By putting all these pieces of information together one concludes that $\Delta(p_\mu)$ must be expressed by a power series in $\sin \frac{ap_\mu}{2}$, with only odd powers and with singularities at $\sin \frac{ap_\mu}{2} = \pm1$ such that the equation
\beq \lim_{p\rightarrow \pm \frac{\pi}{a}} \Delta(p) = \pm \infty, \label{infpoints} \eeq 
is satisfied.
Clearly the presence of only a finite number of powers of $\sin\frac{ap}{2}$ in the expansion of $\Delta(p)$ is not compatible with the condition (\ref{infpoints}), and the presence of arbitrarily large powers of $\sin \frac{ap}{2}$ in the expansion of $\Delta(p)$ means that $\Delta(p)$ defines on the lattice a non local derivative, containing differences between arbitrarily far away points. This is the main difference with the case $\Delta(p)=\Delta_s(p)$ 
considered in \cite{DFKKS} and \cite{DKKS}  where the lattice derivative is the usual local finite difference operator.

There are in principle infinitely many choices of $\Delta(p)$ satisfying eq. (\ref{infpoints}) and all the other conditions discussed above, all of them non local in coordinate representation. However we shall restrict these choices by selecting the most local (or the least non-local) form of $\Delta(p)$ compatible with (\ref{infpoints}).
Consider the power expansion of $\Delta(p)$:
\beq \Delta(p) =  \sum_{k=0}^{\infty} c_k  \left( \sin \frac{ap}{2} \right)^{2k+1}, \label{deltapower} \eeq
normalized with $c_0 = \frac{2}{a}$. 
Suppose that in proximity of $\sin \frac{ap}{2} = 1$ the function  $\Delta(p)$ has a singularity and behaves as
\beq \Delta(p) \approx \left(1-\sin\frac{ap}{2} \right)^{-\alpha}, \label{asymp} \eeq
with $\alpha>0$ so that (\ref{infpoints}) is satisfied. 
Such asymptotic behaviour around the singular point is determined by the large $k$ asymptotic behaviour of $c_k$ according to the relation:
\beq c_k \approx k^{\alpha-1}. \label{asymp2} \eeq

It is clear from (\ref{asymp2}) that the fastest decrease of $c_k$ as $k \rightarrow \infty $ compatible with (\ref{infpoints}) is obtained choosing $\alpha = 0$, to be interpreted as a logarithmic behaviour of $\Delta(p)$ at $\sin \frac{ap}{2} =1$. By imposing the same condition at $\Delta(p)$ at $\sin \frac{ap}{2} =-1$ ( but with a $-\infty$ limiting value) , and using the proper normalization one is led to the following ansatz:
\beq \Delta_G(p) = \frac{2}{a}\gd (\frac{ap}{2}), \label{assdelta} \eeq
where $\gd (x)$ is the inverse of the Gudermannian function \cite{Gudermann}
 and is given by:
\beq \gd(x) = \frac{1}{2} \log \frac{1+ \sin x}{1- \sin x}, \label{gd} \eeq
\begin{figure}
\begin{center}
\mbox{\includegraphics*[width=12.cm]{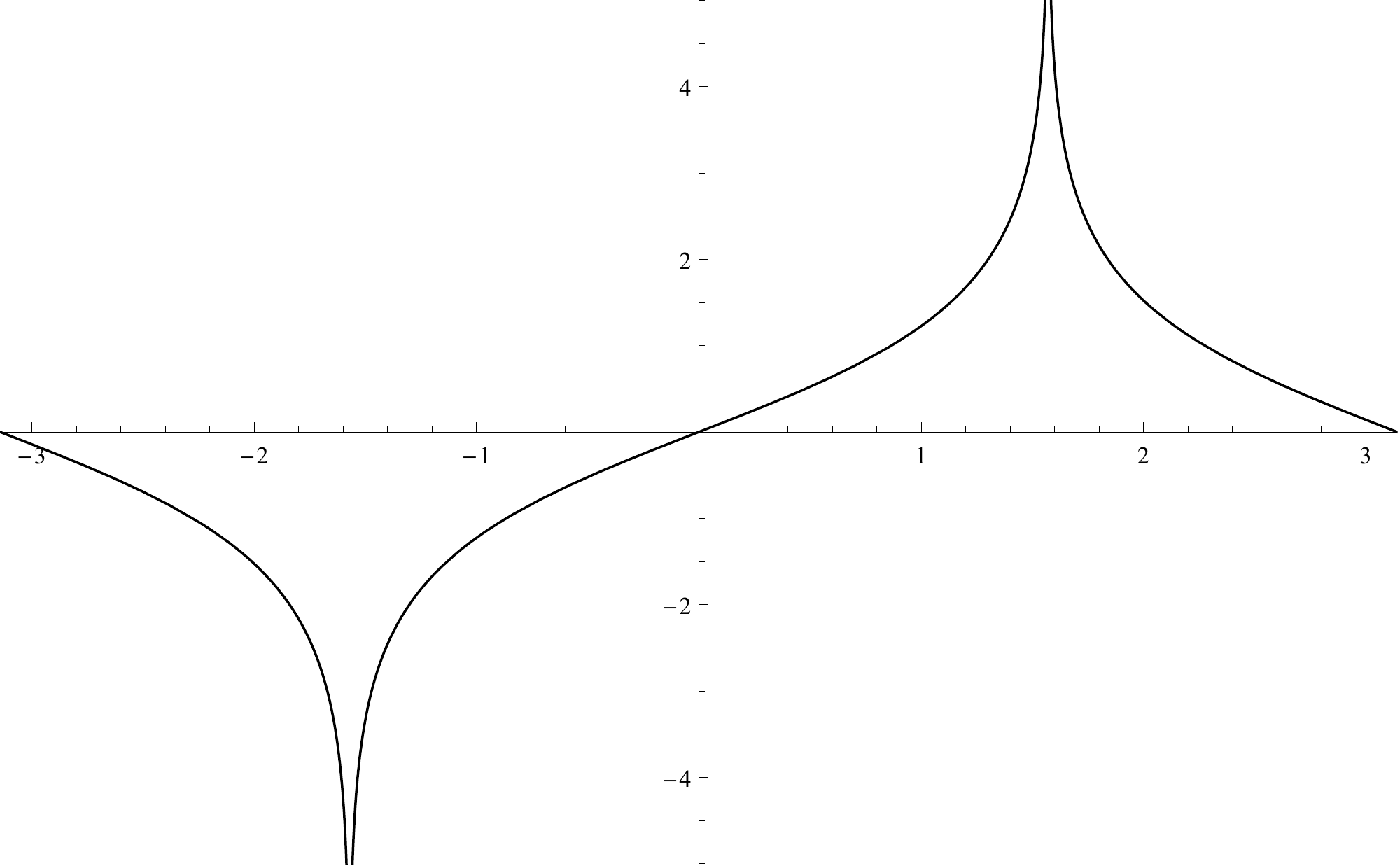}}
\caption{Plot of the Inverse Gudermannian function $\gd(x)$ in the fundamental interval $(-\pi,\pi)$}
\label{guderm}
\end{center}
\end{figure}
If we identify $\Delta_G(p)$ with the conserved and additive momentum $\hat{p}$ in the continuum, eq. (\ref{assdelta}) provides a very natural map between the momentum on the lattice and the momentum of the continuum theory:
\beq \frac{a\hat{p}}{2} = \gd\left(\frac{ap}{2}\right).   \label{assdelta2} \eeq

Notice that in the map described by (\ref{assdelta2}) the momentum $\hat{p}$ of the continuum theory ranges from $-\infty$ to $+\infty$ as $p$ goes from $-\frac{\pi}{a}$ to $\frac{\pi}{a}$  and it goes back to $-\infty$ as $p$ goes from $\frac{\pi}{a}$ to $\frac{3\pi}{a}$, so that the map covers the real axis twice as the angular variable $\frac{ap}{2}$ covers a $2\pi$ period (see Fig.\ref{guderm}).

The Gudermannian function and its inverse are particularly suitable to define a map between a compact and a non-compact one dimensional manifold, as they naturally transform hyperbolic functions into trigonometric functions.
If
\beq y=\gd (x), \label{ygdx} \eeq
then the following relations hold:
\beq \cosh y = \frac{1}{\left| \cos x \right|},~~~~~~~~~~ \tanh y =\sin x ,~~~~~~~~~~\sinh y =\frac{\sin x}{\left| \cos x \right|},  \label{hyptrig} \eeq
and also
\beq dy = \frac{dx}{ \cos x },~~~~~~~~~~dx = \pm \frac{dy}{\cosh y}, \label{dxdy} \eeq
where in the last equation the plus sign holds when $x$ is in the interval $(-\frac{\pi}{2},\frac{\pi}{2})$ and the minus sign for $x$ in $(\frac{\pi}{2},\frac{3\pi}{2})$.

The expansion of $\Delta_G(p)$ in powers of $\sin \frac{ap}{2}$ can be easily derived from the expansion of the logarithm, and reads:
\beq  
\frac{a}{2} \Delta_G(p) = \sin \frac{ap}{2} + \frac{1}{3} \left( \sin \frac{ap}{2}\right)^3 +  \frac{1}{5} \left( \sin \frac{ap}{2}\right)^5 +  \frac{1}{7} \left( \sin \frac{ap}{2}\right)^7 + \cdots.    \label{pwex} \eeq
A different and less trivial expansion in powers of $e^{\imath \frac{ap}{2}}$ can be written as:
\beq \frac{a}{2} \Delta_G(p) = 2 \left[  \sin \frac{ap}{2} -  \frac{1}{3}  \sin \frac{3ap}{2}+  \frac{1}{5}  \sin \frac{5ap}{2}-  \frac{1}{7}  \sin \frac{7ap}{2}+\cdots \label{pwex2}\right] \eeq
This expansion is interesting because it shows immediately how $\Delta_G(p)$ acts on a field in coordinate representation. In fact, if we denote by $\Delta_G \varphi(x)$ the Fourier transform of $\Delta_G(p) \tilde{\varphi}(p)$, eq. (\ref{pwex2}) gives:
\beq \Delta_G\varphi(x) = \frac{2}{a} \sum_{k=1}^{\infty}  \frac{(-1)^{k+1}}{2k-1}  \left[ \varphi\left(x+\frac{(2k-1)a}{2}\right) -  \varphi\left(x-\frac{(2k-1)a}{2}\right)\right]. \label{deltacoord} \eeq
The expansion (\ref{deltacoord}) has a similar structure to the one of the corresponding expansion for the SLAC derivative, that acts however on a lattice with spacing $a$:
\beq \Delta_{\mathrm{SLAC}}\varphi(x) = \frac{1}{a} \sum_{k=1}^{\infty} \frac{(-1)^{k+1}}{k} \left[ \varphi(x+ka)-\varphi(x-ka) \right]. \label{slacdelta} \eeq
Both $\Delta_G$ and $\Delta_{\mathrm{SLAC}}$ reduce to the ordinary derivative in the continuum limit $a \rightarrow 0$ as they both reduce to $p$ for $p \ll \frac{1}{a}$. This can be seen in the case of $\Delta_G$ from the expansion (\ref{pwex}) where only the first term survives in that limit, all the others being proportional to higher derivatives and higher orders of $a$.
A naive continuum limit on (\ref{deltacoord}) and (\ref{slacdelta}) instead gives an undetermined result. In fact in both cases each term, corresponding to a fixed value of $k$, gives in the $a \rightarrow 0$ limit a contribution $\pm \frac{d\varphi(x)}{dx}$ and the total result is
\beq 2  \frac{d\varphi(x)}{dx} \left[ 1 -1+1-1+1-\cdots \right]. \label{undet} \eeq 
In the coordinate space both derivatives behave similar but the momentum representations have a fundamental difference. 
SLAC derivative does not satisfy eq.(\ref{extrinf}) and thus associativity is broken while $\Delta_G$ satisfies associativity 
for the star product. 

The correct result of the infinite alternating series in (\ref{undet}) is $\frac{1}{2}$, but it can only be obtained by resumming the series before taking the $a \rightarrow 0$ limit, that is going back to the momentum space representation.
\begin{figure}
\begin{center}
\mbox{\includegraphics*[width=12.cm]{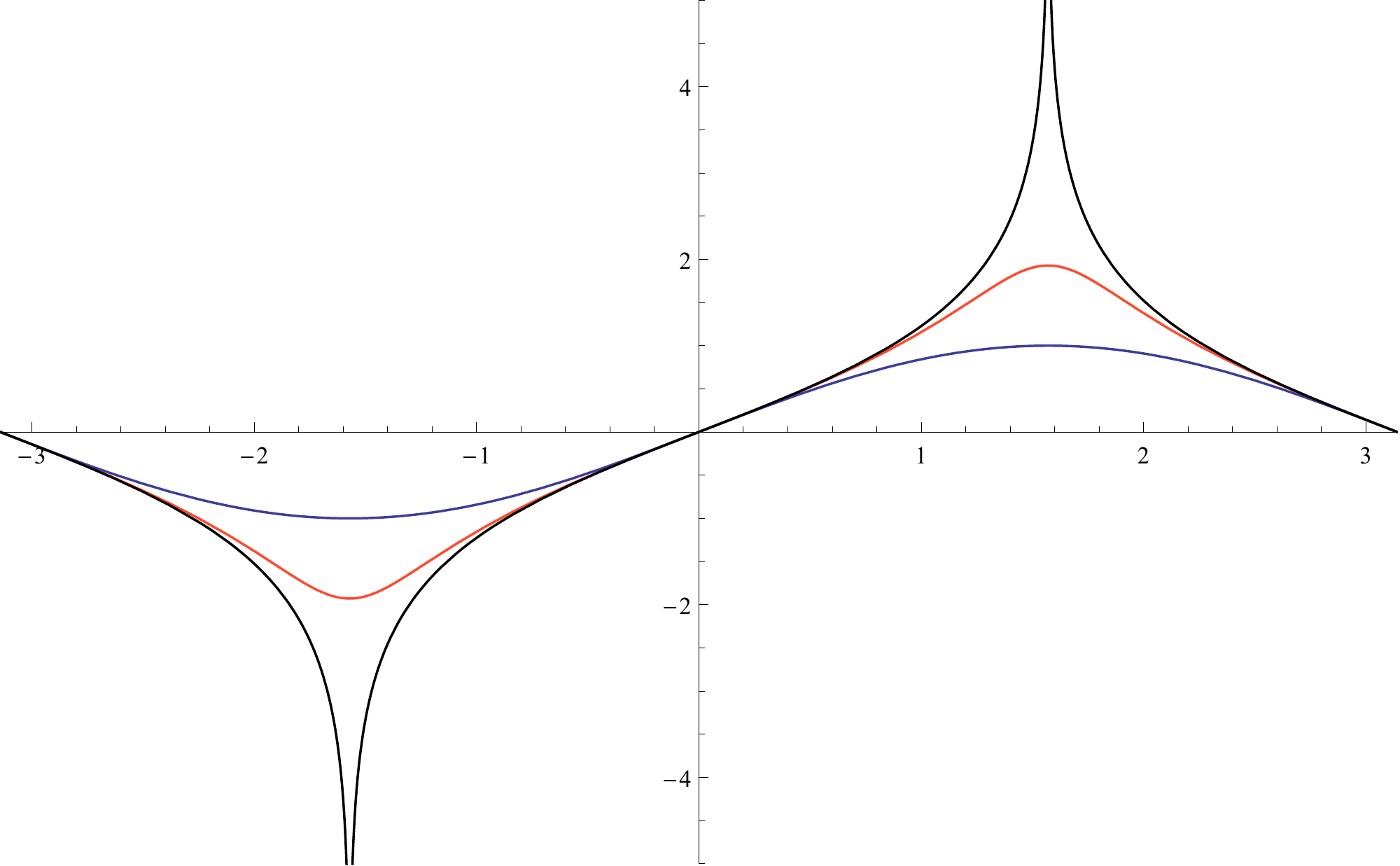}}
\caption{Plot of the smoothed inverse Gudermannian function $\gd(x,\hat{z})$ at $\hat{z}=0.95$, in the fundamental interval $(-\pi,\pi)$. The sine function and the inverse Gudermannian functions are also plotted for comparison.}
\label{guderm2}
\end{center}
\end{figure}
In order to regularize the series at the r.h.s. of (\ref{deltacoord}) we shall introduce a new parameter $z$ and define a regularized derivative operator $\Delta^{(z)}_G$ as:

\beq \Delta^{(z)}_G\varphi(x) = \frac{1+z^2}{a} \sum_{k=1}^{\infty}  \frac{(-1)^{k-1} z^{2(k-1)}}{2k-1}  \left[ \varphi\left(x+\frac{(2k-1)a}{2}\right) -  \varphi\left(x-\frac{(2k-1)a}{2}\right)\right]. \label{deltacoordz} \eeq
Clearly $ \Delta^{(z)}_G\varphi(x)$ coincides with $\Delta_G \varphi(x)$ for $z=1$; for $|z|<1$ the series involved in the continuum limit $a \rightarrow 0$ are convergent and the limit reproduces $\frac{d\varphi(x)}{dx}$ as expected.
The regularization given in (\ref{deltacoordz}) amounts in momentum representation in replacing $\Delta_G(p)$ with its regularized counterpart  $\Delta^{(z)}_G(p)$ given by:
\beq
\Delta^{(z)}_G(p) = \frac{2}{a} \gd(x,\hat{z}),  \label{regD} \eeq
where
\beq
\hat{z} = \frac{2 z}{1+ z^2}, \label{zzhat} \eeq
and $\gd(x,\hat{z})$ is a regularized inverse Gudermannian function given by:
\beq
\gd(x,\hat{z}) = \frac{1}{2\hat{z}} \log\frac{1+\hat{z}\sin x}{1-\hat{z}\sin x}. \label{regguder} \eeq
The function $\gd(x,\hat{z})$ interpolates between the sine function (at $\hat{z}=0$) and the inverse Gudermannian function (at $\hat{z}=1$). This is shown in fig.\ref{guderm2} where  $\gd(x,0.95)$ is plotted together with $\sin(x)$ and $\gd(x)$.

It is clear from (\ref{regD}) and (\ref{regguder}) that an expansion of $\Delta^{(z)}_G(p)$ in powers of $\hat{z}$ is also an expansion in powers of $\sin\frac{ap}{2}$ and it reduces to (\ref{pwex}) in the limit $\hat{z}\rightarrow 1$, while an expansion in powers of $z$ is an expansion in the base of $\sin\frac{nap}{2}$ and it reduces to (\ref{pwex2}) for $z=1$.
For $z<1$ the regularized derivative operator $\Delta^{(z)}_G(p)$ is bounded by
\beq
|\Delta^{(z)}_G(p)| \leq \frac{1}{a\hat{z}} \log\frac{1+\hat{z}}{1-\hat{z}}, \label{bound} \eeq
so that its use in place of $\Delta_G(p)$ in the definition of the star product would lead to a violation of associativity. The momentum cutoff given by the r.h.s. of (\ref{bound}) can be made however very large, indeed much larger than $\frac{1}{a}$, by choosing $z$ or equivalently $\hat{z}$ sufficiently close to $1$ thus providing a natural ultraviolet cutoff independent of $a$.
If the regularization of ultraviolet divergences is done by a cutoff in the momenta the cutoff can be sent to infinity keeping the value of the lattice constant $a$ finite, namely preserving the lattice structure of the theory. This will be discussed more in detail in Sec. \ref{section6}.

A similar regularisation could be introduced in the definition (\ref{slacdelta}) of the SLAC derivative. In momentum representation this would correspond to a smoothing of the saw-tooth function that would eliminate the discontinuity of the function but it would also reintroduce a second zero at $p=\frac{\pi}{a}$ which is responsible for the appearance of the doublers.  
 The previous discussion ultimately shows that in coordinate space both $\Delta_G$ and $\Delta_{\mathrm{SLAC}}$ are intrinsically non local, but as we shall see $\Delta_G$ acts in a completely different framework where the relation between the continuum and the lattice coordinates is not straightforward.

An explicit form of the associative star product is obtained by inserting $\Delta_G(p)$, given by (\ref{assdelta}),  into (\ref{starprodp2}) and 
(\ref{symmetricvolume}). The function $f(p)$ in  (\ref{symmetricvolume}) corresponds simply to a rescaling of the fields in momentum space and can be set equal to $1$ by a field redefinition. However it should be remarked that a rescaling of the fields in momentum space has non trivial effects in coordinate representation where it corresponds to a non-local convolution. Different choices of $f(p)$ may correspond to very different pictures when the fields are represented on the lattice. As we shall see further in this section a special choice for $f(p)$, namely $f(p)=\frac{1}{\sqrt{\left|\cos{\frac{ap}{2}}\right|}}$, results into a more symmetric representation of the associative star product in coordinate representation, and may be needed for a smooth continuum ($a \rightarrow 0$) limit  in coordinate representation. For the moment, while working in the momentum representation, we choose for simplicity $f(p)=1$ and write the associative star product (\ref{assstarprodp}) with $\Delta=\Delta_G$ as:
\beq \left| \cos\frac{ap_{12}}{2}\right|  \widetilde{\varphi_1 \star \varphi_2 }(p_{12})  =\frac{1}{2\pi}  \int_{-\frac{\pi}{a}}^{\frac{3\pi}{a}} dp_1 dp_2  \tilde{\varphi}_1(p_1) \tilde{\varphi}_2(p_2) \delta\left( \Delta_G(p_{12}) -\Delta_G(p_1) - \Delta_G(p_2) \right),  \label{assstarprodp2} \eeq
where the cosine factor on the l.h.s comes from the insertion of (\ref{dxdy}) into the integration volume (\ref{symmetricvolume}). This factor is essential for the associativity, so it cannot be absorbed into the definition of the star product.
Thanks to the symmetry of the fields and of the lattice derivative operator $\Delta_G$ under $p \rightarrow \frac{2\pi}{a}-p$ we can restrict the integration volume in (\ref{assstarprodp}) to the interval $(-\frac{\pi}{a},\frac{\pi}{a})$ and write:
\beq  \cos\frac{ap_{12}}{2}  \widetilde{\varphi_1 \star \varphi_2 }(p_{12})  =\frac{2}{\pi}  \int_{-\frac{\pi}{a}}^{\frac{\pi}{a}} dp_1 dp_2  \tilde{\varphi}_1(p_1) \tilde{\varphi}_2(p_2) \delta\left( \Delta_G(p_{12}) -\Delta_G(p_1) - \Delta_G(p_2) \right),  \label{assstarprodp2} \eeq
where we have assumed that also $p_{12}$ is in the interval $(-\frac{\pi}{a},\frac{\pi}{a})$ so that no absolute value is needed for the cosine factor.
With the change of variable (\ref{assdelta2}) the star product can be written as an integral over the continuum momenta $\hat{p}_i$, and becomes:
\beq
\widetilde{\Phi_1 \star \Phi_2 }(\hat{p}_{12}) = \frac{2}{\pi}  \int_{-\infty}^{\infty} d\hat{p}_1 d\hat{p}_2  \tilde{\Phi}_1(\hat{p}_1)\tilde{\Phi}_2(\hat{p}_2) \delta \left( \hat{p}_{12} - \hat{p}_1 - \hat{p}_2 \right), \label{contpr2} \eeq
where 
\beq
\tilde{\varphi}(p) = \cosh\frac{a\hat{p}}{2} \tilde{\Phi}(\hat{p})~~~~~~~~~~\mbox{or}~~~~~~~~~~~~\tilde{\Phi}(\hat{p}) = \left| \cos\frac{ap}{2} \right| \tilde{\varphi}(p)
,\label{vP} \eeq
and the relation between $\hat{p}$ and $p$ is given by (\ref{assdelta2}).
This result is essentially the same already obtained in subsection \ref{3.1} (see eq.s (\ref{contpr}) and (\ref{Phivarphi})), with $\Delta = \Delta_G$ and $f(p)=1$.
Thanks to the associativity of the product, the star product of an arbitrary number of fields does not depend on the sequence in which the single products are made and is given by
\bea
\left| \cos\frac{ap}{2}\right|  \widetilde{\left( \varphi_1 \star \varphi_2\star \cdots\star \varphi_n\right) }(p) &=&\left(\frac{1}{2\pi}\right)^{n-1}  \int\int_{-\frac{\pi}{a}}^{\frac{3\pi}{a}} dp_1 dp_2\cdots dp_n  \tilde{\varphi}_1(p_1) \tilde{\varphi}_2(p_2)\cdots \tilde{\varphi}_n(p_n) \cdot \nonumber\\  && \cdot \delta\left( \Delta_G(p) -\Delta_G(p_1) - \Delta_G(p_2)-\cdots-\Delta_G(p_n) \right).  \label{multiplestar} \eea
The integration domain can be restricted to the interval $\left(-\frac{\pi}{a},\frac{\pi}{a}\right)$, as in (\ref{assstarprodp2}), using the symmetry $p \rightarrow \frac{2\pi}{a}-p$ and then by using the field identification (\ref{vP}) can be expressed in terms of the continuum fields $\tilde{\Phi}_i$:
\bea
\widetilde{\left(\Phi_1 \star \Phi_2\star \cdots \star \Phi_n \right)}(\hat{p}) &=& 2 \left(\frac{1}{\pi}\right)^{n-1} \int\int_{-\infty}^{\infty} d\hat{p}_1 d\hat{p}_2\cdots d\hat{p}_n  \tilde{\Phi}_1(\hat{p}_1)\tilde{\Phi}_2(\hat{p}_2)\cdots \tilde{\Phi}_n(\hat{p}_n) \cdot \nonumber \\ && \cdot \delta \left( \hat{p} - \hat{p}_1 - \hat{p}_2-\cdots-\hat{p}_n \right). \label{multicont} \eea
As in (\ref{assstarprodp2}) this is just the convolution describing  the ordinary local product of $n$ fields in momentum representation. Therefore the associative star product on the lattice is completely equivalent to the ordinary product in the continuum provided the lattice and continuum fields are identified via eq.s (\ref{vP}).

An $n$-point interaction term can be obtained from (\ref{multiplestar}) by setting in it  $p=0$, namely:
\bea
I_n = \widetilde{\left( \varphi_1 \star \varphi_2\star \cdots\star \varphi_n\right) }(0) &=&\left(\frac{1}{2\pi}\right)^{n-1}  \int\int_{-\frac{\pi}{a}}^{\frac{3\pi}{a}} dp_1 dp_2\cdots dp_n  \tilde{\varphi}_1(p_1) \tilde{\varphi}_2(p_2)\cdots \tilde{\varphi}_n(p_n) \cdot \nonumber\\  && \cdot \delta\left( \Delta_G(p_1) + \Delta_G(p_2)+\cdots+\Delta_G(p_n) \right).  \label{npointlatt} \eea
This corresponds exactly via (\ref{vP}) to setting $\hat{p}=0$ in (\ref{multicont}), giving:
\bea
I_n= \widetilde{\left(\Phi_1 \star \Phi_2\star \cdots \star \Phi_n \right)}(0) &=& 2 \left(\frac{1}{\pi}\right)^{n-1} \int\int_{-\infty}^{\infty} d\hat{p}_1 d\hat{p}_2\cdots d\hat{p}_n  \tilde{\Phi}_1(\hat{p}_1)\tilde{\Phi}_2(\hat{p}_2)\cdots \tilde{\Phi}_n(\hat{p}_n) \cdot \nonumber \\ && \cdot \delta \left(  \hat{p}_1 + \hat{p}_2+\cdots+\hat{p}_n \right). \label{npointcont} \eea
In conclusion, given any field theory in the continuum, one can write a corresponding theory on the lattice simply by replacing in momentum representation the ordinary product with the associative star product (\ref{assstarprodp}) and the derivative operator with $\Delta_G(p)$. As the formal properties of the star product and of the derivative operator on the lattice $\Delta_G(p)$ are the same as the ones of the corresponding entities in the continuum all continuum symmetries are preserved on the lattice.
The degrees of freedom  on the lattice are obtained from the ones in the continuum by the relation (\ref{vP}) which is invertible, which means that the lattice theory has no less information as the original continuum theory, and does not provide on the other hand any regulator.

\subsection{Star product in coordinate representation: the locality issue}
\label{3.3}

The star product of eq.(\ref{starprodp2}) can be expressed in coordinate representation by taking the discrete Fourier transform of the quantities involved, including the delta function at the r.h.s.
For the fields we shall use the following conventions:
\beq
\tilde{\varphi}(p) = \frac{a}{2} \sum_n \varphi\left(\frac{na}{2}\right) e^{-i\frac{na}{2} p}, \label{fourier1} \eeq
\beq \varphi\left(\frac{na}{2}\right) = \frac{1}{2\pi} \int_0^{\frac{4\pi}{a}} dp \tilde{\varphi}(p) e^{i p \frac{na}{2}}, \label{fourier2} \eeq
where notations have been fixed so that in the continuum limit $a \rightarrow 0$ the sum over $n$ becomes the integral over the space-time coordinate $x=\frac{na}{2}$:
\beq
\frac{a}{2} \sum_n \rightarrow  \int dx
.\label{sumn} \eeq

The integration volume $V(p;p_1,p_2)$ in (\ref{starprodp2}) can also be written as a discrete Fourier transform:
\beq
V(p;p_1,p_2) = \frac{a^3}{8} \sum_{m,m_1,m_2} V_{m;m_1,m_2} e^{-i\frac{ma}{2}p+i\frac{m_1a}{2}p_1+i\frac{m_2a}{2}p_2}
\label{intvolx}
. \eeq
By inserting (\ref{fourier1}) and (\ref{intvolx}) into (\ref{starprodp2}) the star product can eventually be written in coordinate representation as:
\beq
\left( \varphi_1 \star \varphi_2 \right)\left(\frac{na}{2}\right) = \frac{a^2}{4} \sum_{n_1,n_2} K_{n;n_1,n_2} \varphi_1\left(\frac{n_1 a}{2}\right) \varphi_2\left(\frac{n_2 a}{2}\right)
\label{starpx}
, \eeq
where the kernel $K_{n;n_1,n_2}$ is given by:
\beq
K_{n;n_1,n_2} =\frac{a^3}{16\pi} \int_{-\infty}^{+\infty} d\xi \sum_{m,m_1,m_2}  V_{m;m_1,m_2} J_\Delta\left( \xi, x_{n-m}\right) J_\Delta\left( \xi,x_{n_1-m_1} \right )J_\Delta\left( \xi, x_{n_2-m_2}\right), \label{Kx} \eeq
where we have defined 
\beq
x_n = \frac{na}{2},  \label{xna2} \eeq
and the function $J_\Delta\left(\xi,x_n\right)$ is given by
\beq
J_\Delta\left(\xi,x_n\right) = \frac{1}{2\pi} \int_0^{\frac{4\pi}{a}} dp e^{i \xi \Delta(p) - i x_n p }
.\label{JDelta} \eeq

The locality properties of the star product are then closely related to the properties of the function $J_\Delta\left(\xi,x_n\right) $. It is interesting to compare the star product (\ref{starpx}) with the product in the standard lattice formulation. In that case the conserved quantity $\Delta(p)$ is the momentum $p$ itself and the momentum is conserved modulo $\frac{2\pi}{a}$.  The parameter $\xi$ is then an integer multiple  of $\frac{a}{2}$, namely $\xi = m \frac{a}{2}$, and the function $J_\Delta\left(\xi,x_n\right) $ is  proportional to $\delta(m - n)$. With a suitable choice of the integration volume, that is $V(p;p_1,p_2)=1$, the star product becomes then the usual local product of functions on the lattice.

The properties of $J_\Delta\left(\xi,x_n\right) $ obviously depend on the choice of the lattice derivative function $\Delta(p)$. Here we shall examine the two most relevant and in a sense extreme cases: the ultra-local symmetric lattice difference operator $\Delta_s(p)$ given in eq. (\ref{finitediff2}) and the non local operator $\Delta_G(p)$, given in (\ref{assdelta}), that appears in the associative star product (\ref{assstarprodp2}). 
Notice that in the first case, namely $\Delta(p)=\Delta_s(p)$ the function $J_{\Delta_s}\left(\xi,x_n\right)$ coincides with a Bessel $J$ function of order $n$:
\beq
J_{\Delta_s}\left(\xi,x_n\right) = J_n\left(\frac{2\xi}{a}\right),      \label{JBessel} \eeq
whose properties are well known and extensively studied\footnote{Notice that in the continuum limit $a \rightarrow 0$ both the order $n = \frac{2 x_n}{a}$ and the argument $\frac{2\xi}{a}$ become very large.}. 
However we prefer to treat the two cases in parallel to emphasize analogies and differences.

As a result  of $\Delta(p)$ being  symmetric under $p \rightarrow \frac{2\pi}{a} -p$, the function  $J_\Delta\left(\xi,x_n\right) $ can be written as:
\beq
J_\Delta\left(\xi,x_n\right) =  J^{(0)}_\Delta\left(\xi,x_n\right) + (-1)^n   J^{(0)}_\Delta\left(\xi,x_{-n}\right), \label{JDelta2} \eeq
where
\beq
 J^{(0)}_\Delta\left(\xi,x_n\right) =   \frac{1}{2\pi} \int_{-\frac{\pi}{a}}^{\frac{\pi}{a}} dp~ e^{i \xi \Delta(p) - i x_n p } = \frac{1}{\pi}\int_{0}^{\frac{\pi}{a}} dp \cos\left( \xi \Delta(p) -  x_n p \right),   \label{J0Delta} \eeq
and the last step follows from $\Delta(p)$ being an odd function. This also implies
\beq
J_\Delta\left(\xi,x_n\right) = (-1)^n J_\Delta\left(\xi,x_{-n}\right). \label{nmenon} \eeq

The arguments $\xi$ and $x_n$ in $J_\Delta(\xi,x_n)$  are respectively the continuum and lattice expressed in term of some unspecified physical units. 
We are interested in the continuum limit $a \rightarrow 0$, namely in the limit where the ratio between $a$ and any physical length goes to zero while $\xi$ and $x_n$ are kept constant, which of course require  $n \rightarrow \infty$ as $a$ goes to zero.
It is then convenient to replace  $x_n$ with a variable $\eta$ which will be treated as a continuum variable and define in the continuum limit a distribution $j^{(0)}_{\Delta}(\xi,\eta)$ given by:
\beq
j^{(0)}_{\Delta}(\xi,\eta) = \lim_{a \rightarrow 0} J^{(0)}_\Delta(\xi, \eta).    \label{distribution2} \eeq
The continuum limit is performed on $J^{(0)}_\Delta(\xi, \eta) $ rather than on  $J_\Delta(\xi, \eta) $  because according to eq. (\ref{JDelta2}) such limit exists separately for odd and even values of $n$, in agreement with the fact that a function on the lattice defines, for each lattice direction, two different functions in the continuum.

In order to define this limit properly let us introduce the large parameter $N$:
\beq
N = \frac{2}{a},     \label{unit} \eeq
and the angular variable $\theta$:
\beq
\theta = \frac{ap}{2}.  \label{angmom} \eeq

We can write then:
\beq
J^{(0)}_\Delta\left( \xi ,\eta\right) = \frac{N}{2\pi} \int_{-\frac{\pi}{2}}^{\frac{\pi}{2}} d\theta e^{i N F_\Delta (\xi,\eta,\theta)}, \label{J0Delta2} \eeq
where the function $ F_\Delta (\xi,\eta,\theta)$ can be easily derived from (\ref{J0Delta}), and  is given in the two cases we are considering by :
\beq
 F_{\Delta_s }(\xi,\eta,\theta) =\xi \sin\theta -\eta \theta,~~~~~~~~~~\mbox{and}~~~~~~~~~~ F_{\Delta_G} (\xi,\eta,\theta) =\xi \gd(\theta) - \eta \theta.  \label{Fxi} \eeq
The continuum limit (\ref{distribution2}) is then the first term of the asymptotic expansion for large $N$ of $J^{(0)}_\Delta\left( \xi ,\eta\right)$ and it can be obtained from (\ref{J0Delta2}) by using the standard saddle point method.

The saddle points are the solutions of $\frac{\partial F_\Delta(\xi,\eta,\theta)}{\partial \theta} = 0$ and the ones corresponding to the two functions in (\ref{Fxi}), which we shall respectively denote as $\theta_s$ and $\theta_G$, are then given by:
\beq
\cos\theta_s = \frac{\eta}{\xi}~~~~~~~~~~\mbox{and}~~~~~~~~~~\cos\theta_G = \frac{\xi}{\eta}.  \label{saddlepoints} \eeq
This result is interesting: for $\xi/\eta>1$ the saddle point $\theta_s$ is real and $\theta_G$ imaginary while the opposite happens for $\xi/\eta<1$. On the other hand a real value of $\theta$ corresponds to a real value of the function $F_\Delta(\xi,\eta,\theta)$ and hence for large $N$ an oscillating behaviour of $J_\Delta^{(0)}(\xi ,\eta)$ as a function of $\xi/\eta$, whereas an imaginary value of   $\theta$ and hence of  $F_\Delta(\xi,\eta,\theta)$  leads to an exponential decay of $J_\Delta^{(0)}(\xi , \eta)$ as $\xi/\eta$ moves away from $1$. This corresponds to the well known behaviour of the Bessel function $J_{N\eta}(N \xi )$ which for large $N$ is rapidly oscillating when the argument is larger than the order $N\eta$, namely when  $\xi/\eta>1$.
The new and rather unexpected result is that with the choice $\Delta(p)=\Delta_G(p)$, which as we have seen is needed to have an associative star product, the situation is reversed and the oscillating behavior occurs for $\xi/\eta<1$ as if the roles of $\xi$ and $\eta$ had been exchanged.

\begin{figure}
\begin{center}
\mbox{\includegraphics*[width=12.cm]{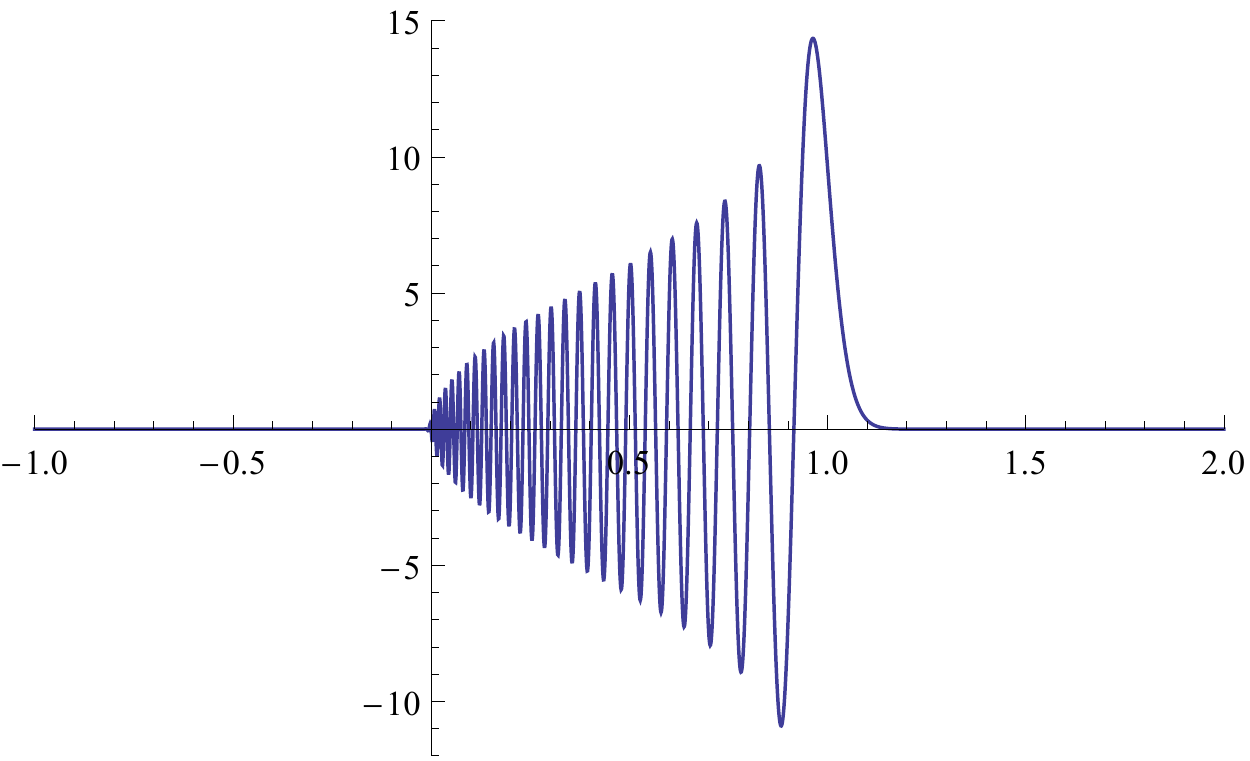}}
\caption{Plot of  $J_{\Delta_G}^{(0)}(\xi ,\eta)$ versus $\xi$ at $N=100$ and $\eta=1$}
\label{Jplot}
\end{center}
\end{figure}

\begin{figure}
\begin{center}
\mbox{\includegraphics*[width=12.cm]{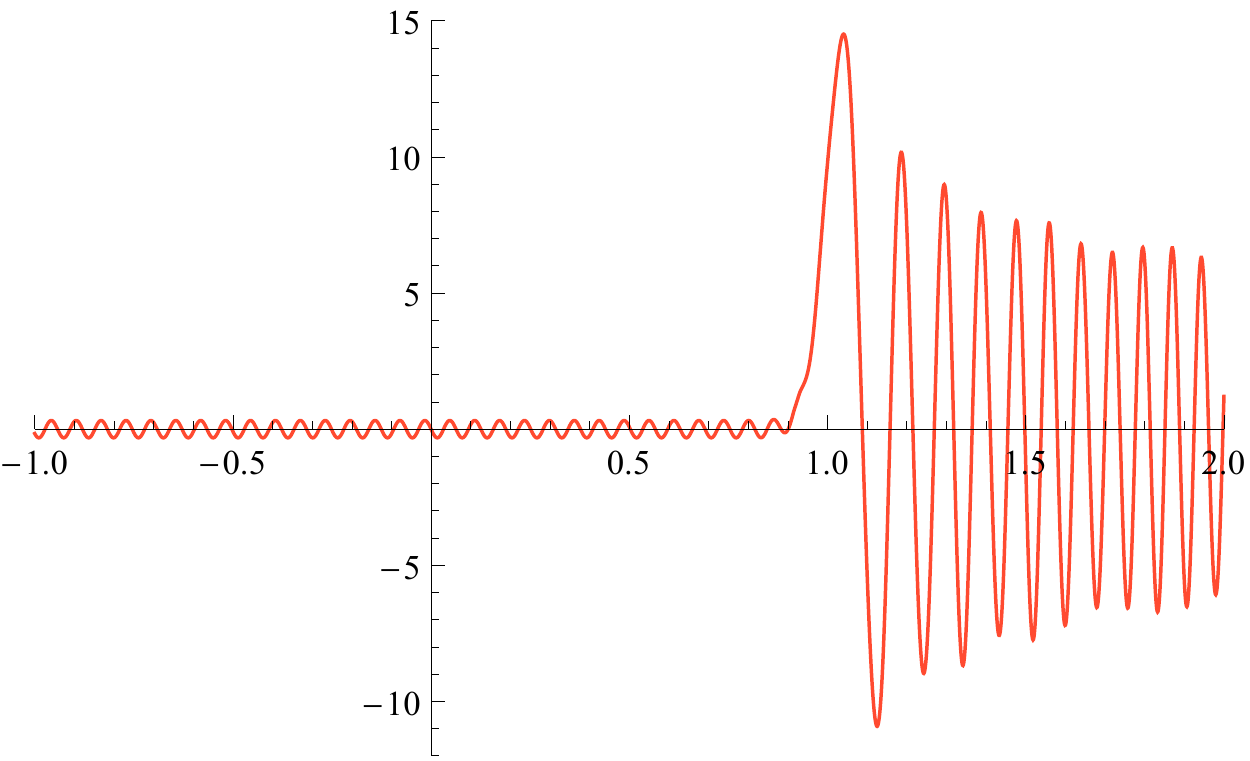}}
\caption{Plot of  $J_{\Delta_s}^{(0)}(\xi, \eta)$ versus $\xi$ at $N=100$ and $\eta=1$. }
\label{Jplot2}
\end{center}
\end{figure}

This can be visualized by plotting with Mathematica  $J_{\Delta_G}^{(0)}(\xi ,\eta)$ and $J_{\Delta_s}^{(0)}( \xi ,\eta)$ for $N=100$, $\eta$ fixed and set to $1$, and $\xi$ ranging from $-1$ to $+2$. The plots are shown respectively in Fig.\ref{Jplot} and Fig.\ref{Jplot2}.

It is also interesting to compare the plot of $J_{\Delta_s}^{(0)}( \xi,\eta)$ at $N=100$  as a function of $\xi$ at $\eta=1$, with the one of  $J_{\Delta_G}^{(0)}( \xi,\eta)$ at $N=100$ as a function of $\eta$ at $\xi=1$.  The latter  is reproduced in Fig. 3: the behaviour is very similar and the two curves are essentially indistiguishable in the proximity of $1$.
\begin{figure}
\begin{center}
\mbox{\includegraphics*[width=12.cm]{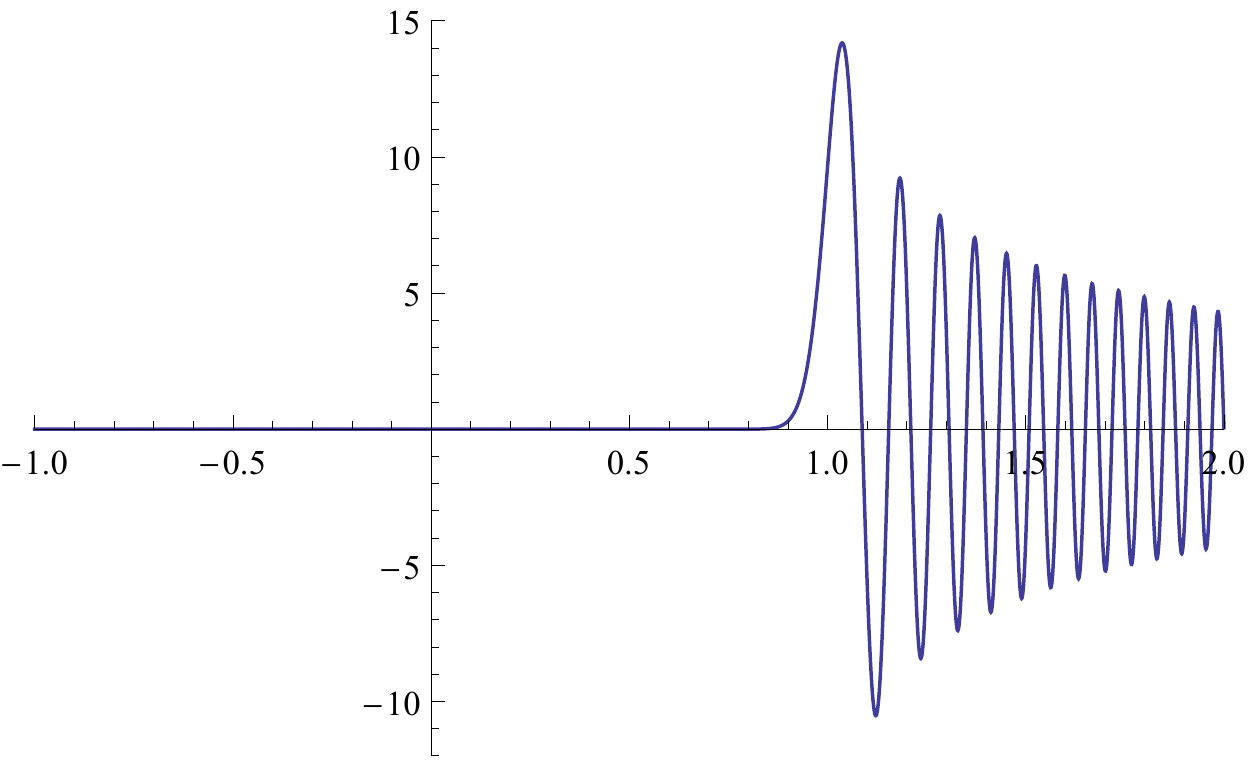}}
\caption{Plot of  $J_{\Delta_G}^{(0)}( \xi, \eta)$ versus $\eta$ at $N=100$ and $\xi=1$.}
\label{Jplot3}
\end{center}
\end{figure}

Analitically the leading term of the asymptotic behaviour for large $N$ of $J_\Delta^{(0)}(\xi ,  \eta)$ can be obtained from the standard formula of the steepest descent method:
\beq
\int_A^B X(z) e^{\rho f(z)} dz = \sqrt{\frac{2\pi}{\rho}} \sum_{i} \frac{e^{\rho f(z_i)}}{\left[-f''(z_i)\right]^{1/2}} \left(X(z_i)+O(1/\rho)\right), \label{steepest} \eeq
where the sum is over the saddle points $z_i$.
For the cases in consideration the results can be summarized by the following equations:

\beq
J_{\Delta_s}^{(0)}( \xi ,\eta) = \sqrt{\frac{2 N}{\pi }}~ \frac{\cos\left[N\left(\sqrt{\xi^2-\eta^2} - \eta \arccos(\eta /\xi)\right) - \pi/4\right]}{\left(\xi^2 - \eta^2\right)^{1/4}} \left(1 + O(1/N) \right)~~~~~~\xi/\eta>1,  \label{JDsosc} \eeq
\beq
 J_{\Delta_s}^{(0)}( \xi ,\eta) = \sqrt{\frac{N}{2\pi }}~\frac{e^{-N\left(\eta \cosh^{-1}(\eta/\xi)-\sqrt{\eta^2-\xi^2}\right)}}{\left(\eta^2-\xi^2 \right)^{1/4}} \left(1 + O(1/N) \right)~~~~~~~~~~~~~~~~~~~~~~~~~~~~~\xi/\eta<1, \label{JDsexp} \eeq
\beq
 J_{\Delta_G}^{(0)}(\xi ,\eta) = \sqrt{\frac{2N\xi}{ \pi\eta }}~\frac{   \cos\left[N\left( \frac{\xi}{2} \log\frac{1+\sqrt{1-(\xi/\eta)^2}}{1-\sqrt{1-(\xi/\eta)^2}}-\eta \arccos (\xi/\eta) \right) + \pi/4\right]}{\left(\eta^2-\xi^2 \right)^{1/4}} \left(1 + O(1/N) \right)~~~~~\xi/\eta<1, \label{JDGosc} \eeq
\beq
 J_{\Delta_G}^{(0)}( \xi ,\eta) = \sqrt{\frac{N\xi}{2\pi \eta}}~\frac{ \xi^{1/2} e^{-N\left(\eta \cosh^{-1}(\xi/\eta)- \xi \arccos (\eta/\xi)\right)}}{\left(\xi^2-\eta^2 \right)^{1/4}} \left(1 + O(1/N) \right)~~~~~~~~~~~~~~~~~~~~~~~~~~~~~\xi/\eta>1. \label{JDGexp} \eeq

A plot of the functions given in eq.(\ref{JDsosc})-(\ref{JDGexp}) shows  that, except for a small interval around $\xi/\eta=1$, they overlap exactly at $N=100$ and $\eta=1$ with the ones given in Fig.3 and Fig.4. 
The asymptotic expansion at $\xi/\eta=1$ cannot  be obtained directly from (\ref{steepest})  because in that case $f''(z)$ vanishes at the saddle point and the corresponding gaussian integral cannot be perfomed.
 
We notice however from (\ref{saddlepoints}) that for  $\xi/\eta=1$ the saddlepoint is $\theta=0$,  and that therefore the first  terms of the  expansion in powers of $\theta$ of $F_{\Delta}(\xi,\eta,\theta)$ will retain all the information of the leading  term in the asymptotic expansion of $J^{(0)}_{\Delta}( \xi ,\eta)$ at  $\xi/\eta=1$.
More precisely, consider the expansions:
\beq
F_{\Delta_s }(\xi,\eta,\theta) =(\xi -\eta) \theta -\frac{\xi}{6} \theta^3 + O(\theta^5) ,~~~~~~~~~~\mbox{and}~~~~~~~~~~ F_{\Delta_G} (\xi,\theta) =(\xi -\eta) \theta +\frac{\xi}{6} \theta^3 + O(\theta^5),  \label{Fxip} \eeq
and notice that they differ only for the sign of the $\theta^3$ term.
We can write then:
\beq
J^{(0)}_{\Delta_{s,G}}( \xi , \eta) = \frac{N}{\pi} \int_0^{\frac{\pi}{2}} d\theta \cos \left[ N (\xi -\eta) \theta \mp \frac{N\xi}{6} \theta^3 + N O(\theta^5) \right],  \label{Ja} \eeq
where the minus and plus sign in front of the cubic term refer respectively to $\Delta_s$ and $\Delta_G$.
With the substitution $z=\left(\frac{N \xi}{2}\right)^{1/3} \theta$ the integral above becomes:
\beq
J^{(0)}_{\Delta_{s,G}}(\xi , \eta ) = \frac{N^{2/3}}{\pi}  \left(\frac{\xi}{2}\right)^{-1/3} \int_0^{\frac{\pi}{2} \left(\frac{N \xi}{2}\right)^{1/3}} dz \cos \left[ \frac{z^3}{3} \mp (\xi-\eta)N^{2/3} \left(\frac{2}{\xi}\right)^{1/3} z + N O\left[\left(\frac{2}{N \xi}\right)^{5/3} z^5\right] \right]. \label{asJ} \eeq
Consider now in (\ref{asJ}) the limit $N \rightarrow \infty$ and $\xi/\eta \rightarrow 1$, but with $(\xi-\eta) N^{2/3}$ kept finite. Then all terms in the cosine vanish except for the linear and the cubic term, and the integral becomes proportional to an Airy function:
\beq
J^{(0)}_{\Delta_{s,G}}(\xi ,  \eta) \approx_{N\rightarrow \infty}  N^{2/3} \left(\frac{\xi}{2}\right)^{-1/3} 
\ai\left(\mp 2^{1/3}\xi^{-1/3} N^{2/3} (\xi-\eta) \right)
.\label{Airy} \eeq

If we choose $|\eta-\xi|  N^{2/3}$ very large, namely $ N^{2/3} \gg |\eta-\xi|^{-1} \gg 1$, then we can use in (\ref{Airy}) the well known asymptotic formulas for the Airy functions. Consider the case with $\Delta=\Delta_G$, which corresponds to the plus sign at the r.h.s. of (\ref{Airy}), then  for $\xi <\eta $ the argument of the Airy function is large and negative and using the  asymptotic formulas for the Airy function one gets:
\beq
J^{(0)}_{\Delta_G}( \xi ,  \eta) \approx \frac{2^{1/4} \xi^{-1/4} N^{1/2}}{\sqrt{\pi} \left(\eta - \xi \right)^{1/4}} \cos\left[\frac{2^{3/2}}{3} N \xi^{-1/2}  \left(\eta - \xi \right)^{3/2} - \frac{\pi}{4}\right]~~~~~~~~~~\xi/\eta <1.  \label{Airyasi} \eeq
For positive arguments  of the Airy function, which for $J^{(0)}_{\Delta_G}$ implies $\xi/\eta >1$, the asymptotic formulas give an exponentially vanishing behaviour, namely:
\beq
J^{(0)}_{\Delta_G}( \xi ,  \eta) \approx  \frac{\xi^{-1/4} N^{1/2}}{2^{3/4} \sqrt{\pi} \left(\xi-\eta \right)^{1/4}} e^{-N \frac{2^{3/2}}{3} \xi^{-1/2} \left(\xi- \eta \right)^{3/2}} ~~~~~~~~~~\xi/\eta >1.     \label{Airyasi2} \eeq

The formulas for $J^{(0)}_{\Delta_s}$ can be obtained from (\ref{Airyasi}) and (\ref{Airyasi2}) by simply exchanging $\xi$ and $\eta$ everywhere.
Equations (\ref{Airyasi}) and  (\ref{Airyasi2}) are valid in the region defined by the inequality  $ N^{2/3} \gg (1-\eta/\xi)^{-1} \gg 1$. This region can also be reached from the domain of validity of (\ref{JDsosc})-(\ref{JDGexp}) by taking the limit where $|1-\eta/\xi| \ll 1$. In fact  by taking in  eq.s (\ref{JDGosc}) and (\ref{JDGexp}) the limit $|1-\eta/\xi| \rightarrow 0 $, and keeping in that limit only the leading terms one reproduces exactly  eqs. (\ref{Airyasi}) and  (\ref{Airyasi2}).

By using the asymptotic formulas from (\ref{JDsosc}) to (\ref{Airy}) we show in Appendix \ref{cont}  that 
provided the test function $\chi(\xi)$ satisfies some weak smoothness condition the large $N$ limit of $J^{(0)}_{\Delta_{G}}$ gives:
\beq
\int_{-\infty}^{+\infty} d\xi~ j^{(0)}_{\Delta_{G}}(\xi,\eta) \chi(\xi) = \lim_{N \rightarrow \infty}\int_{-\infty}^{+\infty} d\xi ~  J^{(0)}_{\Delta_{G}}( \xi, \eta) \chi(\xi) = \chi(\eta),   \label{distribution3} \eeq
namely, for the distribution $j^{(0)}_{\Delta_{G}}(\xi,\eta)$:

\beq
j^{(0)}_{\Delta_{G}}(\xi,\eta) =  \delta(\xi-\eta)
.\label{deltafunc} \eeq 
The exponential damping nature and the oscillating nature of $J^{(0)}_{\Delta_{G}}( \xi, \eta)$ in the both sides of $\eta=1$ 
or $\xi=1$ in Fig.3  and Fig.5  are effective enough to make the non-local nature of the lattice formulation into local 
in the continuum limit. 

The continuum limit $a \rightarrow 0$ of $J_\Delta(\xi, \eta)$ can be obtained from the one of $J^{(0)}_\Delta(\xi, \eta)$  using (\ref{JDelta2}); however due to the $(-1)^n$ factor in (\ref{JDelta2}) the limit has to be done separately for even and odd values of $n$. This is consistent with the fact that in one dimension a lattice field describes two degrees of freedom in the continuum.
So if we define the complete distribution
\beq
j_{\Delta}(\xi,\eta) = \lim_{a \rightarrow 0} J_\Delta(\xi, \eta),    \label{distribution3} \eeq
we have:
\beq 
j_{\Delta_{G}}(\xi,\eta) =  \delta(\xi-\eta) \pm  \delta(\xi+\eta), \label{deltafunc2} \eeq
where the $+$ and $-$ signs apply respectively to the sublattices defined by even and odd values of $n$. 
For $\Delta(p) = \Delta_s(p)$ the limit (\ref{distribution3}) was proved by Dondi and Nicolai in their poineering paper on lattice supersymmetry \cite{Dondi-Nicolai} on the condition that the Fourier transform of test function $\chi(\xi)$ in (\ref{distribution3}) satisfies the integrability condition:
\beq
\int_{-\infty}^{+\infty} dk \left| \tilde{\chi}(k) \right| < \infty.  \label{DNcond} \eeq
The same result could be obtained following the same path used in Appendix \ref{cont} for the case $\Delta=\Delta_G$. 

It is important to recognize at this stage that the non-local nature existing in the star product and the inverse Gudermannian 
derivative operator does not sacrifice the locality of the product in the continuum limit. 

Finally, as an example and application of the limiting procedure described above, we calculate the continuum $a \rightarrow 0$ limit of the $n$-point interaction term $I_n$ given in (\ref{npointlatt}). 
By using the Fourier transforms of the fields and of the  $\delta$-function, the interaction term becomes:
\beq
I_n =\int d\xi~\left(\frac{a}{2}\right)^n~\sum_{r_1,r_2,\cdots,r_n}~ J_{\Delta_G}(\xi,x_{r_1})\cdots J_{\Delta_G}(\xi,x_{r_n})~\varphi(x_{r_1})\varphi(x_{r_2})\cdots\varphi(x_{r_n})
.\label{npointlatt2} \eeq
This is a local $n$-point interaction term in the continuum coordinates $\xi$, namely:
\beq
I_n= \int d\xi~ [\Phi(\xi)]^n,  \label{npointlatt3} \eeq
where the continuum field $\Phi(\xi)$ is given by:
\beq
\Phi(\xi) = \frac{a}{2} \sum_r J_{\Delta_G}(\xi,x_{r}) \varphi(x_{r}).  \label{phicont} \eeq
We consider now the  continuum limit $a \rightarrow 0$ of the r.h.s. of (\ref{phicont}). The discrete variable $x_r=\frac{ra}{2}$ may be replaced in the $a \rightarrow 0$ limit  by a continuum variable $\eta$. In doing that however odd and even values of $r$ must be treated separately. In fact the field $\varphi(x_r)$ satisfies the symmetry condition (\ref{chiralcondx}) and  its odd and even part have separate smooth continuum limits which define two distinct continuum functions $\varphi_e(\eta_i)$ and  $\varphi_o(\eta_i)$ 
\beq
r_i ~\mathrm{even}:~ \varphi(x_{r_i})  \stackrel{a\rightarrow 0}{\Longrightarrow} \varphi_e(\eta_i) ~~~~~;~~~r_i~ \mathrm{odd}:~ \varphi(x_{r_i})     \stackrel{a\rightarrow 0}{\Longrightarrow} \varphi_o(\eta_i),  \label{oddeven} \eeq
which are respectively even and odd functions of $\eta$:
\beq
\varphi_e(-\eta_i) = \varphi_e(\eta_i),~~~~~~~~~~~~~~~~\varphi_o(-\eta_i) = -\varphi_o(\eta_i).     \label{eo2} \eeq
The sum over $r$ at the r.h.s. of (\ref{phicont}) can be split into the sum over the even and the sum over the odd values of $r$, which can be calculated using (\ref{oddeven}), (\ref{eo2}) and (\ref{deltafunc2}). 
The result is:
\beq
\Phi(\xi) = \frac{a}{2} \sum_{r_i} J^{(0)}_{\Delta_G}(\xi, x_{r_i}) \varphi(x_{r_i})~\stackrel{a\rightarrow 0}{\Longrightarrow}  \int d\eta~ \delta(\eta -\xi)~ \varphi_+(\eta) = \varphi_+(\xi),   \label{limitxi} \eeq
where 
\beq
\varphi_+(\xi) = \varphi_e(\xi)+\varphi_o(\xi). \label{pluseo} \eeq
In conclusion: locality of the interaction on the lattice is recovered in the continuum limit.

\subsection{The choice of the function $f(p)$ and the continuum limit.}
\label{3.5}

The associative star product discussed at the beginning of this section (see eq.s (\ref{assstarprodp}) and  (\ref{assstarprodpr})) is not uniquely determined by the lattice derivative operator $\Delta(p)$ as it also depends on an arbitrary function $f(p)$, which corresponds to a  rescaling of the fields $\tilde{\varphi}(p)$ in momentum representation.
A rescaling in momentum space defines a non trivial and non-local transformation in coordinate representation, so that different choices of the function $f(p)$ may correspond to very different representations in coordinate space.
In this subsection we are going to discuss different choices of $f(p)$ from the point of view of the coordinate representation, and discover that strong restrictions of $f(p)$ are required for the continuum limit to be well defined in coordinate space.

Let us consider the associative star product (\ref{assstarprodpr}) which is written in momentum representation  as a convolution over the reduced interval $(-\frac{\pi}{a},  \frac{\pi}{a})$, after making use of  the symmetry of the fields under $p_i \rightarrow \frac{2\pi}{a}-p_i$. 
The derivative operator $\Delta(p)$ is not specified, but we shall assume that $|\Delta(\pm\frac{\pi}{a})|=\infty$ in order for the product to be associative: eventually we shall restrict ourselves to the most interesting case $\Delta(p)=\Delta_G(p)$.
The associative star product (\ref{assstarprodpr}) is equivalent to the usual local product (\ref{contpr}) of the continuum theory  provided the lattice fields $\tilde{\varphi}_i(p)$ are related to the continuum fields $\tilde{\Phi}_i(\Delta(p))$ according to eq.(\ref{Phivarphi}).

We shall now write eq.(\ref{Phivarphi}) in coordinate representation. 
Let $\varphi(x_n)$ be the lattice field in coordinate space, namely the Fourier transform of $\tilde{\varphi}(p)$ over the  $\frac{4\pi}{a}$ period:
 \beq
\varphi(x_n) = \frac{1}{2\pi} \int_{-\frac{\pi}{a}}^{\frac{3\pi}{a}} dp~ \tilde{\varphi}(p) e^{i x_n p}
.\label{varphi} \eeq
Thanks to the symmetry (\ref{chiralcond}) we can write:
\beq
\varphi(x_n) = \varphi_0(x_n) + (-1)^n \varphi_0(-x_n),              \label{nmenon} \eeq 
where
\beq
\varphi_0(x_n) = \frac{1}{2\pi} \int_{-\frac{\pi}{a}}^{\frac{\pi}{a}} dp~ \tilde{\varphi}(p) e^{i x_n p}
.\label{varphi0} \eeq
Notice from (\ref{nmenon}) that the $\varphi_0(x_n)$ determines ${\varphi}(x_n)$ completely and,  unlike $\varphi(x_n)$, it has a smooth dependence on $n$. In the continuum limit we replace the discrete variable $x_n$ with a continuum variable $\xi$ and  $\varphi_0(x_n)$ becomes a continuum function  $\varphi_0(\xi)$. One can see from (\ref{nmenon}) that $\varphi_0(\xi)$ coincides with the sum of the even and the odd part of $\varphi(\xi)$, namely it coincides with  $\varphi_+(\xi)$ defined in (\ref{pluseo}).
It is clear from (\ref{varphi0}) that $\varphi_0(x_n)$ may be regarded as the Fourier transform of a function $\tilde{\varphi}_0(p)$ which coincides with $\tilde{\varphi}(p)$ in the 
interval $(-\frac{\pi}{a},  \frac{\pi}{a})$ and vanishes in $(\frac{\pi}{a},  \frac{3\pi}{a})$.
So in the interval $(-\frac{\pi}{a},  \frac{\pi}{a})$  equation (\ref{varphi0}) can be inverted and gives:
\beq
\tilde{\varphi}(p) = \frac{a}{2} \sum_n \varphi_0(x_n) e^{-i\frac{na}{2} p},~~~~~~~~~~~~-\frac{\pi}{a}\leq p \leq  \frac{\pi}{a}. \label{0varphi} \eeq

Let us denote now by $\Phi(\xi)$  the Fourier transform\footnote{Remember the assumption that $\Delta(p)$ goes from $-\infty$ to $+\infty$ as $p$ goes through the interval $(-\frac{\pi}{a},  \frac{\pi}{a})$.} of $\tilde{\Phi}(\Delta(p))$. 
Then by taking the Fourier transform of (\ref{Phivarphi}) and using (\ref{0varphi}) we obtain:
\beq
\Phi(\xi) = \frac{a}{2} \sum_n~J^{(0)}_{\Delta,f}(\xi,x_n) \varphi_0(x_n) \label{varphiPhi2} 
, \eeq
where the function $J^{(0)}_{\Delta,f}(\xi,x_n)$ is defined as:
\beq
J^{(0)~}_{\Delta,f}(\xi,x_n) = \frac{1}{2\pi} \int_{-\frac{\pi}{a}}^{\frac{\pi}{a}} dp f(p) ~e^{-ix_n p + i \xi\Delta(p)}. \label{JFD} \eeq
The function $J^{(0)}_{\Delta,f}(\xi,x_n) $ is a  generalization of the function $J^{(0)}_{\Delta}(\xi,x_n)$ defined in (\ref{J0Delta})  to which it reduces if one sets $f(p)=1$. 
Eq.(\ref{varphiPhi2}) can be inverted by replacing $\tilde{\varphi}(p)$ at the r.h.s. of (\ref{varphi0}) with its expression given in (\ref{Phivarphi}) and finally expressing $\tilde{\Phi}(\Delta(p))$ as a Fourier transform of  $\Phi(\xi)$. 

The result is:
\beq
\varphi_0(x_n) = \int_{-\infty}^{\infty} d\xi~\bar{J}^{(0)}_{\Delta,f}(x_n,\xi)~\Phi(\xi),  \label{varphi00} \eeq
where the function  $\bar{J}^{(0)}_{\Delta,f}(x_n,\xi) $ \footnote{Notice the inverted arguments} is defined as:
\beq
\bar{J}^{(0)}_{\Delta,f}(x_n,\xi) = \frac{1}{2\pi} \int_{-\frac{\pi}{a}}^{\frac{\pi}{a}} dp \frac{1}{f(p)} \frac{d\Delta}{dp}~e^{ix_n p - i \xi\Delta(p)}= \frac{1}{2\pi} \int_{-\infty}^{\infty} \frac{d\hat{p}}{f(p)}~e^{ix_n \Delta^{-1}(\hat{p}) - i \xi\hat{p}} . \label{JDF} \eeq

Let us consider now the case $\Delta=\Delta_G$. If we choose  $f(p)=1$, eq. (\ref{varphiPhi2}) coincides with the first half of eq. (\ref{limitxi}) and hence it admits a smooth continuum limit as given in the second part of the same equation. 

However some problems arise with the inverse relation (\ref{varphi00}).  If fact the function $\bar{J}^{(0)}_{\Delta,f}(x_n,\xi) $ becomes:
\beq
\bar{J}^{(0)}_{\Delta_G,f=1}(x_n,\xi) = \frac{1}{2\pi} \int_{-\frac{\pi}{a}}^{\frac{\pi}{a}} dp \frac{1}{\cos(\frac{ap}{2})}~e^{ix_n p - i \xi\Delta_G(p)} =  \frac{1}{2\pi} \int_{-\infty}^{+\infty}d\hat{p}~e^{ix_n p - i \xi\hat{p}},\label{JDFcos} \eeq
where in the last term we have set $\hat{p}=\Delta_G(p)$ and chosen $\hat{p}$ as integration variable\footnote{In the last integral $p$ is regarded as a function of $\hat{p}$, namely: $p=\frac{2}{a}\mathrm{gd}(\frac{a\hat{p}}{2})$.}.
The integrals in (\ref{JDFcos}) do not converge at the integration limits, and rather than a function eq. (\ref{JDFcos}) defines a distribution; notice in fact that for $x_n=0$ the r.h.s. is just the integral representation of $\delta(\xi)$ and that $p$ is constant ($\pm\frac{\pi}{a}$) for $\hat{p}\rightarrow \pm\infty$ .
The continuum limit $a\rightarrow 0$ of $\bar{J}^{(0)}_{\Delta_G,f=1}(x_n,\xi)$ can be calculated following the same procedure used in the previous subsection for $J^{(0)}_{\Delta_G}(\xi,x_n)$, 
namely by the saddle point method. The discrete variable $x_n$ is replaced by the continuum variable $\eta$ and eq. (\ref{J0Delta2}) takes now the form:
\beq
\bar{J}_{\Delta_G,f(p)=1}^{(0)}(\eta,\xi) = \frac{N}{2\pi} \int_{-\frac{\pi}{2}}^{\frac{\pi}{2}} \frac{d\theta}{\cos(\theta)}~e^{iNF_{\Delta_G}(\xi,\eta,\theta)}, \label{J0Delta3} \eeq
where  $F_{\Delta_G}(\xi,\eta,\theta)$ is given in eq.(\ref{Fxi}).
Eq. (\ref{J0Delta3}) differs from (\ref{J0Delta2}) for the factor $\frac{1}{\cos(\theta)}$ in the integration volume. This factor is the term $X(z)$ in the steepest descent formula (\ref{steepest}) and according to the second equation in (\ref{saddlepoints}) it produces in the $N\rightarrow \infty$ limit an overall factor $\frac{\eta}{\xi}$.
As a consequence the asymptotic formulas for $\bar{J}_{\Delta_G,f(p)=1}^{(0)}(\eta,\xi)$ are the same as the ones given in (\ref{JDGosc}) and in (\ref{JDGexp}) for $J_{\Delta_G}^{(0)}(\xi,\eta)$, but multiplied by a factor $\frac{\eta}{\xi}$. 
So while $J_{\Delta_G}^{(0)}(\xi,\eta)$ vanishes like $\sqrt{\xi}$ as $\xi \rightarrow 0$,  $\bar{J}_{\Delta_G,f(p)=1}^{(0)}(\eta,\xi)$ diverges like $\frac{1}{ \sqrt{\xi}}$ in the same limit.

This implies that the contribution coming from $\xi =0$ cannot be neglected in the continuum limit. In fact,  one can follow step by step for $\bar{J}_{\Delta_G,f(p)=1}^{(0)}(\eta,\xi)$ the derivation of such limit as given in the Appendix for $J_{\Delta_G}^{(0)}(\xi,\eta)$. We find that  eq. (\ref{inta3}) is modified now by the extra factor $\frac{\eta}{\xi}$ so that  the contribution at $\xi =0$ is  divergent as the factor $\sqrt{\frac{2\xi}{\pi N \eta}}$ is replaced by $\sqrt{\frac{2\eta}{\pi N \xi}}$. 

A vanishing contribution at $\xi=0$ can be obtained in this case only by imposing a rather \textit{ad hoc} restriction on the test function $\chi(\xi)$, namely by requiring that
\beq
\lim_{\xi\rightarrow 0} \frac{\chi(\xi)}{\sqrt{\xi}} = \mathrm{finite}. \label{testcond} \eeq

It is natural at this point to ask the question if a choice of the rescaling function $f(p)$ exists that makes both $\bar{J}^{(0)}_{\Delta_G,f=1}(x_n,\xi) $ and $J^{(0)}_{\Delta_G,f=1}(\xi,x_n) $ well defined at $\xi=0$ so that both (\ref{varphiPhi2}) and (\ref{varphi00}) have a smooth continuum limit without making any \textit{ad hoc} assumption on the test function (namely on $\Phi(\xi)$ in the case of eq.(\ref{varphi00})).
Such a choice does indeed exists as it is rather obvious from the previous discussion. In fact if we choose
\beq
f(p) = f_{\sqrt{}}(p) \equiv \frac{1}{\sqrt{\cos(\frac{ap}{2})}}, \label{fsqrt} \eeq
the functions $\bar{J}_{\Delta_G,f=f_{\sqrt{}}}^{(0)}(\eta,\xi)$  and $J_{\Delta_G,f=f_{\sqrt{}}}^{(0)}(\xi,\eta)$ coincide\footnote{For a generic $\Delta$ this happens with the choice $f(p) = \sqrt{\frac{d\Delta(p)}{dp}}$.} and are given by:
\beq
\bar{J}_{\Delta_G,f=f_{\sqrt{}}}^{(0)}(\eta,\xi) = J_{\Delta_G,f=f_{\sqrt{}}}^{(0)}(\xi,\eta) = \frac{N}{2\pi} \int_{-\frac{\pi}{2}}^{\frac{\pi}{2}} \frac{d\theta}{\sqrt{\cos(\theta)}}~e^{iN\left(\xi \gd(\theta) - \eta \theta \right) }. \label{J0Deltasq} \eeq
The integral on the r.h.s. is convergent and the factor $\frac{1}{\sqrt{\cos(\theta)}}$ in the integral produces an extra factor $\sqrt{\frac{\eta}{\xi}}$ in the large $N$ expansion. This cancels exactly the factor $\sqrt{\frac{\xi}{\eta}}$ that appears in front of the large $N$ expansion in eq.(\ref{JDGosc}) and in eq.(\ref{JDGexp}). 
With this choice of $f(p)$ the derivation of the continuum limit given in the appendix runs smoothly. The crucial point is the contribution from $\xi = 0$ given in (\ref{inta3}). 
With the new choice of $f(p)$ the factor $\sqrt{\frac{\xi}{\eta}}$ is exactly canceled by the factor  $\sqrt{\frac{\eta}{\xi}}$ coming from the  $\frac{1}{\sqrt{\cos(\theta)}}$ factor in the integration measure, but the contribution at $\xi=0$ still vanishes because the function $F'(\xi,\eta)$ at the denominator in (\ref{inta3}) becomes infinite like $\log(\xi)$ as $\xi$ goes to zero.

In conclusion the choice of $f(p)$ given in (\ref{fsqrt}) is the only one for which both $\bar{J}_{\Delta_G,f}^{(0)}(\eta,\xi)$  and $J_{\Delta_G,f}^{(0)}(\xi,\eta)$ reduce to a $\delta(\xi-\eta)$ in the large $N$ limit, with any other choice one of the two functions gives a divergent contribution at $\xi=0$ in that limit and correspondingly requires a vanishing test function at $\xi=0$ (as for instance in eq.(\ref{testcond})) for the limit to be finite.

Another remarkable property of the choice $f(p)=\sqrt{\frac{d\Delta(p)}{dp}}$, which reduces to  (\ref{fsqrt}) for $\Delta=\Delta_G$, is that the star product becomes completely symmetric in coordinate representation. 
In fact from (\ref{assstarprodpr}), using (\ref{varphi0}) and (\ref{0varphi}), a straightforward calculation gives:
\beq
(\varphi_1\star\varphi_2)^{(0)}(x_n) = \frac{a^2}{4} \sum_{n_1,n_2} K_{n,n_1,n_2} \varphi^{(0)}_1(x_{n_1})\varphi^{(0)}_2(x_{n_2}),  \label{asprsymm} \eeq
where the kernel $K_{n,n_1,n_2}$ is completely symmetric in the three indices and is given by
\beq
K_{n,n_1,n_2} = \int_{-\infty}^{+\infty} d\xi ~J^{(0)}_{\Delta,\sqrt{\Delta'}}(\xi,x_n)~J^{(0)}_{\Delta,\sqrt{\Delta'}}(\xi,x_{n_1})~J^{(0)}_{\Delta,\sqrt{\Delta'}}(\xi,x_{n_2})
,\label{symmker} \eeq
with
\beq
J^{(0)}_{\Delta,\sqrt{\Delta'}}(\xi,x_n) = \frac{1}{2\pi} \int_{-\frac{\pi}{a}}^{\frac{\pi}{a}}
dp \sqrt{\frac{d\Delta}{dp}} e^{ix_n p -i\xi \Delta(p)} = \frac{1}{\pi} \int_{0}^{\frac{\pi}{a}}
dp \sqrt{\frac{d\Delta}{dp}} \cos\left[ix_n p -i\xi \Delta(p)\right].  \label{JDeltaprime} \eeq 

\newpage
\section{Lattice actions from blocking transformations}
\label{1.4}

Given a field theory in the continuum it is possible to define a corresponding theory on the lattice by using ``blocking transformations" that associate a field configuration on the lattice to any field configuration in the continuum. Although the term ``blocking transformations" is normally used to denote transformations that map field configurations on a lattice onto field configuration on another coarser lattice we extend here the terminology to denote continuum to lattice transformations. This was already done for instance in ref.  \cite{Bergner-Bruckmann} to which we refer for the general set up.

The explicit form of these continuum-to-lattice blocking transformations is suggested by the correspondence between the continuum and the lattice momentum which, as we have seen, plays a fundamental role in defining the type of lattice theory we are dealing with. 
Indeed we have already seen in sect.\ref{1.1}  that in the framework of the conventional lattice the relation (\ref{momentumrel}) between  the lattice momentum $p_\mu$ and the continuum momentum $\hat{p}_\mu$ naturally induces a map, given in momentum representation by eq. (\ref{bltrp}), from the fields of the continuum theory $\tilde{\Phi}_A(\hat{p})$  onto the fields $\tilde{\varphi}_A(p)$  of the theory on the lattice.
In this section we shall apply this procedure to the new approach to lattice theory described in the present paper.

\subsection{New lattice via blocking transformations}

Let us denote by $\Phi_A(\xi)$ the set of fields (bosonic and/or fermionic) of a field theory in the continuum and by $\tilde{\Phi}_A(\hat{p})$ their Fourier transforms, namely their representation in  momentum space. In a $d$-dimensional theory $\xi$ and $\hat{p}$ denote a set of $d$ coordinates (respectively momenta) $\xi^{\mu}$ (resp. $\hat{p}_{\mu}$) and the label $A$ may include, besides internal symmetry indices, also spinor and vector indices.

Let us also denote by $\varphi_A(x_n)$ the corresponding fields on the lattice, with $x_n$  representing a set of coordinates  $x_n^{\mu}=\frac{a n^{\mu}}{2}$ where $n^\mu$ are $d$ integers labeling the lattice sites.  The discrete Fourier transform of $\varphi_A(x_n)$ will be denoted by $\tilde{\varphi}_A(p)$ and provides the momentum representation of the lattice fields. The fields $\tilde{\varphi}_A(p)$ are periodic with period $\frac{4\pi}{a}$ in all components $p_\mu$ of the momentum.

In order for the correspondence between continuum and lattice fields to be a one to one correspondence we shall also assume that $\tilde{\varphi}_A(p)$ is invariant under any of the $d$ symmetry transformations $p_\mu \rightarrow  \frac{2\pi}{a} - p_\mu$, namely we shall assume that eq. (\ref{chiralcondpd}) and (\ref{chiralcondd}) hold for any of the field species labeled by the index $A$.
Therefore in momentum representation all the independent degrees of freedom of the theory on the lattice are contained, for any value of $\mu$, in the domain $-\frac{\pi}{a} \leq p_\mu \leq \frac{\pi}{a}$, and in  coordinate representation they are all contained in the $d$-dimensional quadrant defined by $n^\mu \geq 0$ for all $\mu$.

In our approach the correspondence between the lattice momenta $p_\mu$ and the momenta $\hat{p}_\mu$ in the continuum is given by eq. (\ref{momentumdelta}), which results from the 
identification of the momenta $\hat{p}_\mu$ in the continuum with the corresponding derivative operator $\Delta(p_\mu)$ on the lattice.

The blocking transformation induced by this correspondence has already been considered in the one dimensional case and it is given by eq. (\ref{Phivarphi}). This can be directly generalized to arbitrary dimensions in the following way:
\beq
F(p)~\tilde{\varphi}_A(p) = \prod_{\mu=1}^d \frac{d\Delta(p_\mu)}{dp_\mu}~\tilde{\Phi}_A(\Delta(p))~~~~~~~~~~~~~~~~-\frac{\pi}{a}\leq p_\mu\leq \frac{\pi}{a},   \label{Phivarphidd} \eeq
where the function $F(p)$ corresponds to an arbitrary rescaling of the lattice fields $\tilde{\varphi}_A(p) $ in momentum space.

A simpler way of writing  (\ref{Phivarphidd}) is obtained by introducing the differentials $dp_\mu$ and $d\hat{p}_\mu$, namely:
\beq
\tilde{\varphi}_A(p) ~F(p)\prod_{\mu=1}^{d} dp_\mu = \tilde{\Phi}_A(\hat{p}) \prod_{\mu=1}^{d} d\hat{p}_\mu~~~~~~~~~~~~~-\frac{\pi}{a}\leq p_\mu\leq \frac{\pi}{a}, \label{Phivarphidiff} \eeq
where $\hat{p}_\mu = \Delta(p_\mu)$. Notice that if the l.h.s. of (\ref{Phivarphidiff}) is to be integrated over a $\frac{4\pi}{a}$ period the product $F(p)~\tilde{\varphi}_A(p)$ must be symmetric under $p_\mu \rightarrow \frac{2\pi}{a} - p_\mu$ or else the integral would vanish.
Since we assumed that the fields  $\tilde{\varphi}_A(p) $ satisfy eq.  (\ref{chiralcondpd}), for consistency we shall assume that the latter is also satisfied by $F(p)$, although in principle it would be enough to require the invariance under  (\ref{chiralcondpd}) of their product.
Eq. (\ref{Phivarphidiff}) also shows that $F(p)$ can be interpreted as a function defining the  integration volume  in the lattice momentum space. 

The blocking transformation (\ref{Phivarphidd}) has been defined in the domain  $-\frac{\pi}{a} \leq p_\mu \leq \frac{\pi}{a}$, but it can be extended, using the symmetry of the fields under $p_\mu \rightarrow \frac{2\pi}{a} - p_\mu$, to the whole $\frac{4\pi}{a}$ interval in each variable. 
However since $\Delta(p_\mu)$  is symmetric under $p_\mu \rightarrow \frac{2\pi}{a} - p_\mu$,  its derivative with respect to $p_\mu$ appearing at the r.h.s. of (\ref{Phivarphidd}) is antisymmetric. So the extension of (\ref{Phivarphidd}) to the $\frac{4\pi}{a}$ interval requires that the absolute value of the derivatives appears on the r.h.s, giving:
\beq
 F(p)~\tilde{\varphi}_A(p) = \left| \prod_{\mu=1}^d \frac{d\Delta(p_\mu)}{dp_\mu}\right|~\tilde{\Phi}_A(\Delta(p))~~~~~~~~~~~~~~~~-\frac{\pi}{a}\leq p_\mu\leq \frac{3\pi}{a}.   \label{Phivarphiddf} \eeq
We remarked in sect.\ref{1.2}  that due to the symmetry  $\Delta(p_\mu)=\Delta(\frac{2\pi}{a}-p_\mu)$ the derivative $ \frac{d\Delta(p_\mu)}{dp_\mu}$ vanishes at $p_\mu=\pm \frac{\pi}{a}$.  It follows then from (\ref{Phivarphiddf}) that either $\tilde{\varphi}_A(p)$ or $F(p)$ vanish whenever any of the momentum components $p_\mu$ takes the value $\pm \frac{\pi}{a}$, and if $p_\mu=\pm \frac{\pi}{a}$ are the only points where  $ \frac{d\Delta(p_\mu)}{dp_\mu}$ vanishes, namely they are the only extremes of $\Delta(p)$, the range of variation of $\hat{p}_\mu=\Delta(p_\mu)$ as $p_\mu$ varies in the $\frac{4\pi}{a}$ period will be limited 
by eq.(\ref{deltainterval}) with $2l=a$.
This means that in the blocking transformation (\ref{Phivarphiddf})  the lattice fields $\tilde{\varphi}_A(p)$ are not affected by the value of the continuum fields $\tilde{\Phi}_A(\hat{p})$ with $\hat{p}_\mu$ outside the interval defined in (\ref{deltainterval}) and the blocking effectively applies a cutoff on the momentum.
The actual value of the cutoff depends on the choice of $\Delta(p)$. For instance if we choose $\Delta(p)\equiv\Delta_G^{(z)}(p)$ (see eq.(\ref{regD})) we have:
\beq
\hat{p}^{(\mathrm{cutoff})} = \frac{1}{a\hat{z}} \log\frac{1+\hat{z}}{1-\hat{z}}, \label{cutoff} \eeq
which can range from $\frac{2}{a}$ for $\hat{z}=0$\footnote{Note that $\Delta_G^{(0)}(p) = \frac{2}{a} \sin\frac{ap}{2}$} to $\infty$ for $\hat{z}=1$.
Only when $\hat{p}^{(\mathrm{cutoff})} =\infty$ the blocking transformation is invertible and the continuum field configuration can be entirely reconstructed from the one on the lattice. 

The local product of fields in a one dimensional continuum theory is represented in momentum space by the convolution (\ref{localproductp}), which can be trivially generalized to arbitrary dimensions. The blocking transformation (\ref{Phivarphiddf}) maps the local product in the continuum onto a star product on the lattice, which is the direct generalization to $d$ dimensions of the associative star product (\ref{assstarprodp}) introduced in the previous section. This is given by:
\bea
\widetilde{\varphi_{A_1}\star\varphi_{A_2}}(p) &=& \left(\frac{1}{2\pi}\right)^d \int_{-\frac{\pi}{a}}^{\frac{3\pi}{a}} d^Dp_1d^Dp_2 \prod_{\mu=1}^{D} \left|\frac{d\Delta(p_\mu)}{dp_\mu}\right| \frac{F(p_1)F(p_2)}{F(p)}\cdot \nonumber \\ && \cdot \tilde{\varphi}_{A_1}(p_1)\tilde{\varphi}_{A_2}(p_2) ~\delta^D\left( \Delta(p) - \Delta(p_1) - \Delta(p_2) \right).   \label{Dstatprod} \eea 
It should be noticed however that for the reasons already given in the previous section the $D$-dimensional star product defined in (\ref{Dstatprod}) is associative only if $\Delta(p)$ ranges from $-\infty$ to $+\infty$.

The blocking transformation (\ref{Phivarphidd}) can also be written in coordinate representation.
In analogy with the one dimensional case (\ref{varphi0}) we introduce a field $\varphi_{0,A}(x_n)$ defined as:
\beq
\varphi_{0,A}(x_n) = \left(\frac{1}{2\pi}\right)^D \int_{-\frac{\pi}{a}}^{\frac{\pi}{a}} d^Dp~ \tilde{\varphi}_A(p) e^{i x_n^\mu p_\mu}.  \label{vAx} \eeq
Thanks to the symmetry of $\tilde{\varphi}_A(p)$ under $p_\mu \rightarrow \frac{2\pi}{a} - p_\mu$ the field $\varphi_{0,A}(x_n)$ determines the lattice field  $\varphi_{A}(x_n)$ completely in spite of the integration domain in (\ref{vAx}) being half period in each momentum variable. In fact the expression of $\varphi_{A}(x_n)$ in terms of $\varphi_{0,A}(x_n)$ is a trivial generalizazion of (\ref{nmenon}) to the $D$-dimensional case.
From (\ref{vAx}) and the blocking transformation (\ref{Phivarphidd}) we obtain with some algebra:
\beq
\varphi_{0,A}(x_n) = \int_{-\infty}^{\infty} \prod_{\mu=1}^{d} d\xi^\mu \bar{J}^{(d)}_{\Delta,F}(x_n,\xi) \Phi_A(\xi), \label{btcoord} \eeq
where the function $\bar{J}^{(d)}_{\Delta,F}(x_n,\xi)$ is given by:
\beq
\bar{J}^{(d)}_{\Delta,F}(x_n,\xi) = \frac{1}{(2\pi)^d} \int_{-\frac{\pi}{a}}^{\frac{\pi}{a}} \prod_{\mu=1}^{d} dp_\mu \frac{d\Delta(p_\mu)}{dp_\mu}~\frac{e^{i x_n^\nu p_\nu - i \xi^\nu \Delta(p_\nu) }}{F(p)}. \label{Jd} \eeq

The form of  $\bar{J}^{(d)}_{\Delta,F}(x_n,\xi)$ simplifies if we assume that the function $F(p)$ is factorized as a product of single variable functions:
\beq
F(p) = \prod_{\mu=1}^D f(p_\mu). \label{factof} \eeq

With this choice of $F(p)$  the function $\bar{J}^{(d)}_{\Delta,F}(x_n,\xi)$ also factorizes and is given by:
\beq
\bar{J}^{(d)}_{\Delta,F}(x_n,\xi) = \prod_{\mu=1}^{d} \bar{J}^{(0)}_{\Delta,f}(x_n^\mu,\xi^\mu),  \label{factJd} \eeq
where $\bar{J}^{(0)}_{\Delta,f}(x_n^\mu,\xi^\mu)$ is the function introduced in (\ref{JDF}).

We shall study now how different symmetries of the continuum theory are mapped onto the lattice by the blocking transformation (\ref{Phivarphiddf}).
Let us consider first a symmetry which is local in the momentum representation, for instance translational invariance. In the continuum this is given by:
\beq \hat{\delta}_\epsilon \tilde{\Phi}_A(\hat{p}) = i \epsilon \hat{p}~ \tilde{\Phi}_A(\hat{p}). \label{translcont} \eeq
By inserting (\ref{translcont}) into the blocking transformation (\ref{Phivarphiddf}) we find that the corresponding variation on the lattice field configuration is given by:
\beq \delta \tilde{\varphi}_A(p) =  i \epsilon \Delta(p) \tilde{\varphi}_A(p). \label{transllat2} \eeq

The fact that the variation $\delta \tilde{\varphi}_A(p)$ induced on the lattice fields by the  symmetry transformations (\ref{translcont}) can be expressed in term of the lattice fields themselves, as shown in eq. (\ref{transllat2}), is a non trivial property that depends on the type of blocking transformation and on the type of symmetry. This point was investigated in ref. \cite{Bergner-Bruckmann} where the general condition for this to happen was studied in detail 
and  given in eq. (13) of that paper.

In the present case the locality of both the blocking transformation and of the symmetry transformations in momentum space assure that this condition is satisfied. The same applies to supersymmetry transformations: the lattice realization of the $D=N=2$ chiral supersymmetry studied in ref. \cite{DKKS} can be viewed in the same way.

The case of gauge transformations is entirely different. An infinitesimal Abelian gauge transformation in momentum space is given by:
\beq  \hat{\delta}_g \tilde{\Phi}(\hat{p}) = i \int_{-\infty}^{\infty} d^D\hat{q}_1 d^D\hat{q}_2~\tilde{\alpha}(\hat{q}_1)~\tilde{\Phi}(\hat{q}_2)~ \delta^D(\hat{p}-\hat{q}_1-\hat{q}_2). \label{gaugetransf} \eeq

where $\tilde{\alpha}(\hat{q})$ is the Fourier transform of the infinitesimal gauge local parameter $\alpha(x)$.
The gauge transformation (\ref{gaugetransf}) would induce on the corresponding  lattice variable the transformation $\delta_g\tilde{\varphi}(p)$ given by:
\beq    \delta_g\tilde{\varphi}(p) = \frac{1}{F(p)}~\left| \prod_{\mu=1}^d \frac{d\Delta(p_\mu)}{dp_\mu}\right| \hat{\delta}_g \tilde{\Phi}(\hat{p}). \label{gaugelatt} \eeq

However the r.h.s of (\ref{gaugelatt}) cannot be expressed in terms of the lattice degrees of freedom. In fact the integral on the r.h.s. of (\ref{gaugetransf}) contains fields $\tilde{\Phi}(\hat{q}_2)$ with arbitrary large momenta which for a generic $\Delta(p)$ have no correspondence on the lattice. 
Only with a choice of $\Delta(p)$ that ranges from $-\infty$ to $+\infty$, such as $\Delta_G(p)$, the r.h.s. of (\ref{gaugelatt}) can be expressed in terms of the lattice degrees of freedom, and gauge transformations can be consistently defined on the lattice.

\subsection{Lattice actions from blocking transformations} \label{4.2}

Given a classical action in the continuum it is possible, by using the blocking transformation (\ref{Phivarphiddf}) with a specific choice of $\Delta(p)$ and of the function $F(p)$, to derive an effective action for the lattice degrees of freedom. A general discussion of this procedure is found in ref. \cite{Bergner-Bruckmann} to which we refer the reader for more details.

Let $S_{cl}[\tilde{\Phi}]$ be the classical action expressed in terms of the fields $\tilde{\Phi}_A(p)$ . The simplest way to construct an effective lattice action $S_\Delta(\tilde{\varphi})$ in terms of the lattice fields $\tilde{\varphi}_A(p)$ is to impose the blocking transformation (\ref{Phivarphiddf}) by a functional delta function and write:
\beq
e^{-S_\Delta(\tilde{\varphi})} = \int \mathcal{D}\tilde{\Phi}_A ~\prod_\mu \prod_{p_\mu=-\frac{\pi}{a}}^{\frac{3\pi}{a}}\prod_A\delta\left( \tilde{\Phi}_A(\Delta(p)) - \frac{F(p)}{\left| \prod_{\mu=1}^d \frac{d\Delta(p_\mu)}{dp_\mu}\right|}~ \tilde{\varphi}_A(p) \right) e^{-S_{cl}(\tilde{\Phi})}
. \label{effect1} \eeq
We can make use of the symmetry $\Delta(\frac{2\pi}{a}-p) = \Delta(p)$ to reduce the $p$-dependence in (\ref{effect1})  to the fundamental region\footnote{We assume here that $p = \pm \frac{\pi}{a}$ are the only extremes of $\Delta(p)$ so that $\frac{d\Delta(p)}{dp} \geq 0$ in $-\frac{\pi}{a} \leq p \leq \frac{\pi}{a}$.}   $-\frac{\pi}{a} \leq p_\mu \leq \frac{\pi}{a}$. Let $\mathcal{A}$  denote a subset of the possible values of the space-time index $\mu$, as in (\ref{subset}) and define:
\beq
\left(T_{\mathcal{A}}p\right)_\mu  = \left\{\begin{array}{ll}  p_\mu & \mbox{if} ~~~\mu \not\in \mathcal{A} \\  \frac{2\pi}{a}-p_\mu  & \mbox{if}~~~ \mu \in \mathcal{A} \end{array} \right.  \label{Tp} 
. \eeq
Then eq. (\ref{effect1}) can be rewritten as:

\beq
e^{-S_\Delta(\tilde{\varphi})} = \prod_\mu \prod_{p_\mu=-\frac{\pi}{a}}^{\frac{\pi}{a}}\prod_A\prod_{\mathcal{A} \neq \emptyset } \delta\left( \frac{F(T_{\mathcal{A}}p)}{ \prod_{\mu=1}^d \frac{d\Delta(p_\mu)}{dp_\mu}}\tilde{\varphi}_A(T_{\mathcal{A}}p) -\frac{F(p)}{ \prod_{\mu=1}^d \frac{d\Delta(p_\mu)}{dp_\mu}}\tilde{\varphi}_A(p) \right) e^{-\hat{S}_\Delta(\tilde{\varphi})}, \label{effect2} \eeq
where
\beq
 e^{-\hat{S}_\Delta(\tilde{\varphi})} =  \int \mathcal{D}\tilde{\Phi}_A \prod_\mu \prod_{p_\mu=-\frac{\pi}{a}}^{\frac{\pi}{a}}\prod_A\delta\left( \tilde{\Phi}_A(\Delta(p)) - \frac{F(p)}{ \prod_{\mu=1}^d \frac{d\Delta(p_\mu)}{dp_\mu}}~ \tilde{\varphi}_A(p) \right) e^{-S_{cl}(\tilde{\Phi})}. \label{effect3} \eeq

The delta functions in (\ref{effect2}) enforce the symmetry of the product $F(p)\tilde{\varphi}(p)$ under $p_\mu \rightarrow \frac{2\pi}{a} - p_\mu$ which can be expressed in the notation of (\ref{Tp}) as $F(T_{\mathcal{A}}p)\,\tilde{\varphi}(T_{\mathcal{A}}p) = F(p)\,\tilde{\varphi}(p)$ and determine the dependence of the lattice fields $\tilde{\varphi}_A(p)$ from $p_\mu$ when  $\frac{\pi}{a} < p_\mu < \frac{3\pi}{a} $  for at least one value of $\mu$. 
 The action $\hat{S}_\Delta(\tilde{\varphi})$ instead only depends on the fields $\tilde{\varphi}_A(p)$ with all momentum components $p_\mu$ in the interval $-\frac{\pi}{a}<p_\mu<\frac{\pi}{a}$.

A key ingredient  in (\ref{effect3}) is the derivative operator $\Delta(p)$. Let us consider the choice $\Delta(p) = \Delta_G^{(z)}(p)$, where  $\Delta_G^{(z)}(p)$ is the regularized derivative operator defined in (\ref{regD}), which interpolates between $\Delta_G(p)$ and $\Delta_0(p)= \frac{2}{a}\sin\frac{ap}{2}$. For $z < 1$ the derivative operator $\Delta_G^{(z)}(p)$ is bounded by $\Delta_G^{(z)}(p) \leq \hat{p}^{(\mathrm{cutoff})}$,  where $\hat{p}^{(\mathrm{cutoff})}$ is given in (\ref{cutoff}). Therefore if for some $\mu$ we have $\hat{p}_\mu > \hat{p}^{(\mathrm{cutoff})}$, the field $\tilde{\Phi}_A(\hat{p})$ has no lattice correspondent and the r.h.s. of (\ref{effect3}) can be written as:
\bea 
 e^{-\hat{S}_{\Delta_G^{(z)}}(\tilde{\varphi})} &=&  \int  \prod_{|\hat{p}_\mu| \leq \hat{p}^{(\mathrm{cutoff})} }\mathcal{D}\tilde{\Phi}_A(\hat{p})\prod_\mu \prod_{p_\mu=-\frac{\pi}{a}}^{\frac{\pi}{a}}\prod_A\delta\left( \tilde{\Phi}_A(\Delta_G^{(z)}(p)) -  \frac{F(p)}{ \prod_{\mu=1}^d \frac{d\Delta_G^{(z)}(p_\mu)}{dp_\mu}}\tilde{\varphi}_A(p) \right)\cdot \nonumber \\ &&  \cdot \int\prod_{\exists \mu :|\hat{p}_\mu| > \hat{p}^{(\mathrm{cutoff})}}\mathcal{D}\tilde{\Phi}_A(\hat{p})  e^{-S_{cl}(\tilde{\Phi})}. \label{effect4} \eea
Eq. (\ref{effect4}) shows that in order to calculate $\hat{S}_{\Delta_G^{(z)}}(\tilde{\varphi})$ one should  first perform the functional integral over the classical fields $\tilde{\Phi}_A(\hat{p})$ with $|\hat{p}_\mu| \geq \hat{p}^{(\mathrm{cutoff})}$  for some $\mu$, then by using the delta functions do the functional integral over the remaining fields $\tilde{\Phi}_A(\hat{p})$   with   $|\hat{p}_\mu| \leq \hat{p}^{(\mathrm{cutoff})}$ for all values of $\mu$, which amounts to replacing  in $S_{cl}(\tilde{\Phi})$  the classical field $\tilde{\Phi}_A(\hat{p})$ with $ \frac{F(p)}{ \prod_{\mu=1}^d \frac{d\Delta_G^{(z)}(p_\mu)}{dp_\mu}} \tilde{\varphi}_A(p)$ where the components of  $p$ are all in the interval $(-\frac{\pi}{a},\frac{\pi}{a})$  and $\hat{p} = \Delta_G^{(z)}(p)$.

No approximation is involved in the functional integral at the r.h.s. of (\ref{effect4}), so the correlation functions obtained from the lattice theory are directly related to the ones of the continuum theory. In fact if we define the normalized generating functional $Z_{\Delta_G^{(z)}}(\tilde{j})$ on the lattice as
\beq
Z_{\Delta_G^{(z)}}(\tilde{j}) = \frac{\int \prod_{|p_\mu|\leq \frac{\pi}{a}} \mathcal{D}\tilde{\varphi}_A(p)  e^{-\hat{S}_{\Delta_G^{(z)}}(\tilde{\varphi}) + \int_{{|p_\mu|\leq \frac{\pi}{a}}} dp ~ \tilde{j}_A(-p) \tilde{\varphi}_A(p)}}{\int \prod_{|p_\mu|\leq \frac{\pi}{a}} \mathcal{D}\tilde{\varphi}_A(p)  e^{-\hat{S}_{\Delta_G^{(z)}}(\tilde{\varphi}) }} ,  \label{Z1} \eeq
we insert (\ref{effect4})  in the r.h.s. of (\ref{Z1}) and we perform the functional integral over $\tilde{\varphi}_A(p)$ using the delta functions we obtain:
\beq
Z_{\Delta_G^{(z)}}(\tilde{j}) = Z^{(cl)}_{\tilde{p}^{(\mathrm{cutoff})}}(\tilde{J}), \label{Z2} \eeq
where $ Z^{(cl)}_{\tilde{p}^{(\mathrm{cutoff})}}(\tilde{J})$ is the normalized generating functional for the continuum theory, but including only source terms  with momenta $\left|\hat{p}_\mu\right| \leq \tilde{p}^{(\mathrm{cutoff})}$ :
\beq
 Z^{(cl)}_{\tilde{p}^{(\mathrm{cutoff})}}(\tilde{J}) =\frac{ \int \mathcal{D} \tilde{\Phi}_A(\hat{p}) e^{-S_{cl}(\tilde{\Phi}) + \int_{\left|\hat{p}_\mu\right| \leq \tilde{p}^{(\mathrm{cutoff})}} d\hat{p}_\mu \tilde{J}_A(-\hat{p}) \tilde{\Phi}_A(\hat{p})}}{\int \mathcal{D} \tilde{\Phi}_A(\hat{p}) e^{-S_{cl}(\tilde{\Phi})}}, \label{Z3} \eeq
and the relation between $\tilde{j}$ and $\tilde{J}$ in (\ref{Z2}) is given by:
\beq
\tilde{J}_A(\Delta_G^{(z)}(p)) = \frac{\tilde{j}_A(p)}{F(p) }. \label{jJ} \eeq



Except for some very special case, like for instance quadratic actions, the functional integral over the classical fields with $|p_\mu| > \tilde{p}^{(\mathrm{cutoff})}$ in eq. (\ref{effect4}) cannot be done explicitely. However an explicit lattice action can be written if one starts from a truncated continuum theory obtained from the original one by imposing the condition
\beq
\tilde{\Phi}_A(\hat{p}) = 0 ~~~~~~~~~~~~~~~~~~\mathrm{if}~~ |\hat{p}_\mu| > \tilde{p}^{(\mathrm{cutoff})} ~~\mathrm{for ~any} ~\mu.   \label{trunc} \eeq
With this truncation the inner functional integral in (\ref{effect4}) is trivial and the last can be done explicitely because of the delta functions.
The result is a lattice action that can be obtained from the one in the continuum by first applying (\ref{trunc}), then  by replacing any derivative $\tilde{p}_\mu$ with its lattice equivalent $\Delta_G^{(z)}(p_\mu)$ and finally the continuum fields  $\tilde{\Phi}_A(\hat{p})$ and the differentials $d\tilde{p}_\mu$  with the corresponding lattice quantities according to the blocking transformation (\ref{Phivarphidiff}).

Unlike the lattice theory obtained from  (\ref{effect4}), the lattice theory obtained after the truncation (\ref{trunc})  is not equivalent to the theory in the continuum,  but it  is expected to reproduce the results of the continuum theory in the limit $a \rightarrow 0$ (although this should be checked case by case). 

Gauge invariance is broken by the the truncation (\ref{trunc}). As already remarked earlier in the paper gauge transformations in momentum space are given by convolutions that involve arbitrarily high momenta, hence any truncation in the  momentum spoils gauge invariance. So gauge invariance is broken in  the lattice theory obtained via the blocking transformation (\ref{Phivarphidiff}) after truncating the continuum theory according to (\ref{trunc}).

On the contrary supersymmetry transformations are local in momentum space, and the truncation (\ref{trunc}) is supersymmetric invariant.
So in supersymmetric non-gauge theories the lattice actions obtained by the combined application of (\ref{trunc}) and (\ref{Phivarphidiff}) have exact supersymmetry on the lattice. This is the case of the $D=N=2$ Wess Zumino model whose supersymmetric lattice action has already been written in ref. \cite{DKKS}. In that paper the action was derived by first writing exact supersymmetry transformations on the lattice for a general $N=D=2$ superfield and then imposing chiral conditions on the lattice. The end result however is the same as we would have obtained by applying directly eq.s (\ref{trunc}) and (\ref{Phivarphidiff}) to the action in the continuum. 
More of this will be discussed in the following section.

As already remarked earlier on,  gauge theories require a choice of the derivative operator $\Delta(p)$ that spans (twice) the whole real axis as $p$ goes over a $\frac{4\pi}{a}$ period. 
As discussed earlier the best choice in this respect is $\Delta_G(p)$  defined in (\ref{assdelta}) which is obtained from the regularized difference operator $\Delta_G^{(z)}(p)$ by taking $z \rightarrow 1$.
In this limit $\hat{p}^{(\mathrm{cutoff})}$ becomes infinite and the second functional integral at the r.h.s. of (\ref{effect4}) becomes $1$ so that (\ref{effect4}) takes the form:
\beq 
 e^{-\hat{S}_{\Delta_G}(\tilde{\varphi})} =  \int  \mathcal{D}\tilde{\Phi}_A(\hat{p})\prod_\mu \prod_{p_\mu=-\frac{\pi}{a}}^{\frac{\pi}{a}}\prod_A\delta\left( \tilde{\Phi}_A(\Delta_G(p)) -  F(p)\prod_{\nu=1}^d \cos\frac{ap_\nu}{2} \tilde{\varphi}_A(p) \right)  e^{-S_{cl}(\tilde{\Phi})}. \label{effect5} \eeq 

With the choice $\hat{p}_\mu=\Delta_G(p_\mu)$ the star product  (\ref{Dstatprod}) is associative, the blocking transformation (\ref{Phivarphidd}) becomes invertible,  and as a consequence also (\ref{effect5}) can be inverted (modulo a constant proportionality factor), and the action in the continuum can be completely reconstructed from the lattice one:
\beq 
 e^{-S_{cl}(\tilde{\Phi})} =  \int  \mathcal{D}\tilde{\varphi}_A(p)\prod_\mu \prod_{p_\mu=-\frac{\pi}{a}}^{\frac{\pi}{a}}\prod_A\delta\left( \tilde{\Phi}_A(\Delta_G(p)) -  F(p)\prod_{\nu=1}^d \cos\frac{ap_\nu}{2} \tilde{\varphi}_A(p) \right) e^{-\hat{S}_{\Delta_G}(\tilde{\varphi})}.  \label{effect6} \eeq 

The generating functional of the correlation functions on the lattice $Z_{\Delta_G}(\tilde{j})$ is obtained from (\ref{Z1}) by  setting $z=1$:
\beq
Z_{\Delta_G}(\tilde{j}) = \frac{\int \prod_{|p_\mu|\leq \frac{\pi}{a}} \mathcal{D}\tilde{\varphi}_A(p)  e^{-\hat{S}_{\Delta_G}(\tilde{\varphi}) + \int_{{|p_\mu|\leq \frac{\pi}{a}}} dp ~ \tilde{j}_A(-p) \tilde{\varphi}_A(p)}}{\int \prod_{|p_\mu|\leq \frac{\pi}{a}} \mathcal{D}\tilde{\varphi}_A(p)  e^{-\hat{S}_{\Delta_G}(\tilde{\varphi}) }} ,  \label{Z1G} \eeq
and its relation with the corresponding functional in the continuum theory is:
\beq
Z_{\Delta_G}(\tilde{j}) = Z^{(cl)}(\tilde{J}), \label{Z2G} \eeq
with
\beq
\tilde{J}_A(\Delta_G(p)) = \frac{\tilde{j}_A(p)}{F(p) }. \label{jJG} \eeq
Notice that in the generating functional $Z^{(cl)}(\tilde{J})$ there is now no restriction on the momenta of the sources $\tilde{J}_A(\hat{p})$, so that the lattice theory contains the full information on the correlation functions of the continuum theory.

\section{Regularization and renormalization in the new lattice: two examples}
\label{section6}
In this section we present two simple examples of field theories formulated in our lattice approach: the massless supersymmetric non interacting Wess-Zumino model and the bosonic theory with $\Phi^4$ interaction in four dimensions.

In the first example we shall show that supersymmetry is not affected and give an explicit realization of the action in coordinate representation. In the second example we shall mostly concentrate on the problem of regularization and renormalization of the theory.

In the standard lattice theory the lattice constant $a$ acts as a regulator for the ultraviolet divergences that plague the continuum theory. On the other hand we have seen in the last section that  in our approach  the lattice action can be obtained from the one in the continuum via a blocking transformation which is invertible if the derivative operator is the one given in (\ref{regD}) with $\hat{z}=1$. This means that there is no loss of information  going from the continuum to the lattice formulation, and that in spite of the introduction of the length scale $a$, the lattice theory is completely equivalent to the original continuum one, and hence contains the same ultraviolet divergences. 
On the other hand, if a different value of $\hat{z}$ is used in (\ref{regD}), the lattice theory is mapped onto a continuum theory whith a cutoff (\ref{cutoff}) in the momenta which  depends on the adimensional parameter $\hat{z}$.  So while the lattice constant $a$ sets a scale for the length the role of regulator of the ultraviolet divergences is not played by the lattice constant by itself but rather by a combination of the lattice constant and of the paramenter $\hat{z}$ which is related to the locality of the derivative operator.
This will be discussed in the present section. Moreover we shall show that by making the interaction in the $\Phi^4$ theory slightly non-local, a different regulator  with similar properties can be introduced leading to a  regularization scheme similar to the momentum cutoff. One loop renormalization is explicitely checked.
  
\subsection{Non interacting supersymmetric Wess-Zumino model in four dimension}
\label{5.1}
As a first application  we shall consider the simplest supersymmetric theory in four dimensions, namely the massless non interacting Wess-Zumino model \cite{Wess:1974tw}.
This is given by just the kinetic terms of a single left handed two component Weyl fermion $\Psi$ and a complex scalar boson $\Phi$:
\beq
S = \int d^4x \left( i \Psi^\dagger \bar{\sigma}^\mu\partial_\mu\Psi - \partial^\mu\Phi^\star\partial_\mu\Phi \right), \label{4dWSmodel} \eeq
 where in our notations $\sigma^i=-\bar{\sigma}^i$ are the Pauli matrices and $\sigma^0 = \bar{\sigma}^0$ is the identity matrix.
The action (\ref{4dWSmodel}) is invariant on shell under the supersymmetry transformations:
\bea
&\delta_\epsilon \Phi = \epsilon \Psi,~~~~~~~~~~~~~~&\delta_\epsilon\Phi^\star = \epsilon^\dagger\Psi^\dagger \label{susytransb}\\ &\delta_\epsilon\Psi_\alpha = -i(\sigma^\mu\epsilon^\dagger)_\alpha \partial_\mu\Phi,~~~~~~~~~~~~~&\delta_\epsilon\Psi^\dagger_{\dot{\alpha}} =  i(\epsilon\sigma^\mu)_{\dot{\alpha}} \partial_\mu\Phi^\star
,   \label{susytransf} \eea
where $\epsilon_\alpha$ is a Grassmann odd Weyl spinor parameter\footnote{In our notations spinors without (resp. with) a dagger always carry  undotted (resp. dotted) indices, and are right (resp. left) handed.}.

The action (\ref{4dWSmodel}) can now be written in momentum representation:
\beq
S = \frac{1}{(2\pi)^4} \int d^4\hat{p}_1 d^4\hat{p}_2~\delta^4 \left( \hat{p}_1+\hat{p}_2 \right) \left[ -\tilde{\Psi}^\dagger(\hat{p}_1)  \bar{\sigma}_\mu \hat{p}_2^\mu \tilde{\Psi}(\hat{p}_2) + \hat{p}_{1\mu} \tilde{\Phi}(\hat{p}_1) \hat{p}_2^\mu \tilde{\Phi}^\dagger(\hat{p}_2)\right],   \label{4dWSmodelp} \eeq
where $\tilde{\Phi}^\dagger(\hat{p})$ is the Fourier transform of $\Phi^\star(x)$\footnote{Note that $\tilde{\Phi}^\dagger(\hat{p})=\tilde{\Phi}^\star(-\hat{p})$.}

The corresponding lattice action is obtained by replacing the fields $\tilde{\Phi}(\hat{p})$ and $\tilde{\Psi}(\hat{p})$ with the lattice fields $\tilde{\varphi}(p)$ and $\tilde{\psi}(p)$ periodic with period $\frac{4\pi}{a}$ in the components $p_\mu$ of the lattice momentum and invariant under the symmetry $p_\mu \rightarrow \frac{2\pi}{a}-p_\mu$ for each value of $\mu$. The derivative operator $i \partial_\mu$, namely $\hat{p}_\mu$ in momentum representation, is replaced on the lattice by the lattice derivative $\Delta(p_\mu)$ discussed in the previous sections. In the remaining part of this section we shall choose $\Delta(p_\mu) = \Delta_G^{(z)}(p_\mu)$ where $ \Delta_G^{(z)}(p_\mu)$ is the regularized operator defined in (\ref{regD}) and discussed there.

Following the prescriptions above we can write the lattice action as:
\bea
S_V^{(z)} &=& \frac{1}{(4\pi)^4} \int_{-\frac{\pi}{a}}^{\frac{3\pi}{a}} d^4p_1~d^4p_2~v(p_1,p_2) \delta^4\left(\Delta_G^{(z)}(p_1)+\Delta_G^{(z)}(p_2) \right)\cdot \nonumber \\ &\cdot& \left[-\tilde{\psi}^\dagger(p_1) \bar{\sigma}^\mu \Delta_G^{(z)}(p_{2\mu}) \tilde{\psi}(p_2) + 
\Delta_G^{(z)}(p_{1\mu})\tilde{\varphi}(p_1) \Delta_G^{(z)}(p_2^\mu)\tilde{\varphi}^\dagger(p_2) \right], \label{4dWSmodellattp} \eea
where $v(p_1,p_2)$ defines the volume of integration in the momentum space and will be specified below.
Let us consider now the delta functions in eq.(\ref{4dWSmodellattp}). Their argument vanish for either $p_{1\mu}+p_{2\mu} =0$ or $p_{1\mu}-p_{2\mu} =\frac{2\pi}{a}$ modulo $\frac{4\pi}{a}$, so we have:
\beq
\delta^2 \left(\Delta_G^{(z)}(p_{1\mu}) + \Delta_G^{(z)}(p_{2\mu}) \right) = \frac{\prod_\mu\left[\delta\left(p_{1\mu} + p_{2\mu} \right)+\delta\left(p_{1\mu} - p_{2\mu}-\frac{2\pi}{a} \right)\right]}{\sqrt{\prod_\mu\left| \frac{d\Delta_G^{(z)}(p_{1\mu})}{dp_{1\mu}} \frac{d\Delta_G^{(z)}(p_{2\mu})}{dp_{2\mu}}\right|} },\label{deltaGG} \eeq
where  the square root in the denominator is the result of a symmetrization with respect to $p_1$ and $p_2$.

Eq. (\ref{deltaGG}) can now be inserted into the action (\ref{4dWSmodellattp}) and the integration volume $v(p_1,p_2)$ can be chosen in such a way to simplify the quadratic action as much as possible. If we choose
\beq
 v(p_1,p_2) = \sqrt{\prod_\mu\left| \frac{d\Delta_G^{(z)}(p_{1\mu})}{dp_{1\mu}} \frac{d\Delta_G^{(z)}(p_{2\mu})}{dp_{2\mu}}\right|}.  \label{simpV} \eeq

the lattice action, which we shall now simply denote as $S^{(z)}$,  becomes:
\bea
S^{(z)} &=& \frac{1}{(4\pi)^4} \int_{-\frac{\pi}{a}}^{\frac{3\pi}{a}} d^4p_1~d^4p_2~\prod_\mu\left[\delta\left(p_{1\mu} + p_{2\mu} \right)+\delta\left(p_{1\mu} - p_{2\mu}-\frac{2\pi}{a} \right)\right]\cdot \nonumber \\ &\cdot& \left[-\tilde{\psi}^\dagger(p_1) \bar{\sigma}^\mu \Delta_G^{(z)}(p_{2\mu}) \tilde{\psi}(p_2) + 
\Delta_G^{(z)}(p_{1\mu})\tilde{\varphi}(p_1) \Delta_G^{(z)}(p_2^\mu)\tilde{\varphi}^\dagger(p_2) \right]. \label{4dWSmodellattps} \eea

With the choice of $v(p_1,p_2)$ given in (\ref{simpV}) the non-locality of $S_{\mathrm{latt}}$ in coordinate representation is entirely due to the non-locality of the lattice derivative $\Delta_G^{(z)}(p_{\mu})$ (which is in fact local for $z=0$ and very non-local in the limit $z \rightarrow 1$) and to the symmetry $p_\mu \rightarrow \frac{2\pi}{a} - p_\mu$ which couples the points of coordinates $\frac{a n_\mu}{2}$ and $\frac{-a n_\mu}{2}$ in the $\mu$ direction.
Any other choice of $v(p_1,p_2)$ would introduce extra dependence on the lattice momenta, and hence extra non-locality in coordinate representation. So we can say that the integration volume (\ref{simpV}) gives the most local action in this context.

We shall now make use of the fact that all fields are invariant under $p_\mu \rightarrow \frac{2\pi}{a} - p_\mu$ (separately for each $\mu$) and that the sum of delta functions is also invariant under such symmetry. As a result, each momentum integration over  the $\frac{4\pi}{a}$  interval $(-\frac{\pi}{a}, \frac{3\pi}{a})$ is equal to twice the integral over the $\frac{2\pi}{a}$ interval  $(-\frac{\pi}{a}, \frac{\pi}{a})$ and the action (\ref{4dWSmodellattps}) can be written as:
\bea
S^{(z)} &=& \frac{1}{\pi^4} \int_{-\frac{\pi}{a}}^{\frac{\pi}{a}} d^4p_1~d^4p_2~\prod_\mu \delta\left(p_{1\mu} + p_{2\mu} \right)\cdot \nonumber \\ &\cdot& \left[-\tilde{\psi}^\dagger(p_1) \bar{\sigma}^\mu \Delta_G^{(z)}(p_{2\mu}) \tilde{\psi}(p_2) + 
\Delta_G^{(z)}(p_{1\mu})\tilde{\varphi}(p_1) \Delta_G^{(z)}(p_2^\mu)\tilde{\varphi}^\dagger(p_2) \right]. \label{4dWSmodellattpss} \eea

A more direct relation between the lattice action (\ref{4dWSmodellattpss}) and the action in the continuum (\ref{4dWSmodelp}) can be obtained by setting in (\ref{4dWSmodellattpss})  $\hat{p}_\mu = \Delta_G^{(z)}(p_{\mu})$ and by using  $\hat{p}_\mu$, which can be identified with the  momentum of a theory in the continuum,  as independent integration variable. 
With a few algebraic manipulations (essentially by tracing our steps back from lattice to continuum) we find:
\beq
S^{(z)} = \frac{1}{(2\pi)^4} \int_{|\hat{p}_\mu|\leq \hat{p}^{(\mathrm{cutoff})}} d^4\hat{p}_1 d^4\hat{p}_2~\delta^4 \left( \hat{p}_1+\hat{p}_2 \right) \left[  -\tilde{\Psi}^\dagger(\hat{p}_1)  \bar{\sigma}_\mu \hat{p}_2^\mu \tilde{\Psi}(\hat{p}_2) + \hat{p}_{1\mu} \tilde{\Phi}(\hat{p}_1) \hat{p}_2^\mu \tilde{\Phi}^\dagger(\hat{p}_2)
 \right],   \label{4dWSmodelpct} \eeq
with
\beq
\tilde{\Psi}(\Delta_G^{(z)}(p_{\mu}))=\frac{2\tilde{\psi}(p)}{\prod_\mu\sqrt{\frac{d\Delta_G^{(z)}(p_\mu)}{dp_\mu}}}~~~~~~,~~~~~~\tilde{\Phi}(\Delta_G^{(z)}(p_{\mu}))=\frac{2\tilde{\varphi}(p)}{\prod_\mu\sqrt{\frac{d\Delta_G^{(z)}(p_\mu)}{dp_\mu}}}. \label{cl} \eeq

Eq. (\ref{cl}) coincides with the blocking transformation (\ref{Phivarphidd}), but with a specific choice of the function $F(p)$, namely with $F(p)$ of the factorized form (\ref{factof}) and $f(p)$ given by:
\beq
f(p) = \sqrt{\left| \frac{d\Delta_G^{(z)}(p)}{dp}\right|} = \sqrt{\left|\frac{\cos\frac{ap}{2}}{1-\hat{z}^2 \sin^2\frac{ap}{2}}\right|}. \label{fspecial} \eeq
For $z \rightarrow 1$, namely for $\Delta_G^{(z)}(p) \rightarrow \Delta_G(p)$ this choice of $f(p)$ coincides with the one given in (\ref{fsqrt}) that guarantees a smooth continuum limit in coordinate representation. 

The action (\ref{4dWSmodelpct}) is the action of a free theory in the continuum where a cutoff on the components of the momenta has been introduced. It is important to notice that in (\ref{4dWSmodelpct}) the dependence on the lattice constant $a$ and the parameter $\hat{z}$, which was explicit in (\ref{4dWSmodellattpss}), is all contained in the value of  $\hat{p}^{(\mathrm{cutoff})}$ which is given in (\ref{cutoff}). 
As a consequence\emph{ lattice actions  (\ref{4dWSmodellattpss}) with different values of the lattice constant $a$ and of the parameter $\hat{z}$ but corresponding to the same value of the cutoff $\hat{p}^{(\mathrm{cutoff})}$ according to eq. (\ref{cutoff}) are physically equivalent, as they correspond to the same continuum theory (\ref{4dWSmodelpct}).}

This means that it is possible via a (proper) blocking transformation to map \emph{exactly} the action (\ref{4dWSmodelpct}) with a given lattice constant $a$ and regulator parameter $\hat{z}$ onto the same action with a different constant $a'$ and regulator parameter $\hat{z}'$ provided the value of the cutoff  $\hat{p}^{(\mathrm{cutoff})}$ is the same in the two cases. 

In order to give a specific example let us start from a lattice with lattice spacing $a$ and a local derivative
\beq
\Delta(p_\mu) = \frac{2}{a} \sin{\frac{ap_\mu}{2}}, \label{localder} \eeq
that corresponds to $\Delta_G^{(z)}(p)$  with  $z=\hat{z}=0$. The momentum cutoff is in this case $\frac{2}{a}$.  Consider now a lattice with a lattice spacing $a'$ which is the double of the original one: $a'=2 a$.  We can determine the value of the regulator parameter $\hat{z}'$ for which the value of the momentum cutoff  $\hat{p}^{(\mathrm{cutoff})}$ calculated from (\ref{cutoff}) is the same in the two cases. This leads to the following equation:
\beq
\frac{1}{\hat{z}'} \log\frac{1+\hat{z}'}{1-\hat{z}'} = 4, \label{zprime} \eeq
whose solution has the numerical value $\hat{z}' = 0.9575\dots$~ which corresponds to a value of $z$ in eq.(\ref{deltacoordz}) equal to 0.74316\dots~.
\footnote{It is interesting that one could also choose $a'<a$, for instance $a'=\frac{a}{2}$. In order to keep the value of the cutoff unchanged this requires $\hat{z}'$ imaginary: for $a'=\frac{a}{2}$ one finds for instance $\hat{z}' = 2.3311.. i$ which corresponds to $z' =0.2827 i$. A purely imaginary value of  $z$ in eq.(\ref{deltacoordz}) leads to a series with all positive signs so that the contributions of far away terms (i.e. large $k$) are enhanced.} 

Let us denote by $p_\mu$ and $p'_\mu$ the components of the momentum on the two lattices. Then the lattice derivatives in the two cases are given by (\ref{localder}) and by:
\beq
\Delta'(p'_\mu) = \frac{1}{2a\hat{z}'}\log\frac{1+\hat{z}'\sin ap'_\mu}{1-\hat{z}'\sin ap'_\mu}, \label{derprime} \eeq
where in the last equation we have made use of the fact that $a'=2a$. Notice in particular that the periodicity in $p'_\mu$ is $\frac{4\pi}{a'} = \frac{2\pi}{a}$.
The two derivative operators $\Delta(p_\mu) $ and $\Delta'(p'_\mu) $ are to be identified as they both represent the momentum $\hat{p}_\mu$ of the continuum theory. So we can write:
\beq
 \frac{2}{a} \sin{\frac{ap_\mu}{2}} = \frac{1}{2a\hat{z}'}\log\frac{1+\hat{z}'\sin ap'_\mu}{1-\hat{z}'\sin ap'_\mu} = \hat{p}_\mu. \label{pipiprime} \eeq
Eq. (\ref{pipiprime}) establishes a relation between the lattice momenta $p_\mu$ and $p'_\mu$ and it is consistent because thanks to (\ref{zprime}) the range of variation of $\Delta(p_\mu) $ and $\Delta'(p'_\mu) $ coincide.
Let us denote with $\tilde{\varphi}_o(p)$ and $\tilde{\psi}_o(p)$ the lattice fields of the theory with lattice spacing $a$ and $\hat{z}=0$ and with $\tilde{\varphi}'_o(p')$, and $\tilde{\psi}'_o(p')$ the lattice fields of the theory with lattice spacing $a'=2a$ and $\hat{z}'$ given by (\ref{zprime}).
The blocking transformations (\ref{cl}) can be written in both cases and give for the bosonic fields\footnote{Analogous equations are found for the fermionic fields.}:
\beq
\tilde{\Phi}(\hat{p}) = \frac{2\tilde{\varphi}_o(p)}{\prod_\mu \sqrt{\cos\frac{ap_\mu}{2}}} =
 \frac{2\tilde{\varphi}'_o(p')}{\prod_\mu \sqrt{\frac{\cos ap'_\mu}{1 - \hat{z}^{'2} \sin^2 ap'_\mu}}}, \label{btlatlat} \eeq
where the relation between $\hat{p}$,$p$ and $p'$ is given in (\ref{pipiprime}).

The last equality in (\ref{btlatlat}) together with the similar one for the fermionic fields defines a blocking transformation that maps \emph{exactly} the free theory defined by the action (\ref{4dWSmodellattpss}) with $\hat{z}=0$ and lattice constant $a$ into one defined by the same action but with a lattice constant $a'=2a$ and with the parameter $\hat{z}'$ given by (\ref{zprime}) and numerically close to the limiting value of $1$: $\hat{z}' = 0.9575...$.

The conventional continuum action is reached by letting $\hat{p}^{(\mathrm{cutoff})} \rightarrow  \infty$. This can be obtained in two ways, namely by either keeping $\hat{z}'$ fixed (for instance $\hat{z}'=0$) and taking the limit where the lattice spacing $a$ goes to zero, or by keeping the lattice spacing fixed and taking the limit $\hat{z}' \rightarrow 1$.

The first case corresponds to the standard continuum limit of a lattice theory, and the lattice structure disappears in the limit; the second case is specific to our approach: the lattice constant remains finite in the limit and the resulting theory is still defined on a lattice in spite of being equivalent to the continuum theory. The lattice derivative in this limit is given by $\Delta_G(p)$ and the equivalence to the continuum theory is related to the existence of an \emph{invertible} blocking transformation, as discussed in the previous section. 

The action $S^{(z)}$, defined in (\ref{4dWSmodellattps}), can be expressed in coordinate representation. In order to do that let us notice that both the bosonic and the fermionic terms in 
(\ref{4dWSmodellattps}) are of the form:
\beq
I = \frac{1}{(4\pi)^4} \int_{-\frac{\pi}{a}}^{\frac{3\pi}{a}} d^4p_1~d^4p_2~\prod_\mu\left[\delta\left(p_{1\mu} + p_{2\mu} \right)+\delta\left(p_{1\mu} - p_{2\mu}-\frac{2\pi}{a} \right)\right] \tilde{H}_1(p_1) \tilde{H}_2(p_2), \label{HH} \eeq
where $\tilde{H}_1(p_1)$ and $\tilde{H}_2(p_2)$ are both invariant under $p_{i\mu} \rightarrow \frac{2\pi}{a} - p_{i\mu}$
By expressing  $\tilde{H}_i(p_i)$ in terms of their Fourier transform as in (\ref{fourier1}) and performing the integration over the momenta we easily obtain:
\bea &&I = \frac{1}{16} \sum_{n^\mu} \left[ H_1(x_n)H_2(x_n) +\sum_\mu (-1)^{n^\mu} H_1(x_n)H_2(P_\mu x_n) + \sum_{\mu \neq \nu} (-1)^{n^\mu+n^\nu} H_1(x_n)H_2(P_\mu P_\nu x_n)\right. \nonumber \\ &&+\left. \sum_{\mu \neq \nu \neq \rho} (-1)^{n^\mu+n^\nu +n^\rho} H_1(x_n)H_2(P_\mu P_\nu P_\rho x_n) + \sum_{\mu \neq \nu \neq \rho \neq \sigma} (-1)^{\sum_\rho n^\rho}  H_1(x_n)H_2(P_\mu P_\nu P_\rho P_\sigma x_n) \right] \nonumber \\ &&,  \label{Icoord} \eea
where $x_n$ denotes the coordinate of the lattice site with components $x_n^{\mu} = \frac{a n^{\mu}}{2}$ and $P_\mu$ is an operator that changes the sign of $n^{\mu}$. 
Since $H_2(x_n)$ satisfies the symmetry (\ref{chiralcondd}), namely $ H_2(P_\mu x_n) =(-1)^{n^\mu} H_2(x_n)$, eq. (\ref{Icoord}) can be rewritten in the apparently local form:
\beq
I= \sum_{n^\mu} H_1(x_n)H_2(x_n). \label{Icoord2} \eeq
Thanks to the same symmetry (see eq.(\ref{chiralcondd})) the sum can be restricted to a single quadrant of the lattice coordinate space, namely:
\beq
I= 16 \sum_{n^\mu>0} H_1(x_n)H_2(x_n) + \mathrm{boundary ~~ terms},  \label{Icoord3} \eeq
where the boundary terms are the ones with one or more $n^\mu$ set to zero. They have the same form as the r.h.s. of (\ref{Icoord3}) but with the coefficient in front equal to $2^{d_0}$ where $d_0$ is the dimensionality of the boundary.

All the previous forms of $I$ (eq.s (\ref{Icoord}), (\ref{Icoord2}) and (\ref{Icoord3})) can be used to write the action $S^{(z)}$ in coordinate representation by simply replacing $H_1$ and $H_2$ with the corrsponding quantities in the bosonic and fermionic term of the Lagrangian. If we consider for simplicity the form (\ref{Icoord2}) we get:
\beq
S^{(z)} = \sum_{n^\mu} \left[ i\psi^\dagger(x_n) \bar{\sigma}^\mu \Delta_{G \mu}^{(z)}\psi(x_n)  -\Delta_G^{(z) \mu}\varphi^\star(x_n)  \Delta_{G \mu}^{(z)}\varphi(x_n) \right], \label{szcoord} \eeq
where the operator $\Delta_{G \mu}^{(z)}$ is defined in (\ref{deltacoordz}) with the index $\mu$ denoting that it acts only on the $\mu$ component of $x_n$.
It is easy to check that the action (\ref{szcoord}) is invariant under supersymmetry transformations obtained from (\ref{susytransb}) and  (\ref{susytransf}) by replacing the quantities of the continuum theory with the corresponding one on the lattice:
\bea
&\delta_\epsilon \varphi(x_n) = \epsilon \psi(x_n) ,~~~~~~~~~~~~~~&\delta_\epsilon\varphi^\star(x_n)  = \epsilon^\dagger\psi^\dagger(x_n) \label{susytranslattb}\\ &\delta_\epsilon\psi_\alpha(x_n) = -i(\sigma^\mu\epsilon^\dagger)_\alpha \Delta_{G \mu}^{(z)}\varphi(x_n),~~~~~~~~~~~~~&\delta_\epsilon\psi^\dagger_{\dot{\alpha}} =  i(\epsilon\sigma^\mu)_{\dot{\alpha}} \Delta_{G \mu}^{(z)}\varphi^\star(x_n)
.   \label{susytranslattf} \eea

 The prescription adopted in this section for constructing the lattice action can be applied in any number of  dimensions and extended to interacting theories. Supersymmetry, when present in the original theory, is exactly preserved on the lattice. 

Ward-Takahashi identities for lower dimensional (D=1,2) N=2 Wess-Zumino models with interactions for $\Delta_{G\mu}^{(z=0)}$ 
were investigated to check quantum level consistencies of the exact supersymmetry\cite{WTid}. 
It was shown that the identities are perfectly consistent up to two loops. These consistencies are based on the 
typical features of the formulation that the lattice version of fermion and boson propagators have a simple relation 
without doublers as in the continuum theory. We expect that the vacuum energy exactly vanishes due to exact fermion 
boson cancellation.   

\subsection{$\Phi^4$  theory in four dimensions}

As a second example we consider in this subsection the four dimensional action of a real scalar field with a quartic interaction. We shall work in momentum representation, in which the Euclidean  action can be written in the continuum theory as:
\bea
S_c &=& \int d^4\hat{p}_1 d^4\hat{p}_2~ \delta^{(4)}(\hat{p}_1+\hat{p}_2) \left[-\hat{p}_1^\mu \tilde{\Phi}(\hat{p}_1)\hat{p}_{2\mu}\tilde{\Phi}(\hat{p}_2) + m_0^2 \tilde{\Phi}(\hat{p}_1)\tilde{\Phi}(\hat{p}_2)\right] \nonumber \\ &+&\lambda_0 \int \prod_{i=1}^4 d^4\hat{p}_i~\delta^{(4)}(\hat{p}_1+\hat{p}_2+\hat{p}_3+\hat{p}_4)~  \tilde{\Phi}(\hat{p}_1)\tilde{\Phi}(\hat{p}_2) \tilde{\Phi}(\hat{p}_3)\tilde{\Phi}(\hat{p}_4), \label{phi4c} \eea
where as usual $\hat{p}_\mu$ and $\tilde{\Phi}(\hat{p})$ are the momenta and the fields in the continuum, whereas the corresponding lattice quantities will be denoted by $p_\mu$ and $\tilde{\varphi}(p)$.
Following the same prescription used in the Wess-Zumino case  of the previous subsection (see the discussion following eq.(\ref{4dWSmodelp})) we can write the action on the lattice as:

\bea
S_{l}&=&-\int_{-\frac{\pi}{a}}^{\frac{3\pi}{a}} d^4p_1d^4p_2~v_k(p_1,p_2)~\delta^{(4)}\left(\Delta_G^{(z)}(p_1^\mu)+\Delta_G^{(z)}(p_2^\mu)\right)~\Delta_G^{(z)}(p_1^\mu)\Delta_G^{(z)}(p_{2\mu}) ~\tilde{\varphi}(p_1)\tilde{\varphi}(p_2)  \nonumber \\
&+&m_0^2 ~\int_{-\frac{\pi}{a}}^{\frac{3\pi}{a}} d^4p_1d^4p_2~v_m(p_1,p_2)~\delta^{(4)}\left(\Delta_G^{(z)}(p_1^\mu)+\Delta_G^{(z)}(p_2^\mu)\right)~\tilde{\varphi}(p_1)\tilde{\varphi}(p_2) \nonumber \\ &+& \lambda_0 \int_{-\frac{\pi}{a}}^{\frac{3\pi}{a}} \prod_{i=1}^4 d^4 p_i v_i(p_1,p_2,p_3,p_4) \delta^{(4)}\left( \sum_{i=1}^4 \Delta_G^{(z)}(p_i)\right)\tilde{\varphi}(p_1)\tilde{\varphi}(p_2)\tilde{\varphi}(p_3)\tilde{\varphi}(p_4), \label{Phi4l} \eea
where the functions $v_k$, $v_m$ and $v_i$ define respectively the integration volumes in momentum space for the kinetic, mass and interaction term. These functions are not determined by our continuum-lattice correspondence, and we shall fix them here   
by requiring that the lattice action (\ref{Phi4l}) is derived from the one in the continuum (\ref{phi4c}) by the blocking transformation  (\ref{cl}) already used in the Wess-Zumino model. 
An alternative  choice for the integration volumes, which is not associated to any blocking transformation and that leads to a different regularization scheme will be considered later in the section.

In our case the blocking transformation  (\ref{cl})  reduces simply to:
\beq
\tilde{\Phi}(\Delta_G^{(z)}(p_{\mu}))=\frac{2\tilde{\varphi}(p)}{\prod_\mu\sqrt{\frac{d\Delta_G^{(z)}(p_\mu)}{dp_\mu}}}
.\label{btl} \eeq
By replacing (\ref{btl}) into the continuum action (\ref{phi4c}) we obtain\footnote{Notice that the blocking transformation is invertible only for $z=1$, so for $z<1$ continuum fields with momenta higher than $|\Delta_G^{(z)}(\frac{\pi}{a})|$ have no correspondence on the lattice.} the lattice action (\ref{Phi4l}) with the following integration volumes:
\beq
v_k(p_1,p_2)=v_m(p_1,p_2)= 2^{-8} \left[ \prod_\mu \left| \frac{d\Delta_G^{(z)}(p_{1\mu})}{dp_{1\mu}}\frac{d\Delta_G^{(z)}(p_{2\mu})}{dp_{2\mu}}\right| \right]^{1/2}, \label{vol1/2} \eeq
and
\beq
v_i(p_1,p_2,p_3,p_4) = 2^{-16} \left[ \prod_\mu \left| \frac{d\Delta_G^{(z)}(p_{1\mu})}{dp_{1\mu}} \frac{d\Delta_G^{(z)}(p_{2\mu})}{dp_{2\mu}}\frac{d\Delta_G^{(z)}(p_{3\mu})}{dp_{3\mu}} \frac{d\Delta_G^{(z)}(p_{4\mu})}{dp_{4\mu}}\right| \right]^{1/2}. \label{volint} \eeq
The one used here is a very specific form of blocking transformation:
in Sec.\ref{1.4} a more general form of blocking transformation was introduced  (see eq.(\ref{Phivarphidd}))  that contains an arbitrary function $F(p)$. The function $F(p)$ is the multi-dimensional analogue of the arbitrary  function $f(p)$ that appears in the definition of the associative star product (\ref{Phivarphi}) in the one dimensional case. 
Different choices of $F(p)$ correspond to different rescalings  of the field $\tilde{\varphi}(p)$ in momentum space and they are essentially equivalent.  The choice of the square root in (\ref{btl}) however has a double advantage: it corresponds to the simplest form of the kinetic term (see eq. (\ref{4dWSmodellattps})),  and to a definition of the associative star product that has a smooth limit $a \rightarrow 0 $ in coordinate representation as discussed in section \ref{1.3}. 

By replacing (\ref{vol1/2}) and (\ref{volint}) into (\ref{Phi4l}) we obtain the lattice action:
\bea
&S^{(z)} = \int_{-\frac{\pi}{a}}^{\frac{\pi}{a}} d^4p_1 d^4p_2 \prod_{\mu=1}^4 \delta(p_{1\mu}+p_{2\mu}) \left[-\Delta^{(z)}_G(p_{1\mu})\Delta^{(z)}_G(p_{2\mu}) + m_0^2 \right] \tilde{\varphi}(p_1) \tilde{\varphi}(p_2) + \nonumber \\& + \lambda_0  \int_{-\frac{\pi}{a}}^{\frac{\pi}{a}}  d^4p_i \delta^{(4)}\left( \sum_{i=1}^4 \Delta^{(z)}_G(p_{i}) \right) \sqrt{ \prod_{\mu,i=1}^4  \frac{d\Delta^{(z)}_G(p_{i\mu})}{dp_{i\mu}} }  \tilde{\varphi}(p_1) \tilde{\varphi}(p_2) \tilde{\varphi}(p_3) \tilde{\varphi}(p_4),  \label{Sz} \eea
where it should be noticed that $\frac{d\Delta^{(z)}_G(p_{\mu})}{dp_{\mu}}$ can be written explicitely and is given by 
\beq
\frac{d\Delta^{(z)}_G(p_{\mu})}{dp_{\mu}} = \frac{\cos\frac{ap_\mu}{2}}{1-z^2 \sin^2\frac{ap_\mu}{2}} \stackrel{z\rightarrow 1}{\longrightarrow} \frac{1}{\cos\frac{ap_\mu}{2}}. \label{gdder} \eeq
The Feynman diagrams generated by the perturbative expansion of (\ref{Sz}) can easily be constructed, and for $z<1$ they are all finite. On the other hand we can see from (\ref{gdder}) that in the limit $z \rightarrow 1$ a divergence arises at the extremes of integration, i.e. at $|p_\mu| = \frac{\pi}{a}$, due to the vanishing of the cosine at the denominator.
This is better understood by going  back to the continuum notation, namely by using $\hat{p}_{i\mu} = \Delta^{(z)}_G(p_{i\mu})$ as independent momenta and $\Phi(\hat{p}_i)$, defined in (\ref{btl}), as fundamental fields.
The action $S^{(z)}$ then reads:
\bea
S^{(z)} &=& \int_{-\hat{p}^{(\mathrm{cutoff})}}^{\hat{p}^{(\mathrm{cutoff})}} d^4\hat{p}_1 d^4\hat{p}_2~ \delta^{(4)}(\hat{p}_1+\hat{p}_2) \left[-\hat{p}_1^\mu \tilde{\Phi}(\hat{p}_1)\hat{p}_{2\mu}\tilde{\Phi}(\hat{p}_2) + m_0^2 \tilde{\Phi}(\hat{p}_1)\tilde{\Phi}(\hat{p}_2)\right] \nonumber \\ &+&\lambda_0 \int_{-\hat{p}^{(\mathrm{cutoff})}}^{\hat{p}^{(\mathrm{cutoff})}} \prod_{i=1}^4 d^4\hat{p}_i~\delta^{(4)}(\hat{p}_1+\hat{p}_2+\hat{p}_3+\hat{p}_4)~  \tilde{\Phi}(\hat{p}_1)\tilde{\Phi}(\hat{p}_2) \tilde{\Phi}(\hat{p}_3)\tilde{\Phi}(\hat{p}_4), \label{phi4ctoff} \eea
where $\hat{p}^{(\mathrm{cutoff})}$ is given in (\ref{cutoff}) and is a function  of the lattice spacing $a$ and of the parameter $z$.  

The action (\ref{phi4ctoff}) is the original action (\ref{phi4c}) of the continuum theory, regularized by the introduction of a cutoff $\hat{p}^{(\mathrm{cutoff})}$ on each momentum component and the Feynman diagrams of the regularized continuum theory (\ref{phi4ctoff}) and of the lattice theory (\ref{Sz}) coincide, modulo the momentum dependent rescaling of the external lines determined by the blocking transformation (\ref{btl}).
That is, correlation functions calculated from  (\ref{Sz}) and  (\ref{phi4ctoff}) are related by:
\beq
\langle \tilde{\varphi}(p_1)\tilde{\varphi}(p_2)\dots\tilde{\varphi}(p_n) \rangle = 2^{-n} \prod_{j=1}^n\prod_\mu \sqrt{\frac{d\Delta_G^{(z)}(p_{j\mu})}{dp_{j\mu}}} \langle\tilde{\Phi}(\hat{p}_1)\tilde{\Phi}(\hat{p}_2)\dots\tilde{\Phi}(\hat{p}_n) \rangle, \label{crfunc} \eeq
with $\hat{p}_\mu = \Delta_G^{(z)}(p_{\mu})$.

However in the continuum theory the external momenta $\hat{p}_\mu$ are not restricted and the momentum cutoff on the intermediate states violates unitarity, whereas in the lattice action  (\ref{Sz}) external and intermediate states all have $p_\mu$ in the fundamental region and unitarity is not violated\footnote{This is of course true also for standard lattice theories, which can be regarded as a sofisticated way of introducing a momentum cutoff without violating unitarity but breaking space-time symmetries}.

On the other hand the action (\ref{phi4ctoff}) depends on $z$ and on the lattice spacing $a$ only through the cutoff $\hat{p}^{(\mathrm{cutoff})}$, so that all pairs $(a,z)$ that correspond to the same value of  $\hat{p}^{(\mathrm{cutoff})}$ describe the same physical system.
More precisely, if  $(a,z)$ and $(a',z')$ are two pairs corresponding to the same value of $\hat{p}^{(\mathrm{cutoff})}$ and we denote the  fields of the corresponding actions by $\tilde{\varphi}(p)$ and $\tilde{\varphi}'(p')$ then we have from (\ref{crfunc}):
\beq
\frac{\langle \tilde{\varphi}(p_1)\tilde{\varphi}(p_2)\dots\tilde{\varphi}(p_n) \rangle }{\prod_{j=1}^n\prod_\mu \sqrt{\frac{d\Delta_G^{(z)}(p_{j\mu})}{dp_{j\mu}}}}=\frac{\langle \tilde{\varphi}'(p'_1)\tilde{\varphi}'(p'_2)\dots\tilde{\varphi}'(p'_n) \rangle }{\prod_{j=1}^n\prod_\mu \sqrt{\frac{d\Delta_G^{'(z')}(p'_{j\mu})}{dp'_{j\mu}}}} \label{crff}
, \eeq
where the primed quantities contain $a'$ in place of $a$ and the relation between $p_\mu$ and $p'_\mu$ is given by:
\beq
\Delta_G^{(z)}(p_{\mu}) = \Delta_G^{'(z')}(p'_{\mu}). \label{ppprime} \eeq
Eq. (\ref{crff}) provides a non trivial relation between correlation functions of two lattice theories defined by (\ref{Sz}) with the same momentum cutoff $\hat{p}^{(\mathrm{cutoff})}$ but with different values of the lattice spacing $a$ and of the parameter $z$.

In the limit where the cutoff is sent to infinity the ultraviolet divergences appear and the theory needs to be renormalized by absorbing the divergences into the bare mass and coupling constant. The $\Phi^4$ theory in four dimensions is a textbook example of  renormalizable theory, and the introduction of a cutoff on the momenta in the loop integrations is a standard regularization procedure.   The exact correspondence between the Feynman diagrams of the lattice theory (\ref{Sz}) and of the regularized continuum theory guarantees that also in the lattice theory the divergences arising in the  $\hat{p}^{(\mathrm{cutoff})} \rightarrow \infty$ limit can be removed by exactly the same renormalization procedure used in the continuum theory.

In conventional lattice theories  the momentum cutoff is directly related to the inverse of the lattice constant, and the limit where the cutoff is sent to infinity coincides with the limit where the lattice constant $a$ is sent to zero, namely with the continuum limit.

The novel feature of our approach is that this limit can be reached in different ways: in fact $\hat{p}^{(\mathrm{cutoff})} \rightarrow \infty$ can be obtained either by letting $a \rightarrow 0$ or by letting $z \rightarrow 1$. The first possibility corresponds to the standard continuum limit of a lattice theory; in fact if we set for instance $z=0$ the derivative on the lattice is the local symmetric finite difference operator, the cutoff is simply $\hat{p}^{(\mathrm{cutoff})}=\frac{2}{a}$ and the continuum theory is reached as usual by letting the lattice spacing go to zero.
The second possibility is more interesting: we keep the lattice spacing fixed at an arbitrary value and let the parameter $z$ go to one.
The lattice structure of the theory is then preserved in the limit but the derivative on the lattice  becomes non-local and at the limiting value $z=1$  it is the one defined in eq.(\ref{deltacoord}) in terms of the inverse Gudermannian function.
Correlation functions of the renormalized continuum theory and correlation functions of the renormalized lattice theory are then related in the $z \rightarrow 1$ limit at fixed $a$ by eq. (\ref{crfunc}) which at $z=1$ reads:
\beq
\langle \tilde{\varphi}(p_1)\tilde{\varphi}(p_2)\dots\tilde{\varphi}(p_n) \rangle_R = 2^{-n} \prod_{j=1}^n\prod_\mu \sqrt{\cosh \hat{p}_{j\mu}} \langle\tilde{\Phi}(\hat{p}_1)\tilde{\Phi}(\hat{p}_2)\dots\tilde{\Phi}(\hat{p}_n) \rangle_R, \label{crfuncR} \eeq
with $\hat{p}_\mu = \frac{2}{a} \gd\left(\frac{ap}{2}\right)$.
 
\subsection{$\Phi^4$  theory in four dimensions: a more general lattice action}

In this subsection we shall consider for the integration volumes in the action  (\ref{Phi4l}) a choice which  is  more general  than the one  obtained by simply replacing (\ref{vol1/2}) and (\ref{volint}) into (\ref{Phi4l}). 
We shall keep for the integration volume $v_k(p_1,p_2)$ of the kinetic term the expression given in (\ref{vol1/2}), so that the kinetic term has its simplest form, as in (\ref{Sz}). This amounts to fixing the arbitrary rescaling of the lattice fields in momentum representation.
Instead we shall introduce in the definition of  $v_m(p_1,p_2)$ and $v_i(p_1,p_2,p_3,p_4)$ a new parameter $\alpha$, by replacing  the square root  in eq.(\ref{vol1/2}) and (\ref{volint}) with an arbitrary power $\alpha$ :
\bea
v_m(p_1,p_2)&=& 2^{-8} \left[ \prod_\mu \left| \frac{d\Delta_G^{(z)}(p_{1\mu})}{dp_{1\mu}}\frac{d\Delta_G^{(z)}(p_{2\mu})}{dp_{2\mu}}\right| \right]^{\alpha}  \nonumber \\
v_i(p_1,p_2,p_3,p_4) &=& 2^{-16} \left[ \prod_\mu \left| \frac{d\Delta_G^{(z)}(p_{1\mu})}{dp_{1\mu}} \frac{d\Delta_G^{(z)}(p_{2\mu})}{dp_{2\mu}}\frac{d\Delta_G^{(z)}(p_{3\mu})}{dp_{3\mu}} \frac{d\Delta_G^{(z)}(p_{4\mu})}{dp_{4\mu}}\right| \right]^{\alpha}. \label{volalpha} \eea 

Notice that the power $\alpha$ is the same for $v_m(p_1,p_2)$ and $v_i(p_1,p_2,p_3,p_4)$: as we shall see that this is needed for a consistent  renormalizability if $\alpha$ is used as a regulator. 
 
The action obtained by inserting  the integration volumes  (\ref{volalpha}) into  (\ref{Phi4l})  does not follow from the continuum theory  via a  blocking transformation;  so the correspondence, and in the $z=1$ limit the equivalence, between lattice and continuum theory that we discussed in the previous subsection is lost. 

The continuum theory is reached in the limit $z \rightarrow 1$ \textit{and} $\alpha \rightarrow 1/2$. In the last subsection we had $\alpha$ set to $1/2$ and we let $z$ act as a regulator for the ultraviolet divergences. In the remaining part of this subsection we shall instead set $z=1$ and use the parameter $\alpha$ to regularize the ultraviolet divergences of the continuum theory.
The corresponding action $S^{(\alpha)}$ is obtained from (\ref{Phi4l}) by using (\ref{volalpha}) and, for the kinetic term only, (\ref{vol1/2}) and by keeping (\ref{gdder}) into account for the $z\rightarrow1$ limit.
The result is:
\bea
&S^{\alpha}=\int_{-\frac{\pi}{a}}^{\frac{\pi}{a}} d^4p_1 d^4p_2 \prod_{\mu=1}^4 \delta(p_{1\mu}+p_{2\mu}) \left[-\Delta_G(p_{1\mu})\Delta_G(p_{2\mu})+ m_0^2 \left(\prod_\mu \cos \frac{ap_{1\mu}}{2}\right)^{1-2\alpha} \right] \tilde{\varphi}(p_1) \tilde{\varphi}(p_2) \nonumber \\&+\lambda_0  \int_{-\frac{\pi}{a}}^{\frac{\pi}{a}} \prod_{i=1}^4 d^4p_i \prod_{\mu=1}^4\delta\left( \sum_{i=1}^4 \Delta_G(p_{i\mu}) \right) \left( \prod_{i,\mu=1}^4  \cos\frac{ap_{i\mu}}{2} \right)^{-\alpha}  \tilde{\varphi}(p_1) \tilde{\varphi}(p_2) \tilde{\varphi}(p_3) \tilde{\varphi}(p_4),   \label{slatttot} \eea
where in each term the integration over the momentum components $p_{i\mu}$ has been reduced to the interval $(-\frac{\pi}{a}, \frac{\pi}{a})$ by using the symmetry of the action under $p_{i\mu} \rightarrow \frac{2\pi}{a} - p_{i\mu}$. This produces a factor $2^d$ for each momentum integration, thus canceling the powers of $2$ introduced in (\ref{vol1/2}) and (\ref{volalpha}).

Notice also that $\cos\frac{ap_{\mu}}{2}$ is positive in the interval $(-\frac{\pi}{a}, \frac{\pi}{a})$ so that the absolute values in  (\ref{vol1/2}) and (\ref{volalpha}) can now be dropped.

 It is  convenient also in this case to go back to the continuum representation, namely to use $\hat{p}_{i\mu} = \Delta_G(p_{i\mu})$ as independent momenta and $\Phi(\hat{p}_i)$, defined in (\ref{btl}), as fundamental fields.
With this change of variables we find:
\bea
S^{(\alpha)} &=& \int_{-\infty}^{\infty} d^4\hat{p}_1 d^4\hat{p}_2~ \delta^{(4)}(\hat{p}_1+\hat{p}_2) \left[-\hat{p}_1^\mu \hat{p}_{2\mu} +\frac{m_0^2}{\left(\prod_{\mu} \cosh\frac{a\hat{p}_{1\mu}}{2}\right)^{1-2\alpha}} \right]\tilde{\Phi}(\hat{p}_1)\tilde{\Phi}(\hat{p}_2) \nonumber \\ &+&\lambda_0 \int_{-\infty}^{\infty} \prod_{i=1}^4 d^4\hat{p}_i~\delta^{(4)}(\hat{p}_1+\hat{p}_2+\hat{p}_3+\hat{p}_4)~ \frac{ \tilde{\Phi}(\hat{p}_1)\tilde{\Phi}(\hat{p}_2) \tilde{\Phi}(\hat{p}_3)\tilde{\Phi}(\hat{p}_4)}{\left(\prod_{\mu}\prod_{i=1}^4 \cosh\frac{a\hat{p}_{i\mu}}{2}\right)^{1/2-\alpha}}. \label{phi4alpha} \eea
The kinetic term in (\ref{phi4alpha}) is the same as in the standard continuum theory, but the mass and the interaction terms are modified by the presence of the hyperbolic cosine factors which  provide a smooth cutoff in the momenta: in fact for $\alpha<1/2$ each hyperbolic cosine denominator becomes very large for
\beq
\hat{p}_{i\mu} \gg \frac{1}{a(1/2-\alpha)}. \label{othercutoff} \eeq
The quantity on the r.h.s. is an effective cutoff in the momenta; this cutoff becomes very large either in the standard continuum limit $a \rightarrow 0$ or in the limit where $\alpha \rightarrow 1/2$ keeping $a$ fixed.
Because of this cutoff all Feynman diagrams in the perturbative expansion of either (\ref{slatttot}) or (\ref{phi4alpha}) are finite, and the corresponding correlation functions are proportional in analogy to (\ref{crfuncR}): 
\beq
\langle \tilde{\varphi}(p_1)\tilde{\varphi}(p_2)\dots\tilde{\varphi}(p_n) \rangle_\alpha = 2^{-n} \prod_{j=1}^n\prod_\mu \sqrt{\cosh \hat{p}_{j\mu}} \langle\tilde{\Phi}(\hat{p}_1)\tilde{\Phi}(\hat{p}_2)\dots\tilde{\Phi}(\hat{p}_n) \rangle_\alpha. \label{crfuncalpha} \eeq
Ultraviolet divergences appear as singularities in $\alpha$ at $\alpha=1/2$ in the Feynman diagrams which are ultraviolet divergent in the continuum limit. 
As in the case of the standard momentum cutoff regularization we expect that these singularities can be absorbed by a redefinition of the mass and of the coupling constant. We are going to check this explicitely at the one loop level.

We shall work in the continuum representation of action (\ref{phi4alpha}) which is more directly related to the continuum theory at $\alpha=1/2$.

\begin{figure}
	\begin{center}
		\includegraphics[scale=0.9]{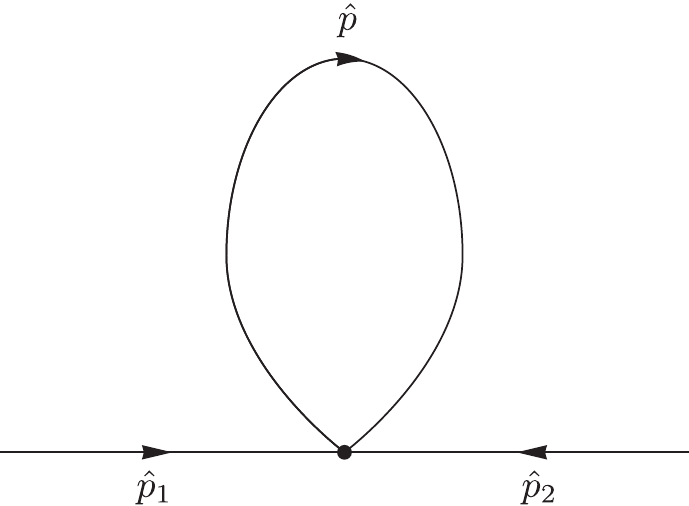}
	\end{center}
	\caption{One-loop correction to the propagator.}
	\label{fd2pt}
\end{figure}

\begin{figure}
	\begin{center}
		\includegraphics[scale=0.9]{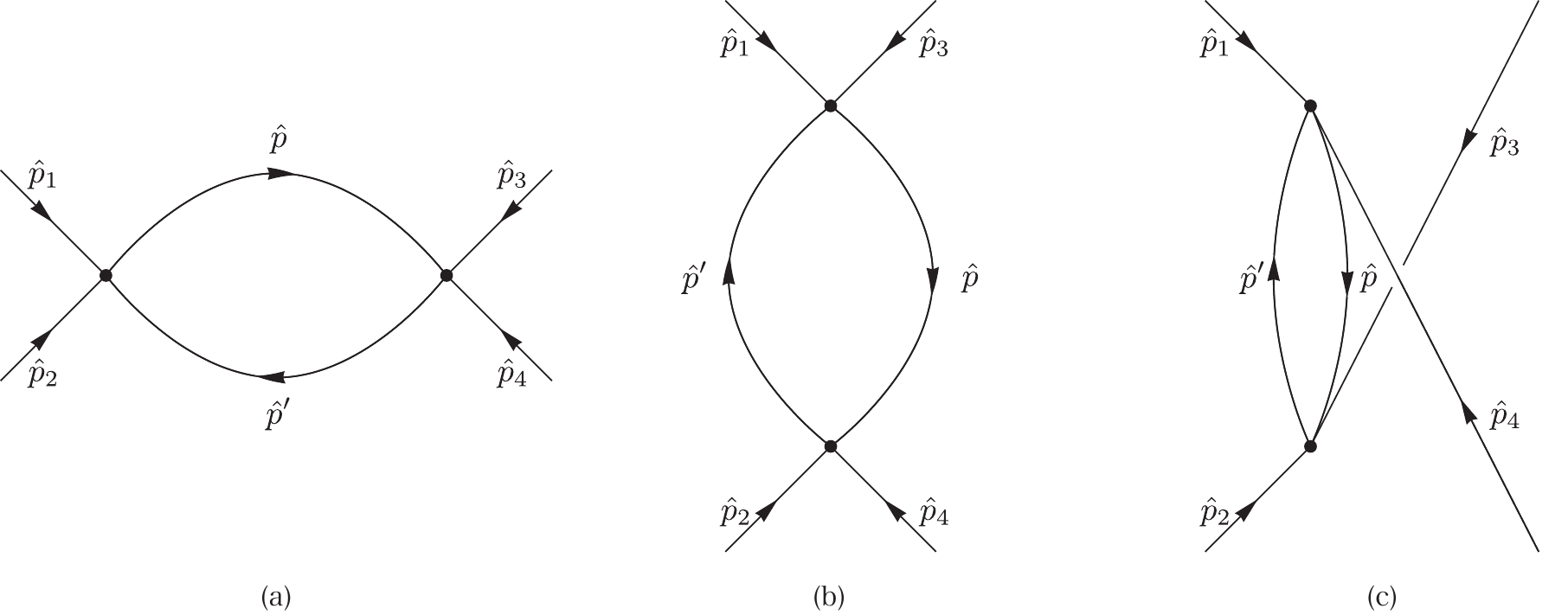}
	\end{center}
	\caption{One-loop corrections to the four point vertex.}
	\label{fd4pts}
\end{figure}

At one loop level the ultraviolet divergent diagrams are the one loop mass renormalization diagram, shown in fig.\ref{fd2pt}, and the one-loop coupling renormalization diagrams illustrated  in fig.\ref{fd4pts}.

The building blocks of the Feynmann diagrams are the full propagator $D^{(\alpha)}(p_1,p_2)$, namely the propagator including the contributions of the mass term insertions, and the four point  vertex  $V_4(p_1,p_2,p_3,p_4)$.
They are 
given respectively  by:
\beq
D^{(\alpha)}(\hat{p}_1,\hat{p}_2) = \frac{\prod_\mu\delta\left(\hat{p}_{1\mu}+\hat{p}_{2\mu}\right)}{\sum_\mu \hat{p}_1^\mu \hat{p}_{1\mu} + \left(\prod_{\mu} \cosh\frac{a\hat{p}_{1\mu}}{2}\right)^{2\alpha-1} m_0^2},  \label{fullprop} \eeq
and
\beq
V_4(\hat{p}_1,\hat{p}_2,\hat{p}_3,\hat{p}_4) = \lambda_0~ \delta^{(4)}\left(\hat{p}_1+\hat{p}_2+\hat{p}_3+\hat{p}_4\right) \left(\prod_{\mu}\prod_{i=1}^4 \cosh\frac{a\hat{p}_{i\mu}}{2}\right)^{\alpha-1/2}. \label{fint} \eeq

Let us consider first  the mass renormalization diagram, which we shall denote as $V_2^{(1\textrm{loop})}(\hat{p}_1,\hat{p}_2)$.

 Using (\ref{fullprop}) and (\ref{fint}) we have:
\beq
V_2^{(1\textrm{loop})}(\hat{p}_1,\hat{p}_2)= \lambda_0 ~\delta^{(4)}\left(\hat{p}_1+\hat{p}_2\right)\left(\prod_{\mu} \cosh\frac{a\hat{p}_{1\mu}}{2}\right)^{2\alpha-1} I_\alpha,   \label{1loopmass} \eeq
where

\beq
 I_\alpha = \int_{-\infty}^{\infty} d^4\hat{p}  \frac{ \left(\prod_{\mu} \cosh\frac{a\hat{p}_{\mu}}{2}\right)^{2\alpha-1}}{\sum_\mu \hat{p}_\mu\hat{p}_\mu + m_0^2 \left(\prod_{\mu} \cosh\frac{a\hat{p}_{\mu}}{2}\right)^{2\alpha-1}}. \label{Ialpha} \eeq

For $\alpha=1/2$ the integral $I_\alpha$ is divergent and coincides with the corresponding loop integral of the continuum theory, so the introduction of the parameter $\alpha$ may be regarded as a way to regularize the ultraviolet divergences of the original continuum theory.

The singularity structure of $I_\alpha$ at $\alpha=1/2$  is what one would expect from a momentum cutoff regularization with a cutoff $\Lambda = \frac{1}{a(1/2-\alpha)}$. In fact a direct (although non trivial) calculation gives:
\beq
I_\alpha = b_2 \left(\frac{1}{a(1/2-\alpha)}\right)^2 + b_0 ~ m_0^2 \log(1/2-\alpha) + \textrm{regular terms},  \label{singIa} \eeq 
where $b_2$ and $b_0$ are numbers that can be calculated.

The crucial point to observe at this stage  is that  the mass renormalization diagram $V_2^{(1\textrm{loop})}(\hat{p}_1,\hat{p}_2)$ given in (\ref{1loopmass}) and the diagram corresponding to a single mass insertion, which coincides with the  mass term in (\ref{phi4alpha}), have the same form. 
This is not a trivial result, and in fact it requires that the power $\alpha$ is the same in the mass and interaction term as we assumed from the beginning.

As a result, if one includes the one loop corrections,  the two point function is still given by (\ref{fullprop}) but with  $m_0^2$ simply  replaced by $m_0^2+\lambda_0 I_\alpha$:
\beq
D_{1\textrm{loop}}^{(\alpha)}(\hat{p}_1,\hat{p}_2) = \frac{\prod_\mu\delta\left(\hat{p}_{1\mu}+\hat{p}_{2\mu}\right)}{\sum_\mu\hat{p}_1^\mu \hat{p}_{1\mu} + \left(\prod_{\mu} \cosh\frac{a\hat{p}_{1\mu}}{2}\right)^{2\alpha-1} \left(m_0^2+\lambda_0 I_\alpha\right)}.  \label{fullprop1loop} \eeq

Renormalization of the mass at one loop now follows by writing the bare mass as the sum of a renormalized mass $m_R^2$ plus a counterterm $\delta m_0^2$:
\beq
m_0^2 = m_R^2 + \delta m_0^2,  \label{counterterm} \eeq
where the $\alpha$ dependence of the counterterm can be tuned to cancel the singularities of $\lambda_0 I_\alpha$.

Let us consider next the first of the diagrams of fig.\ref{fd4pts} which provide the one loop coupling constant renormalization. It will suffice to consider the first one, as the others are obtained just by crossing symmetry.
This is given by:

\beq
V_4^{\textrm{1loop}}(\hat{p}_1,\hat{p}_2;\hat{p}_3,\hat{p}_4) = \lambda_0^2~ \delta^{(4)}\left(\hat{p}_1+\hat{p}_2+\hat{p}_3+\hat{p}_4\right) \left(\prod_{\mu}\prod_{i=1}^4 \cosh\frac{a\hat{p}_{i\mu}}{2}\right)^{\alpha-1/2}  I_\alpha(\hat{p}_1+\hat{p}_2), \label{crenor12} \eeq
where
\beq
 I_\alpha(\hat{p}_1+\hat{p}_2) =\int_{-\infty}^{\infty} d^4p~ d^4p'~\prod_\mu\left(\cosh\frac{a\hat{p}_{i\mu}}{2}\right)^{2\alpha-1}   \frac{\delta^{(4)}\left(\hat{p}_1+\hat{p}_2+\hat{p}+\hat{p}'\right)}{\Sigma(p)\Sigma(p')}, \label{I12} \eeq
with
\beq
\Sigma(p) = \sum_\mu \hat{p}_\mu\hat{p}_\mu + \left(\prod_\mu\cosh\frac{a\hat{p}_{\mu}}{2}\right)^{2\alpha-1} m_0^2. \label{Sigmap} \eeq

Again for $\alpha=1/2$ the integral $ I_\alpha(\hat{p}_1+\hat{p}_2)$ is the same as the corresponding one of the continuum theory and is logarithmically divergent, but it is convergent for $\alpha<1/2$. Its singularity at $\alpha=1/2$ is, as expected, a logarithmic one:
\beq
 I_\alpha(\hat{p}) = c_0 \log(1/2-\alpha) + \textrm{regular terms},  \label{singIap} \eeq
where $c_0$ does not depend on the momentum $\hat{p}_\mu$. In fact it can be easily checked that the partial derivatives of $I_\alpha(p)$ with respect to $p_\mu$ produce integrals that are convergent at $\alpha=1/2$, so that the singular part is $p_\mu$ independent. 

By putting (\ref{fint}) and  (\ref{crenor12}) together one finds that the one loop corrections to the coupling constant amount to the following replacement:
\beq
\lambda_0 \longrightarrow \lambda_0 + \lambda_0^2 \left( I_\alpha(\hat{p}_1+\hat{p}_2)+ I_\alpha(\hat{p}_1+\hat{p}_3)+ I_\alpha(\hat{p}_3+\hat{p}_2)\right). \label{lambdacorr} \eeq

The integrals on the r.h.s. depend on the outside momenta as well as on  $\alpha$. However   we have seen that the part of $I_\alpha(p)$ which is singular at $\alpha=1/2$ does not depend on $p_\mu$. 

The singularities of the one loop integrals at the r.h.s. of (\ref{lambdacorr}) can then be absorbed, at one loop level,  into a redefinition of the coupling constant $\lambda_0$, and a renormalized coupling constant $\lambda_R$ can be consistently defined.

Renormalization beyond one loop follows here the same pattern as in the standard momentum cutoff regularization. We shall not discuss the details here.

\section{ Conclusion and Discussions}
\label{1.7}
In this paper we propose a new and unconventional approach to lattice theories. Although originally motivated and developed in quest of exact supersymmetry on the lattice, it is quite general  and it solves some long standing problems of conventional lattice theories, such as the chiral fermion problem and the associated doubling problem, and the problem of the violation of the Leibniz rule for the derivative operator on the lattice. 
These were indeed the main obstacles on the path of having exact supersymmetry on the lattice and  in fact exact lattice supersymmetry is naturally implemented in this approach. 
In the case of some extended supersymmetries different members of the same supermultiplet  may be identified with doublers on the lattice and supersymmetry has a simpler and  more economical formulation on the lattice than in the continuum.

All these successes however come at a price. Locality is lost on the lattice: the star product defined in sec.\ref{1.3} is non local and in general it is not associative, although it becomes local in the continuum limit. That induces a violation of gauge invariance, except in the very special case in which associativity is satisfied. This particular case is most interesting also for another reason: the degrees of freedom of the theory on the lattice are in one-to-one correspondence with the ones in the continuum, so that the blocking transformation from the continuum to the lattice is completely invertible and does not involve any loss of information. In other words the lattice theory is in this case just a reshuffling of the degrees of freedom of the continuum theory. It is non local on the lattice, but its non locality is only a lattice artifact as the theory is completely equivalent to the local continuum theory. For the same reason one expects causality to be exactly preserved.

As a consequence in that case  all symmetries, including gauge invariance, are exactly preserved in the lattice formulation. However, for the same reason, the lattice spacing $a$ does not act as a regulator and the theory on the lattice has the same ultraviolet divergences as the continuum one.

In the case of non-gauge theories, like the ones discussed in sec.\ref{section6}, a regulator can be naturally introduced by replacing for instance the lattice derivative  $\Delta_G(p)$ with its 
regularized version $\Delta_G^{(z)}(p)$ defined in eq.(\ref{regD}). The parameter $z$ can be used as a regulator while keeping $a$ fixed, namely while preserving the lattice structure.  The renormalization procedure can then be carried on in analogy with the continuum and a renormalized lattice theory can be defined.

The continuum-lattice duality introduced at the classical level by the reversible blocking transformation can in this case be extended to the quantum level, and the lattice actions obtained in this way may be regarded as 
{\it perfect actions} \cite{Hasenfratz, Bietenholz}, with no lattice artifact in spite of finiteness of $a$. 

The $z$ regulator however breaks the associativity of the star product, and hence in gauge theories it violates gauge invariance. This means that a lattice theory formally exactly equivalent to a gauge theory in the continuum can be defined but cannot be regularized, at least at the present stage of our knowledge, in a gauge invariant way while keeping the lattice structure. This poses severe limitations to practical applications of the present formalism to gauge theories.

Finite lattices, which you  would want for  lattice computer simulations, also present a problem. Consider lattice fields that satisfy a periodic boundary condition in coordinate space, say in one dimension:
\beq
\varphi\left(\frac{na}{2}\right) = \varphi\left(\frac{na}{2} + Na\right)  \label{perboudcond}
. \eeq
 This is compatible with (\ref{chiralcondx}) provided $N$ is even. In momentum representation eq.(\ref{perboudcond}) implies that momentum components  on the lattice are quantized and are integer multiples of $\frac{2\pi}{Na}$, which in turn means that the momentum conservation can never be satisfied since the conserved momentum $\hat{p} = \Delta(p)$ is not a rational function of the lattice momentum $p$. 
 A way out to this problem may involve tricks like replacing the delta function of momentum conservation with a gaussian whose width goes like $\frac{1}{N}$ in such a way to always allow an overlap between the continuum  and the  lattice momentum conservation. This may be useful for simulations, but at the price of introducing a breaking of order $\frac{1}{N}$ in some relevant symmetries, for instance in all symmetries that, like supersymmetry, rely on the exactness of the Leibniz rule.
 
In spite of these limitations there are some possible interesting developments which we have not fully investigated yet. We shall briefly discuss here what seems to be a promising one.

The lattice action was obtained from the continuum action in (\ref{effect1}) by implementing the blocking transformation (\ref{Phivarphiddf}) by means of a functional delta function. This is only a limiting case of a more general procedure where the delta function is replaced by a gaussian whose width may be related to the lattice spacing and sent eventually to zero. The effect of this gaussian smearing, particularly on the symmetries of the lattice fields, has been extensively studied \cite{Bergner-Bruckmann}.

To illustrate this point consider for instance a blocking transformation from the continuum to the lattice where the lattice fields are defined as the avarage of the continuum fields over an hypercube of size $a$ . Chiral invariance is well defined on the lattice, and  the lattice action of a free fermion obtained by implementing the blocking transformation with a functional delta function is chiral invariant. However the fermion propagator becomes non-local decaying like $|r|^{1-d}$ at large distances\cite{Bietenholz:1995kr,Bietenholz:1997sf}, thus avoiding contradiction with the Nielsen-Ninomiya theorem. The gaussian term breaks chirality, but in a controlled way, and the remnant of the original symmetry is expressed by the Ginsparg-Wilson relation. The breaking of the chiral symmetry allows the fermion propagator to become local, with an exponential decay in the distance $|r|$.

The new feature of our approach is that the blocking transformation introduced in the previous sections, being invertible,  does not destroy any of the symmetries present in the original continuum theory and at the classical level such symmetries, including gauge symmetries, are also symmetries of the lattice theory resulting from the blocking transformation with a delta function kernel.
The gaussian kernel will in general break the symmetries, but in the controlled way described in the general formula of ref.\cite{Bergner-Bruckmann}, thus providing a kind of generalization of the Ginsparg-Wilson relation to other symmetries, including perhaps gauge symmetries. Locality of the lattice theory may well be also restored\footnote{Some preliminary calculations show that in spite of the non locality of the derivative operator $\Delta_G(p)$, at least the free fermion propagator becomes local with the gaussian smearing.}, while it is not clear if the width of the gaussian can be used a regulator for the ultraviolet divergences.


\label{section7}
\setcounter{equation}{0}

\vspace{1cm}

{\bf{\Large Acknowledgments}}
We thank P. H. Damgaard, I. Kanamori, H. B. Nielsen and Hiroshi Suzuki,  
for useful discussions over long years of our investigations and M. Panero for carefully reading the manuscript. 
We also wish to thank JSPS and INFN for supporting our long standing collaboration.

\begin{center}
{\bf{\Large Appendix}}
\end{center}

\appendix
\section{The large $N$ limit of $ J_\Delta(\xi,\eta)$}
\label{cont}

In this Appendix we give the proof of eq. (\ref{distribution3}), namely of the identity:
\beq
 \lim_{N \rightarrow \infty}\int_{-\infty}^{+\infty} d\xi ~  J^{(0)}_{\Delta_{s,G}}( \xi, \eta) \chi(\xi) = \chi(\eta),   \label{distribution4} \eeq
for any function $\chi(\xi)$ continuous and bounded on the real axis.
We shall give the details of the proof only for the case $\Delta=\Delta_G$; for $\Delta=\Delta_s$ the proof works along the same lines. 

We shall divide the integration domain in eq. (\ref{distribution4}) into four intervals corresponding to the domains in which the three eq.s (\ref{JDGosc}), (\ref{JDGexp}) and (\ref{Airy}) hold, plus the domain $\xi<0$. 

For $\xi<0$ the saddle points given by eq.(\ref{saddlepoints}) are outside the interval of integration $(-\pi/2,\pi/2)$, and their contribution to $J^{(0)}_{\Delta_G}( \xi, \eta)$ in (\ref{J0Delta2}) is vanishing in the large $N$ limit.

We shall now divide the region $\xi>0$ into three intervals, namely:
\begin{enumerate}[(a)]
\item
$\mathcal{D}_a \equiv  \left[0 , \eta-\left(\frac{\Lambda}{N}\right)^{2/3}\eta^{1/3}\right] $
\item
$ \mathcal{D}_b \equiv \left[\eta-\left(\frac{\Lambda}{N}\right)^{2/3}\eta^{1/3} , \eta+\left(\frac{\Lambda'}{N}\right)^{2/3}\eta^{1/3}\right] $
\item
$ \mathcal{D}_c \equiv \left[\eta+\left(\frac{\Lambda'}{N}\right)^{2/3} \eta^{1/3},\infty\right]$
\end{enumerate}
where both $\Lambda$ and $\Lambda'$ are chosen to satisfy the strong inequalities $1 \ll \Lambda (\Lambda') \ll N$.
Let us consider first the interval (b), where $|\xi -\eta| \ll 1$ and hence the large $N$ limit (\ref{Airy}) holds. 
Let us define: 
\beq
I_{(b)}[\chi] = \lim_{N \rightarrow \infty}\int_{\mathcal{D}_b} d\xi ~  J^{(0)}_{\Delta_{G}}( \xi, \eta) \chi(\xi). \label{If} \eeq
Replacing the limit with the expression given in (\ref{Airy}) we get:
\beq 
I_{(b)}[\chi] = 2^{1/3} N^{2/3}\eta^{-1/3} \int_{\eta-(\Lambda/N)^{2/3}\eta^{1/3}}^{\eta+(\Lambda'/N)^{2/3}\eta^{1/3}} d\xi~ \ai\left(2^{1/3}(\xi-\eta) N^{2/3}\eta^{-1/3} \right)\chi(\xi).  \label{distribution5} \eeq
With the substitution $\xi =\eta+\frac{x \eta^{1/3}}{2^{1/3}N^{2/3}}$ this becomes:
 \beq
I_{(b)}[\chi] = \int_{-2^{1/3}\Lambda^{2/3}}^{2^{1/3}\Lambda'^{2/3}} dx \ai(x) \chi\left( \eta+ \frac{x \eta^{1/3}}{2^{1/3}N^{2/3}} \right). \label{ibf} \eeq
We show now that given an arbitrarily small positive real number $\epsilon$ we can find $\Lambda$ and $N$ such that
\beq
I_{(b)}[\chi] - \chi(\eta) < \epsilon.      \label{ineq} \eeq
The Airy function $\ai(x)$ is normalized in such a way that its integral over the real axis is $1$, so we can write:
\beq
I_{(b)}[\chi] - \chi(\eta) = \int_{-2^{1/3}\Lambda^{2/3}}^{2^{1/3}\Lambda'^{2/3}} dx \ai(x) \left[
\chi\left( \eta  + \frac{x \eta^{1/3}}{2^{1/3}N^{2/3}} \right) - \chi(\eta)\right] - \chi(\eta)\int_{\mathcal{D}_{\Lambda}} dx \ai(x),   \label{aidiff} \eeq
with
\beq
\mathcal{D}_\Lambda \equiv  \left[ -\infty, -2^{1/3}\Lambda^{2/3}\right] \cup \left[ 2^{1/3}\Lambda'^{2/3}, \infty \right]. \label{reg} \eeq
We shall assume that the test function $\chi(\eta)$ is bounded and continuous on the real axis. 

Consider now the second term at the r.h.s. of (\ref{aidiff}). Given the boundedness of $\chi(\eta)$ and the convergence of the integral over the Airy function it is possible to find a $\Lambda$ and $\Lambda'$  large enough to have:
\beq
\left| \chi(\eta) \int_{\mathcal{D}_\Lambda} dx \ai(x) \right| < \frac{\epsilon}{2}. \label{bound2} \eeq
The continuity of $\chi(\eta)$ ensures that in the first term at the r.h.s. of (\ref{aidiff}), for any given $\Lambda$ and $\Lambda'$ we can choose $N$ large enough to have:
\beq
\left|
\chi\left( \eta  + \frac{x \eta^{1/3}}{2^{1/3}N^{2/3}} \right) - \chi(\eta)\right| < \frac{\epsilon}{2~ 2^{1/3}\left(\Lambda^{2/3}+\Lambda'^{2/3}\right)},~~~~~~~~~\forall x : -2^{1/3}\Lambda^{2/3} \leq x \leq 2^{1/3}\Lambda'^{2/3}.    \label{bound2} \eeq
In this way, taking into account that $|\ai(x)|<1$, we find that also the absolute value of the first term at the r.h.s. of (\ref{aidiff}) can be reduced to be smaller that $\frac{\epsilon}{2}$, end eq.(\ref{ineq}) is proved.

Let us consider now the interval $\mathcal{D}_a$ and define $I_{(a)}[\chi]$ as in eq.(\ref{If}).
 In  $\mathcal{D}_a$ $ J^{(0)}_{\Delta_{G}}( \xi, \eta)$ is represented in the large $N$ limit by eq.(\ref{JDGosc}), so by replacing   (\ref{JDGosc}) into  $I_{(a)}[\chi]$ we find an expression of the form:
\beq
I_{(a)}[\chi] =\lim_{N\rightarrow \infty} \sqrt{\frac{2N}{ \pi }}~\int_{\mathcal{D}_a} d\xi  \cos\left[N F(\xi,\eta) + \pi/4\right] \psi(\xi,\eta),  \label{inta} \eeq
where
\beq
F(\xi,\eta) = \frac{\xi}{2} \log\frac{1+\sqrt{1-(\xi/\eta)^2}}{1-\sqrt{1-(\xi/\eta)^2}}-\eta \arccos (\xi/\eta),  \label{Fxieta} \eeq
and
\beq
\psi(\xi,\eta) = \sqrt{\frac{\xi}{\eta}} \frac{\chi(\xi)}{\left(\eta^2-\xi^2\right)^{1/4}}. \label{psixieta} \eeq
In order to find the leading term of (\ref{inta}) in the large $N$ limit let us make a change of variable in the integral, defining:
\beq
t = N F(\xi,\eta) + \pi/4.  \label{variablet} \eeq
By partial integration we get:
\beq
I_{(a)}[\chi] =\lim_{N\rightarrow \infty} \sqrt{\frac{2N}{ \pi }} \left\{ \left. \frac{\psi(\xi,\eta)}{ N F'(\xi,\eta)} \sin(t)  \right|_{\xi=0}^{ \xi=\eta-\left(\frac{\Lambda}{N}\right)^{2/3}\eta^{1/3}} - \int_{\mathcal{D}_a} dt \sin(t) \frac{d}{dt} \left( \frac{\psi(\xi,\eta)}{N F'(\xi,\eta)}\right) \right\}, \label{inta2} \eeq
where $F'(\xi,\eta)$ is the partial derivative with respect to $\xi$.
By repeating the partial integration we find that each term has an extra $1/N$ factor, so that the first term in the curly bracket in eq.(\ref{inta2}) is the leading term in the large $N$ asymptotic expansion. 
So, by replacing (\ref{variablet}) and (\ref{psixieta}) back into (\ref{inta2}) we get:
\beq
I_{(a)}[\chi] = \sqrt{\frac{2\xi}{ \pi N\eta}} \left. \frac{\chi(\xi)}{\left(\eta^2-\xi^2\right)^{1/4} F'(\xi,\eta)} \sin( N F(\xi,\eta) + \pi/4 )  \right|_{\xi=0}^{ \xi=\eta-\left(\frac{\Lambda}{N}\right)^{2/3}\eta^{1/3}} 
\label{inta3}
. \eeq
The integration limit $\xi=0$ gives a vanishing contribution because of the $\sqrt{\xi}$ term and because $F'(\xi,\eta)$ becomes infinite as $\xi$ goes to zero.
At the other limit of integration $\eta - \xi \rightarrow 0$  as $N\rightarrow \infty$, so we can replace $F(\xi,\eta)$ and $F'(\xi,\eta)$ with the first term of their asymptotic expansion in $\eta-\xi$, namely:
\beq
F(\xi,\eta) = -\frac{2^{3/2}}{3} \eta^{-1/2} \left(\eta - \xi \right)^{3/2} + O\left(\left(\eta - \xi \right)^{5/2} \right), \label{asyF} \eeq
\beq
F'(\xi,\eta) = \sqrt{2} \eta^{-1/2} \left(\eta - \xi \right)^{1/2}+ O\left(\left(\eta - \xi \right)^{3/2} \right).\label{asyF2} \eeq
By inserting (\ref{asyF}) and (\ref{asyF2}) into (\ref{inta3}) we get:
\beq
I_{(a)}[\chi] = \frac{\chi(\eta)}{2^{1/4}\sqrt{\pi \Lambda}}\sin\left( -\frac{2^{3/2}}{3} \Lambda + \frac{\pi}{4} \right) + \mbox{terms vanishing as} ~ N \rightarrow \infty.  \label{asyI} \eeq 
Given the boundedness of $\chi(\eta)$ it is then always possible to find a $\Lambda$ large enough that 
\beq
\left|I_{(a)}[\chi] \right| < \epsilon,    \label{Iaeps} \eeq
for any given $\epsilon$, however small.
If two different values of $\Lambda$ are obtained to satisfy (\ref{bound2}) and (\ref{Iaeps}) the largest of the two should be taken, in order to satisfy both at the same time.

The large $N$ limit of $I_{(c)}[\chi]$ follows the same lines, with an exponential replacing the cosine. We have:
\beq
I_{(c)}[\chi] =\lim_{N\rightarrow \infty} \sqrt{\frac{2N}{ \pi }}~\int_{\mathcal{D}_c} d\xi  e^{-N F_c(\xi,\eta)}  \psi_c(\xi,\eta),  \label{intc} \eeq
where
\beq
F_c(\xi,\eta) = \xi \arccos\left(\frac{\eta}{\xi}\right) - \eta \cosh^{-1}\left(\frac{\xi}{\eta}\right), \label{Fcxieta} \eeq
and
\beq
\psi_c(\xi,\eta) = \sqrt{\frac{\xi}{\eta}} \frac{\chi(\xi)}{\left(\xi^2-\eta^2\right)^{1/4}}. \label{psicxieta} \eeq
By doing in (\ref{intc}) the change of variable
\beq
t = N F_c(\xi,\eta), \label{tc} \eeq
and doing a partial integration as in the previous case we find that the large $N$ limit of $I_{(c)}[\chi]$ is given by:
\beq
I_{(c)}[\chi] = -\sqrt{\frac{2\xi}{ \pi N\eta}} \left. \frac{\chi(\xi)}{\left(\xi^2-\eta^2\right)^{1/4} F_c'(\xi,\eta)} e^{-NF_c(\xi,\eta)} \right|_{ \xi=\eta+\left(\frac{\Lambda'}{N}\right)^{2/3}\eta^{1/3}}^{\xi=\infty} 
\label{intc3}
. \eeq
The integration limit $\xi=\infty$ gives a vanishing contribution because in the exponential $F_c(\xi,\eta)$ goes to $+\infty$ as $\xi$ goes to $\infty$.
At the other limit of integration  we can replace $F_c(\xi,\eta)$ and $F_c'(\xi,\eta)$ with the first term of their asymptotic expansion in $\eta-\xi$, namely:
\beq
F_c(\xi,\eta) = -\frac{2^{3/2}}{3} \eta^{-1/2} \left(\xi -\eta \right)^{3/2} + O\left(\left(\xi -\eta \right)^{5/2} \right), \label{asyFc} \eeq
\beq
F_c'(\xi,\eta) = \sqrt{2} \eta^{-1/2} \left(\xi -\eta \right)^{1/2}+ O\left(\left(\xi -\eta \right)^{3/2} \right).\label{asyFc2} \eeq
By inserting (\ref{asyFc}) and (\ref{asyFc2}) into (\ref{intc3}) we get:
\beq
I_{(c)}[\chi] = \frac{\chi(\eta)}{2^{1/4}\sqrt{\pi \Lambda'}}e^{ -\frac{2^{3/2}}{3} \Lambda'} + \mbox{terms vanishing as} ~ N \rightarrow \infty,  \label{asyIc} \eeq 
which becomes smaller than any prescribed $\epsilon$ for sufficiently large $\Lambda'$.

%


\begin{thebibliography}{99}


\bibitem{NiNi}
H.B. Nielsen and M. Ninomiya, Absence Of Neutrinos On A Lattice. 1. Proof By Homotopy Theory,
\textit{Nucl}.\ \textit{Phys}.\ \textbf{B 185} (1981) 20 [Erratum-ibid. \textbf{B 195} (1982) 541].
%
\bibitem{NiNi2}
H.B. Nielsen and M. Ninomiya, Absence Of Neutrinos On A Lattice. 2. Intuitive Topological Proof,
\textit{Nucl}.\ \textit{Phys}.\ \textbf{B 193} (1981) 173.
%
\bibitem{KarSm} L.H. Karsten and J. Smit, Lattice Fermions :Species Doubling, Chiral Invariance, 
And The Triangle Anomaly,
\textit{Nucl}.\ \textit{Phys}.\ \textbf{B 183} (1981) 103. 
%
\bibitem{chiral-ferm} 
P. Hasenfratz, \textit{Nucl}.\ \textit{Phys}.\ \textbf{B525} (1988) 401.
%
\bibitem{chiral-ferm2} 
H. Neuberger, \textit{Phys}.\ \textit{Lett}.\ \textbf{417B} (1988) 141.
%
\bibitem{chiral-ferm3} 
H. Neuberger, \textit{Phys}.\ \textit{Lett}.\ \textbf{427B} (1988) 353.
%
\bibitem{chiral-ferm4} 
M. L\"{u}scher, \textit{Phys}.\ \textit{Lett}.\ \textbf{428B} (1988) 342.
%
\bibitem{GW-relation}
P. H. Ginsparg and K. G. Wilson, A Remnant of Chiral Symmetry on the Lattice, 
\textit{Phys}.\ \textit{Rev}.\ \textbf{D25} (1982) 2649.
%
\bibitem{Neuberger}
N. Neuberger, Bounds on the Wilson Dirac Operator, 
\textit{Phys}.\ \textit{Rev}.\ \textbf{D61} (2000) 085015.
%
\bibitem{HJL}
P. Hernandes, K. Jansen and M. L\"{u}scher, Locality properties of Neuberger's 
lattice Dirac operator, 
\textit{Nucl}.\ \textit{Phys}.\ \textbf{B552} (1999) 363.
%
\bibitem{Dondi-Nicolai} P. Dondi and H. Nicolai, Lattice Supersymmetry, \textit{Nuovo Cimento} A 41 
(1977) 1.
%
\bibitem{Nojiri}
 S.~Nojiri,
 \textit{Prog}.\ \textit{Theor}.\ \textit{Phys}.\ {\bf 74} (1985) 819.
%
\bibitem{Nojiri2}
 S.~Nojiri,
 \textit{Prog}.\ \textit{Theor}.\ \textit{Phys}.\ {\bf 74} (1985) 1124.
%
\bibitem{Fujikawa1}
 K.~Fujikawa,
``Supersymmetry on the lattice and the Leibniz rule,''
\textit{Nucl}.\ \textit{Phys}.\  B {\bf 636} (2002) 80, 
[arXiv:hep-th/0205095].
%
\bibitem{Kato-Sakamoto-So1} 
M. Kato, M. Sakamoto and H. So, Taming the Leibniz Rule on the Lattice, 
\textit{JHEP} 05 (2008) 057, [arXiv:hep-lat/0803.3121].
%
\bibitem{Bergner-Bruckmann} G. Bergner, F. Bruckmann and J. M. Pawlowski, 
Generalising the Ginsparg-Wilson relation: Lattice Supersymmetry from Blocking 
Transformations, \textit{Phys}.\ \textit{Rev}.\ \textbf{D79} (2009) 115007. 
%
\bibitem{SLAC}
S. D. Drell, M. Weinstein and S. Yankielowicz, Strong Coupling Field Theories. 2. Fermions and 
Gauge Fields on a Lattice, \textit{Phys}.\ \textit{Rev}.\ \textbf{D14} (1976) 1627.
%
\bibitem{DFKKS} A. D'Adda, A. Feo, I. Kanamori, N. Kawamoto and J. Saito, 
Species Doublers as Super Multiplets in Lattice Supersymmetry: Exact Supersymmetry with interactions for $D=1, N=2$,
\textit{JHEP} 1009 (2010) 059, [arXiv:1006.2046].
%
\bibitem{DKKS} A. D'Adda, I. Kanamori, N. Kawamoto and J. Saito, 
Species Doublers as Super Multiplets in Lattice Supersymmetry: Chiral Conditions of Wess-Zumino Model for $D=N=2$, 
\textit{JHEP} 1203 (2012) 043, [arXiv:1107.1629].
%
\bibitem{Nicolai-map}
H. Nicolai, 
On a New Characterization of Scalar Supersymmetric Theories, 
\textit{Phys}.\ \textit{Lett}.\ \textbf{89B} (1980) 341.
%
\bibitem{Nicolai-map2}
H. Nicolai, 
N. Sakai and M. Sakamoto, 
Lattice Supersymmetry and the Nicolai Mapping, 
\textit{Nucl}.\ \textit{Phys}.\ \textbf{B229} (1983) 173.
%
\bibitem{Kaplan1} D.B. Kaplan, E. Katz and M. Unsal, Supersymmetry on a spatial lattice, \textit{JHEP} 0305 (2003) 037, [arXiv:hep-lat/0206019].
%
\bibitem{Kaplan1-2}
	A.G. Cohen, D.B. Kaplan, E. Katz and M. Unsal, Supersymmetry on a Euclidean spacetime lattice. I: A target theory with four supercharges,
	\textit{JHEP} 0308 (2003) 024, [arXiv:hep-lat/0302017].
%
\bibitem{Kaplan1-3}
	A.G. Cohen, D.B. Kaplan, E. Katz and M. Unsal, Supersymmetry on a Euclidean spacetime lattice. II: Target theories with eight supercharges,
	\textit{JHEP} 0312 (2003) 031, [arXiv:hep-lat/03070120].
%
\bibitem{phyrep} 
S. Catterall, D.B. Kaplan and M. \"{U}nsal, Exact lattice supersymmetry, \textit{Phys}.\ \textit{Rept}.\ \textbf{484} (2009) 71, [arXiv:0903.4881[hep-lat]].
%
\bibitem{Catterall1}
S. Catterall and S. Karamov, 
Exact lattice supersymmetry: The Two-dimensional N=2 Wess-Zumino model, 
\textit{Phys}.\ \textit{Rev}.\ \textbf{D65} (2002) 094501, [arXiv:hep-lat/0108024].
%
\bibitem{Sugino}
F. Sugino, A lattice formulation of super Yang-Mills theories with exact supersymmetry, 
\textit{JHEP} 0401 (2004) 015 [arXiv:hep-lat/0311021].
%
\bibitem{Sugino2}
	F. Sugino, Super Yang-Mills theories on the two-dimensional lattice with exact supersymmetry, \textit{JHEP} 0403 (2004) 067, [arXiv:hep-lat/0401017]; 
%
\bibitem{Sugino3}
	F. Sugino, Various super Yang-Mills theories with exact supersymmetry on the lattice , 
	\textit{JHEP} 0501 (2005) 016, [arXiv:hep-lat/0410035]. 
%
\bibitem{DKKN}
A.~D'Adda, I.~Kanamori, N.~Kawamoto and K.~Nagata,
Twisted superspace on a lattice,
Nucl.\ Phys.\ {\bf B707} 100 (2005)
[arXiv:hep-lat/0406029].
\bibitem{DKKN2}
A.~D'Adda, I.~Kanamori, N.~Kawamoto and K.~Nagata,
Exact extended supersymmetry on a lattice: Twisted N = 2 super  Yang-Mills
in two dimensions,
\textit{Phys}.\ \textit{Lett}.\  B {\bf 633}, 645 (2006)
[arXiv:hep-lat/0507029].
\bibitem{DKKN3}
A.~D'Adda, I.~Kanamori, N.~Kawamoto and K.~Nagata,
Exact Extended Supersymmetry on a Lattice: Twisted N=4 Super Yang-Mills in
Three Dimensions,
\textit{Nucl}.\ \textit{Phys}.\  B {\bf 798}, 168 (2008)
[arXiv:0707.3533 [hep-lat]].
\bibitem{Bruckmann-Kok} F. Bruckmann and M. de Kok, Noncommutativity approach to supersymmetry on the lattice: 
SUSY quantum mechanics and an inconsistency, \textit{Phys}.\ \textit{Rev}.\ \textbf{D73} (2006) 074511, [arXiv:hep-lat/0603003].
\bibitem{Bruckmann-Kok2}
F. Bruckmann, S. Catterall and M. de Kok, Critique of the link approach to exact lattice supersymmetry, 
\textit{Phys}.\ \textit{Rev}.\ \textbf{D 75} (2007) 045016, [arXiv:hep-lat/0611001].
\bibitem{Dadda-Kawamoto-Saito1} A. D'Adda, N. Kawamoto and J. Saito, Formulation of Supersymmetry on a Lattice as a Representation of a Deformed Superalgebra,
\textit{Phys}.\ \textit{Rev}.\ \textbf{D 81} (2010) 065001, [arXiv:0907.4137].
%
%
\bibitem{Damgaard-Matsuura} P. H. Damgaard and S. Matsuura, 
Classification of Supersymmetric Lattice Gauge Theories by Orbifolding,
\textit{JHEP} {\bf 0707} (2007) 051
[arXiv:0704.2696 [hep-lat]].
%
\bibitem{Damgaard-Matsuura2}
P.~H.~Damgaard and S.~Matsuura,
Relations among Supersymmetric Lattice Gauge Theories via Orbifolding,
\textit{JHEP} {\bf 0708} (2007) 087
[arXiv:0706.3007 [hep-lat]].
%
\bibitem{Damgaard-Matsuura3}
P.~H.~Damgaard and S.~Matsuura,
Lattice Supersymmetry: Equivalence between the Link Approach and
Orbifolding,
\textit{JHEP} {\bf 0709} (2007) 097
[arXiv:0708.4129 [hep-lat]].
%
\bibitem{Damgaard-Matsuura4}
P.~H.~Damgaard and S.~Matsuura,
Geometry of Orbifolded Supersymmetric Lattice Gauge Theories,
\textit{Phys}.\ \textit{Lett}.\  B {\bf 661} (2008) 52
[arXiv:0801.2936 [hep-th]].
%
\bibitem{twist-susy}
N. Kawamoto and T. Tsukioka, 
N=2 supersymmetric model with Dirac-Kaehler fermions from generalized gauge theory in 
two-dimensions,
\textit{Phys}.\ \textit{Rev}.\ \textbf{D 61} (2000) 105009, 
[hep-th/0502119].
%
\bibitem{twist-susy2}
J. Kato, N. Kawamoto and Y. Uchida, 
Twisted superspace for N=D=2 super BF and Young-Mills with Dirac-Kahler fermion mechanism, \textit{Int}.\ \textit{J}. \textit{Mod}.\ \textit{Phys}.\ \textbf{A 19} (2004) 2149-2182.
%
\bibitem{twist-susy3}
J. Kato, N. Kawamoto and A. Miyake, 
N=4 twisted superspace from Dirac-Kahler twist and off-shell SUSY invariant 
actions in four dimensions, \textit{Nucl}.\ \textit{Phys}.\ \textbf{B 721} (2005) 229-286. 
%
\bibitem{Kato-Sakamoto-So2}
M. Kato, M. Sakamoto and H. So, A criterion for lattice supersymmetry: cyclic Leibniz rule, 
\textit{JHEP} 1305 (2013) 089, [arXiv:hep-lat/1303.4472].
%
\bibitem{So-Ukita}
T. Aoyama and Y. Kikukawa, Overlap formula for the chiral multiplet, 
\textit{Phys}.\ \textit{Rev}.\ {\bf D59}(1999)054507.
%
\bibitem{So-Ukita2}
H. So and N. Ukita, Ginsparg-Wilson relation and lattice supersymmetry, 
\textit{Phys}.\ \textit{Lett}.\ \textbf{B457} (1999) 314. 
%
%
\bibitem{Kikukawa-Nakayama}
Y. Kikukawa and Y. Nakayama, 
Nicolai mapping versus exact chiral symmetry on the lattice, 
\textit{Phys}.\ \textit{Rev}.\ \textbf{D66} (2002) 094508, [arXiv:hep-lat/0207013]. 
%
\bibitem{Bartels-Kramer} J. Bartels and G. Kramer, A lattice version of the wess-zumino model, \textit{Z}.\ \textit{Phys}.\ C 20 (1983) 159.
%
\bibitem{GKPY} J. Giedt, R. Koniuk, E. Poppitz and T. Yavin, Less naive about supersymmetric lattice quantum mechanics, 
\textit{JHEP} 0412 (2004) 033, [arXiv:hep-th/0410041].
%
\bibitem{Kadoh-Suzuki}
D. Kadoh and H. Suzuki, 
Supersymmetric nonperturbative formulation of the WZ model in lower dimensions, 
\textit{Phys}.\ \textit{Lett}.\ \textbf{B684} (2010) 167, [arXiv:hep-th/0909.3686]. 
%
\bibitem{Bergner}
  G.~Bergner,
  Complete supersymmetry on the lattice and a No-Go theorem: A simulation
  with intact supersymmetries on the lattice,
  \textit{JHEP} {\bf 1001} (2010) 024, 
  [arXiv:hep-lat/0909.4791].
%
\bibitem{WTid}
K. Asaka , A. D'Adda, N. Kawamoto, Y. Kondo, 
Exact lattice supersymmetry at the quantum level for N=2 Wess–Zumino models 
in 1- and 2-dimensions,   
\textit{Int}.\ \textit{J}.\ \textit{Mod}.\ \textit{Phys}.\ A31 (2016) 1650125, [arXiv:hep-lat/1607.04371]. 
%
\bibitem{Gudermann}
Abramowitz and Stegun, ``Handbook of Mathematical Functions'', p.365, 366.
%
\bibitem{Hasenfratz}
P. Hasenfratz, 
Lattice QCD without tuning, mixing and current renormalization, 
\textit{Nucl}.\ \textit{Phys}.\ \textbf{B525} (1998) 401, 
[arXiv:hep-lat/9802007].

\bibitem{Wess:1974tw}
  J.~Wess and B.~Zumino,
  \textit{Nucl}.\ \textit{Phys}.\ B {\bf 70} (1974) 39.
  doi:10.1016/0550-3213(74)90355-1
%
\bibitem{Bietenholz}
W. Bietenholz, 
Exact supersymmetry on the lattice,
Mod. \textit{Phys}.\ \textit{Lett}.\ \textbf{A14} (1999) 51.
%
\bibitem{Bietenholz:1995kr}
  W.~Bietenholz and U.~J.~Wiese,
  \textit{Nucl}.\ \textit{Phys}.\ \textit{Proc}.\ \textit{Suppl}.\  {\bf 47} (1996) 575
  doi:10.1016/0920-5632(96)00125-9
  [hep-lat/9509052].
\bibitem{Bietenholz:1997sf}
  W.~Bietenholz,
  In *Buckow 1997, Theory of elementary particles* 466-482
  [hep-lat/9802014].


\end{thebibliography}
\end{document}